\documentclass[fleqn,usenatbib]{mnras}

\usepackage{graphicx}	% Including figure files
\usepackage{amsmath}	% Advanced maths commands
\usepackage{amssymb}	% Extra maths symbols
\usepackage{placeins}
\usepackage{gensymb}
\usepackage{caption}
\usepackage[labelfont=bf]{caption} 
\usepackage{lscape}

\def\SPSB#1#2{\rlap{\textsuperscript{\textcolor{black}{#1}}}\SB{#2}}
\def\SP#1{\textsuperscript{\textcolor{black}{#1}}}
\def\SB#1{\textsubscript{\textcolor{black}{#1}}}
\title[\Large{Angular Momentum in High-$z$ Star--Forming Galaxies}]
      {The Dynamics and Distribution of Angular Momentum in HiZELS Star\,--\,Forming Galaxies at $z$\,=\,0.8\,--\,3.3}

\author[S. Gillman et al.]{S. Gillman,$^{1}$\thanks{E-mail: steven.r.gillman@durham.ac.uk}
A. M. Swinbank,$^{1,2}$
A. L.  Tiley,$^{1}$
C. M. Harrison,$^{3}$
 Ian Smail,$^{1,2}$ 
\and U. Dudzevi\v{c}i\={u}t\.{e},$^{1}$
R. M. Sharples,$^{1,4}$
P. N. Best,$^{5}$
R. G. Bower$^{1,2}$
R. Cochrane,$^{5,6}$
\and D. Fisher,$^{7}$
J. E. Geach,$^{8}$
K. Glazebrook,$^{7}$
Edo Ibar,$^{9}$
J. Molina,$^{10}$
\and D. Obreschkow,$^{11,12}$
M. Schaller,$^{2,13}$
D. Sobral,$^{14}$
S. Sweet,$^{7}$
J. W. Trayford$^{2,13}$
\and T. Theuns$^{2}$
\\
$^{1}$Centre for Extragalactic Astronomy, Durham University, South Road, Durham, DH1 3LE UK\\
$^{2}$Institute for Computational Cosmology, Durham University, South Road, Durham DH1 3LE UK\\
$^{3}$European Southern Observatory, Karl-Schwarzschild-Str. 2, 85748 Garching bei München, Germany \\
$^{4}$Centre for Advanced Instrumentation, Durham University, South Road, Durham DH1 3LE UK\\
$^{5}$SUPA, Institute for Astronomy, Royal Observatory Edinburgh, EH9 3HJ, UK \\
$^{6}$Isaac Newton Group of Telescopes, E-38700 Santa Cruz de La Palma, Canary Islands, Spain \\
$^{7}$Centre for Astrophysics and Supercomputing, Swinburne University of Technology, PO Box 218, Hawthorn, VIC 3122, Australia \\
$^{8}$School of Physics, Astronomy $\&$ Mathematics, University of Hertfordshire, College Lane, Hatfield, AL10 9AB, UK \\
$^{9}$Instituto de F\'isica y Astronom\'ia, Universidad de Valpara\'iso, Avda. Gran Breta\~na 1111, Valpara\'iso, Chile\\
$^{10}$Departamento de Astronomía, Universidad de Chile, Casilla 36-D, Santiago, Chile\\
$^{11}$International Centre for Radio Astronomy Research (ICRAR), University of Western Australia,
Crawley WA 6009, Australia \\
$^{12}$Australian Research Council Centre of Excellence for All-Sky Astrophysics (CAASTRO), 44 Rosehill Street Redfern, NSW 2016, Australia \\
$^{13}$Leiden Observatory, Leiden University, P.O. Box 9513, 2300 RA Leiden, The Netherlands \\
$^{14}$Department of Physics, Lancaster University, Lancaster, LA1 4BY, UK \\
}
\date{Accepted 2019 March 7. Received 2019 March 7; in original form 2018 December 14}

\pubyear{2018}

\begin{document}
\label{firstpage}
\pagerange{\pageref{firstpage}--\pageref{lastpage}}
\maketitle

\begin{abstract}
We present adaptive optics assisted integral field spectroscopy of 34
star--forming galaxies at $z$\,=\,0.8--3.3 selected from the HiZELS narrow-band survey.
We measure the kinematics
of the ionised interstellar medium on $\sim$1 kpc scales, and show that
the galaxies are turbulent, with a median ratio of rotational to
dispersion support of $V$\,/\,$\sigma$\,=\,0.82\,$\pm$\,0.13.  We
combine the dynamics with high-resolution rest-frame optical
imaging and extract emission line rotation curves. 
We show that high--redshift star--forming galaxies follow a 
similar power-law trend in specific angular momentum with stellar mass as that of local late type galaxies.
We exploit the high resolution of our data and examine the radial distribution of angular momentum within each galaxy by constructing total angular momentum profiles.  Although the stellar mass of a typical star--forming galaxy is expected to grow by a factor $\sim$\,8 in the $\sim$5 Gyrs between $z$\,$\sim$\,3.3  and $z$\,$\sim$\,0.8, we show that  the internal distribution of angular momentum becomes less centrally concentrated in this period i.e the angular momentum  grows outwards.   
To interpret our observations, we exploit the EAGLE
simulation and trace the angular momentum evolution of star--forming
galaxies from $z$\,$\sim$\,3 to $z$\,$\sim$\,0, identifying a similar trend of
decreasing angular momentum concentration.  This change is attributed
to a combination of gas accretion in the outer disk, and feedback
that preferentially arises from the central regions of the galaxy.  We
discuss how the combination of the growing bulge and angular momentum
stabilises the disk and gives rise to the Hubble sequence.
\end{abstract}

\begin{keywords}
galaxies: evolution -- galaxies: high redshift -- galaxies: kinematics and dynamics
\end{keywords}

%%%%%%%%%%%%%%%%%%%%%%%%%%%%%%%%%%%%%%%%%%%%%%%%%%

%%%%%%%%%%%%%%%%% BODY OF PAPER %%%%%%%%%%%%%%%%%%

\section{Introduction}

The galaxy population in the local Universe is dominated by two
distinct populations, with $\sim$70 per cent spirals, and
$\sim$25 per cent spheroidal and elliptical galaxies \citep{Abraham2001}.  These two populations
make up the long-defined classes of the Hubble sequence defined as late-- and early--type galaxies
\citep{Hubble1926,Sandage1986}. The differences are also reflected in
many properties, including the galaxy integrated colours, star formation rates, rotation velocity, and velocity dispersion
\citep[e.g.][]{Tinsley1980,Kauffmann2003,Delgado2010,Zhong2010,Whitaker2012,Aquino2018,Eales2018}

The two populations can be separated fundamentally
by differences in the baryonic angular momentum.  In a lambda cold dark matter $(\Lambda$CDM) Universe
angular momentum originates from tidal torques between dark matter
haloes in the early Universe \citep{Hoyle1956}.  The amount of halo angular momentum acquired
has a strong dependence on the halo mass (J\,$\propto$\,M$^{5/3}_{\rm halo}$) as predicted from tidal torque theory, as well as the epoch of formation (J\,$\propto$\,t)  \citep[e.g.][]{Catelan1996}. As the baryonic material within the
halo cools and collapses, it should weakly (within a factor of 2)
conserve angular momentum, due to tensor invariance, and form a star--forming disc. Subsequent gas accretion, star formation and feedback will redistribute the angular momentum within
the disc, whilst mergers will preferentially remove angular momentum
from the system  \citep{Mo1998}. 

\cite{Fall1980} demonstrated that the baryons in today's 
spiral galaxies must have lost $\sim$\,30 per cent of their initial angular momentum,
most likely through secular processes and viscous angular momentum redistribution 
\citep{Bertola1975,Burkert2009,Romanowsky2012}. In contrast, in early types (spheroids) 
the initial angular momentum of the baryons must have been redistributed (or lost) to 
the halo, most efficiently through major mergers.
As first suggested by \cite{Fall1983}, stellar angular
momentum in galaxies is predicted to follow a power-law scaling between
specific stellar angular momentum ($j_\star$\,=\,J$_{*}$/M$_{*}$) and stellar mass (M$_\star$) where local spiral galaxies follow a scaling with
$j_\star\propto M_\star^{2/3}$ 
%whereas early-type galaxies lie significantly below this trend 
%due to their high velocity dispersion and low rotation velocities
\citep[e.g.][]{Romanowsky2012,Cortese2016}.  

Recent studies of low-redshift galaxies have expanded upon these works showing that 
the specific angular momentum and mass also correlate with total bulge
to disc ratio (B/T) of the galaxy \citep[e.g.][]{Glazebrook2014,Fall2018,Sweet2018}. 
Indeed, galactic discs and spheroidal galaxies occupy independent regions of the $j_\star$--M$_\star$--B/T
plane, suggesting they were formed via distinct physical processes.
Major mergers play a minimal role in disc galaxies' evolution, whilst
elliptical galaxies' histories are often dominated by major mergers,
stripping the galaxy of gas required for star formation and disc
creation, as shown in observational studies \citep{Cortese2016,Posti2018,Rizzo2018} 
and hydro-dynamical simulations \citep{Lagos2017,Trayford2018}.

Two of the key measurements required to follow the formation of 
today's disc galaxies are: how is the angular momentum within a baryonic galaxy
(re)distributed; and which physical processes drive the evolution such 
that the galaxies evolve from turbulent systems at high redshift into 
rotation-dominated, higher angular momentum, low redshift galaxies.

At high redshift star--forming galaxies are clumpy and turbulent,
and whilst showing distinct velocity gradients \citep[e.g.][]{Forster2009,Forster2011,Wisnioski15},
they are typically dominated by `thick' discs and irregular morphologies. Morphological surveys \citep[e.g.][]{Conselice2011,Elmegreen2014}, as well as hydro-dynamical 
simulations \citep[e.g.][]{Trayford2018} highlight that a critical epoch 
in galaxy evolution is $z$\,$\sim$\,1.5. This is when the 
spiral galaxies (that would lie on a traditional Hubble classification)  become as common as peculiar galaxies.
If one of the key elements that dictate the morphology of a galaxy is angular momentum, as suggested by the studies of local galaxies  \citep[e.g.][]{Shibuya2015,Cortese2016,Elson2017}  then this would imply that this is the epoch when the internal angular momentum of star--forming 
galaxies is becoming sufficiently high to stabilize the disc \citep{Mortlock2013}.
%so that star--forming galaxies at the present day are neatly classified using the Hubble diagram. 

Observationally we can test whether the emergence of galaxy morphology at 
this epoch is driven by the increase in the specific angular momentum of
the young stars and star--forming gas. A star--forming galaxy 
with a given rotation velocity but lower angular momentum will have
a smaller stellar disc and high surface density and assuming the gas
is Toomre unstable, the gaseous disc will have a higher Jeans mass \citep{Toomre1972}.
This results in more massive star--forming clumps, which can be observed
in the ionized-gas (e.g. H\,$\alpha$) morphology \citep[e.g.][]{Genzel2011,Livermore2012,NFS2014}.

%To constrain how the angular momentum within galaxy discs evolves
%with cosmic time, constraints at higher redshift are required.
Integral field spectroscopy studies of $z$\,=\,1\,--\,2 star--forming galaxies
also show that galaxies with 
increasing S\'ersic index have lower specific angular momentum, where 
sources with the highest specific angular momentum, for a given mass,
have the most disc-dominated morphologies \citep[e.g.][]{Burkert2016,Swinbank2017,Harrison2017}.
%Since most of the galaxies at this epoch 
%are rotationally supported, it must be the internal redistribution of 
%angular momentum and bulge formation that causes the morphology to evolve . 
Measuring the resolved dynamics of galaxies at high redshift on $\sim$\,1\,kpc scales allows us to go beyond a measurement of size and asymptotic rotation speed, examining the radial distribution of the angular momentum, comparing it to the distribution of the stellar mass. 

Numerical studies \citep[e.g.][]{Bosch2002,Lagos2017} further motivate the need to study the 
internal (re)distribution of angular momentum of gas discs with redshift, and suggest that 
the majority of the evolution occurs within the half stellar mass radius of the galaxy.
Resolving galactic discs on kpc scales in the distant Universe presents an observational 
challenge. At $z$\,$\sim$\,1.5 galaxies have smaller half--light
radii ($\sim$\,2\,--\,5\,kpc; \citealt{Ferguson2004,Stott2013}), which equate to
$\sim$\,0\farcs{2}\,--\,0.5$''$. The typical resolution of seeing-limited
observations is $\sim$\,0.7$''$. 
To measure the internal dynamics 
on kilo-parsec scales (which are required to derive the shape and 
normalization of the rotation curve within the disc, with minimal beam--smearing effects)
requires very high resolution, which, prior to the James Webb Space Telescope (JWST; \citealt{Garcia2018}), can only be achieved with adaptive optics.
The advent of adaptive optics (AO) integral field observations at high
redshift allows us to map the dynamics and distribution of star formation
on kpc scales in distant galaxies \citep[e.g.][]{Genzel2006,Cresci2007,Wright2007,Genzel2011,Swinbank2012b,Livermore2015,Molina2017,Schreiber2018,Circosta2018,Perna2018}.

%This is the epoch when spirals/spheroids become as common as peculiar galaxies, 
In this paper we investigate the dynamics and both total and radial distribution of angular momentum in high--redshift galaxies, and explore how this evolves with cosmic time. The data comprises of adaptive optics observations of 34 star--forming galaxies from 0.8\,$\leq$\,\emph{z}\,$\,\leq$\,3.3 observed with the OH-Suppressing Infrared Integral Field Spectrograph (OSIRIS; \citealt{Larkin2006}), the Spectrograph for INtegral Field Observations in the Near Infrared (SINFONI; \citealt{Bonnet2004a}), and the Gemini Northern Integral Field Spectrograph (Gemini-NIFS; \citealt{McGregor2003}). Our targets lie in the SA22 \citep{Steidal1998}, UKIDSS Ultra-Deep Survey (UDS; \citealt{Lawrence2007}), and Cosmological Evolution Survey (COSMOS; \citealt{Scoville2007}) extra-galactic fields (Appendix \ref{App:B}, Table \ref{Tab:Obs}). The sample brackets the peak in cosmic star formation and the high--resolution $\lesssim$0.1 arcsec observations allow the inner regions of the galaxies to be spatially resolved. Just over two--thirds of the sample have H\,$\alpha$ detections whilst the remaining third were detected at \emph{z}\,$\sim$\,3.3 via [O\,{\sc{iii}}] emission. All of the galaxies lie in deep extragalactic fields with excellent multiwavelength data, and the majority  were selected from the HiZELS narrow--band survey \citep{Sobral2013}, and have a nearby natural guide  or tip--tilt star to allow adaptive optics capabilities.

In Section \ref{Sec:Observations and Data Reduction} we describe the observations and the data reduction. In Section \ref{Sec:Analysis} we present the analysis used to derive stellar masses, galaxy sizes, inclinations, and dynamical properties. In Section \ref{Sec:AnD} we combine stellar masses, sizes, and dynamical measurements to infer the redshift evolution of the angular momentum in the sample. We derive the radial distributions of angular momentum within each galaxy and compare our findings directly to a stellar mass and star formation rate selected sample of {\sc{eagle}} galaxies. We discuss our findings and give our conclusions in Section \ref{Sec:Conc}. 

Throughout the paper, we use a cosmology with $\Omega_{\Lambda}$\,=\,0.73, $\Omega_{\rm m}$\,=\,0.30 and H$_{\rm 0}$\,=\,70\,km\,s$^{-1}$Mpc$^{-1}$ \citep{Planck2018}. In this cosmology a spatial resolution of 1 arcsecond corresponds to a physical scale of 8.25\,kpc at a redshift of $z$\,=\,2.2 (the median redshift of the sample.) All quoted magnitudes are on the AB system and stellar masses are calculated assuming a Chabrier IMF \citep{Chabrier2003}.

\section{Observations and Data Reduction} \label{Sec:Observations and Data Reduction}

The majority of the observations (31 targets; 90 percent of the sample)\footnote{Three galaxies are taken from the KMOS Galaxy Evolution Survey (KGES; Tiley et al, in prep), a sample of  $\sim$300 star--forming galaxies at $z$\,$\sim$\,1.5. Their selection was based on  H\,$\alpha$ detections in the KMOS observations and the presence of a tip--tilt star of $M_H$\,$<$14.5 within 40.0 arcsec of the galaxy to make laser guide star adaptive optics corrections possible.}, were obtained from follow--up spectroscopic observations of the High Redshift emission--line Survey (HiZELS; \citealt{Geach2008,Best2013}), which targets H\,$\alpha$--emitting galaxies in five narrow ($\Delta z$\,=\,0.03) redshift slices: $z$\,=\,0.40, 0.84, 1.47, 2.23 and 3.33 \citep{Sobral2013}. This panoramic survey provides a luminosity-limited sample of H\,$\alpha$ and [O\,{\sc{iii}}] emitters spanning $z$\,=\,0.4--3.3.

\begin{figure*}
	\centering
	\includegraphics[width=1\linewidth]{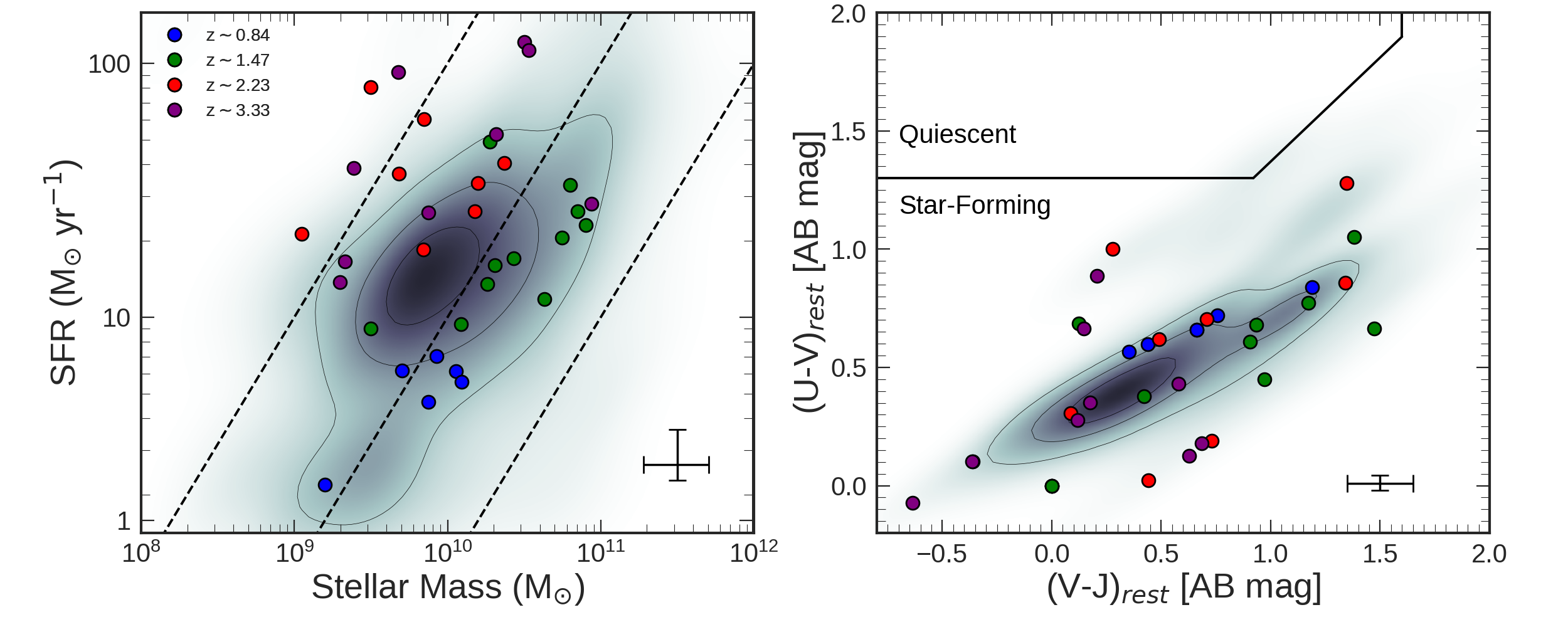}
	\caption{\emph{Left:} the H\,$\alpha$ and [O\,{\sc{iii}}] dust-corrected star formation rate of each galaxy as function of stellar mass derived from {\sc{magphys}}. The  HiZELS sample is shown as the grey shaded region whilst our sample is coloured by redshift. The adopted 0.2 dex stellar mass uncertainty and median fractional star formation rate uncertainties are indicated by black lines. We show tracks  of constant specific star formation rate (sSFR) with sSFR = 0.1, 1, and 10\,Gyr$^{-1}$. This shows that our sample covers a broad range of stellar mass and star formation rates. %but are consistent with sSFR$\sim$0.1\,--\,10\,Gyr$^{-1}$. 
	\emph{Right:} the rest-frame ($U-V$) colour as a function of rest-frame ($V-J$) colour for our sample and galaxies in the HiZELS survey, demonstrating that the galaxies in our sample cover the full range of HiZELS galaxy colour-colour parameter space. Median uncertainties in ($V-J$) and ($U-V$) colour are indicated by black lines. The \citet{Williams2009} boundary (black wedge) separates quiescent galaxies (top left) from star--forming galaxies (bottom right).}
	\label{Fig:MS}
\end{figure*}

Exploiting the wide survey area, the targets from the HiZELS survey were selected to lie within 25.0 arcsec of a natural guide star to allow for adaptive optics capabilities. The sample spans the full range of the rest-frame ($U-V$) and rest-frame ($V-J$) colour space as well as the stellar mass and star formation rate plane of the HiZELS parent sample (Appendix \ref{App:A}, Table \ref{Table:Sample}, and Figure \ref{Fig:MS}). The data were collected from 2012 August to 2017 December from a series of observing runs on SINFONI (VLT), NIFS (Gemini North Observatory), and OSIRIS (Keck) integral field spectrographs (see Appendix \ref{App:B}, Table \ref{Tab:Obs} for details).

Our sample includes the galaxies first studied by \cite{Swinbank2012a} and \cite{Molina2017}, who analysed the dynamics and metallicity gradients in 20 galaxies from our sample. In this paper we build upon this work and include 14 new sources, of which 9 galaxies are at $z$\,$>$\,3. We also combine observations of the same galaxies from different spectrographs in order to maximize the signal to noise of the data.

\subsection{VLT/SINFONI}
To map the H\,$\alpha$  and [O\,{\sc{iii}}] emission in the galaxies in our sample, we undertook a series of observations using the Spectrograph for INtegral Field Observations in the Near Infrared (SINFONI; \citealt{Bonnet2004a}). SINFONI is an integral field spectrograph mounted at the Cassegrain focus of UT4 on the VLT and can be used in conjunction with
a curvature sensing adaptive optics module (MACAO; \citealt{Bonnet2004}).
SINFONI's wavelength coverage is from 1.1\,--\,2.45$\mu$m, which is ideally suited for mapping high redshift H\,$\alpha$ and [O\,{\sc{iii}}] emission.

SINFONI employs an image slicer and mirrors to reformat a field of 3.0 arcsec $\times$ 3.0 arcsec with a pixel scale of 0.05 arcsec. At $z$\,=\,0.84, 1.47, and 2.23 the H\,$\alpha$ emission--line is redshifted to $\sim$\,1.21$\mu$m, 1.61$\mu$m, and 2.12$\mu$m, into the $J$, $H$, and $K$ bands, respectively. The [O\,{\sc{iii}}] emission--line at $z$\,$\sim$\,3.33 is in the $K$ band at 2.16$\mu$m. The spectral resolution in each band is $\lambda / \Delta \lambda\,\sim$\,4500. Each observing block (OB) was taken in an ABBA observing pattern (A\,=\,Object frame, B\,=\,Sky frame)  with 1.5 arcsec chops to sky, keeping the target in the field of view. We undertook observations between 2009 September 10 and 2016 August 01 
with total exposure times ranging from 3.6ks to 13.4ks (Appendix \ref{App:B}, Table \ref{Tab:Obs}) where each individual exposure was 600s. All observations were carried out in dark time with good sky transparency and with a closed--loop adaptive optics correction using natural guide stars.

In order to reduce the SINFONI data the ESOREX pipeline was used to extract, wavelength calibrate, and flat--field each spectra and form a data cube from each observation. The final data cube was generated by aligning the individual observing blocks, using the continuum peak, and then median combining them and sigma clipping the average at the 3$\sigma$ level to reject pixels with cosmic ray contamination. For flux calibration, standard stars were observed each night either immediately before or after the science exposures. These were reduced in an identical manner to the science observations.

\subsection{Gemini/NIFS}
The Gemini Northern Integral Field Spectrograph (Gemini-NIFS; \citealt{McGregor2003}) is a single object integral field spectrograph mounted on the 8\,m Gemini North telescope, which we used in conjunction with the adaptive optics system ALTAIR. NIFS has a 3.0 arcsec $\times$ 3.0 arcsec field of view and 
an image slicer which divides the field into 29 slices with angular sampling of 0.1 arcsec $\times$ 0.04 arcsec. The dispersed spectra from the slices are reformatted on the detector to provide two-dimensional spectra imaging using the $K$--band grism covering a wavelength range of 2.00\,--\,2.43$\mu$m. All of our observations were undertaken using an ABBA sequence in which the `A' frame is an object frame and the `B' frame is a 6 arcsecond chop to blank sky to enable sky subtraction. Individual exposures were 600s and each observing block 3.6ks, which was repeated four times resulting in a total integration time of 14.4ks per target.

The NIFS observations were reduced with the standard Gemini IRAF NIFS pipeline which includes extraction, sky-subtraction, wavelength calibration and flat-fielding. Residual OH sky emission lines were removed using sky subtraction techniques described in \cite{Davies2007}. The spectra were then flux calibrated by interpolating a black body function to the spectrum of the telluric standard star. Finally data cubes for each individual exposure were created with an angular sampling of 0.05 arcsec $\times$ 0.05 arcsec. These cubes were then mosaicked using the continuum peak as reference and median combined to produce a single final data cube for each galaxy. The average Full Width Half Maximum (FWHM) of the point spread function (PSF) measured from the telluric standard star in the NIFS data cubes is 0.13 arcsec with a spectral resolution of $\lambda / \Delta \lambda \sim$ 5290.

The three galaxies in our sample observed with NIFS also have SINFONI AO observations. We stacked the observations from different spectrographs, matching the spectral resolution of each, in order to maximize the signal to noise. In the stacking procedure, each observation was weighted by its signal to noise. The galaxy SHIZELS--21 is made up of two NIFS (14.6ks, 15.6ks) and one SINFONI (9.6ks) observation  whilst SHIZELS--23 and SHIZELS--24 are the median combination of one NIFS (15.6ks) and one SINFONI (12.0ks) observation. On average the median signal to noise per pixel increased by a factor of $\sim$\,2 as a result of stacking the frames and the redshift of the H\,$\alpha$ emission lines in the individual and stack data cubes agreed to within $\leq$0.01 per cent.

\subsection{Keck/OSIRIS}
We also include in our sample three galaxies observed with the OH-Suppressing Infrared Integral Field Spectrograph (OSIRIS; \citealt{Larkin2006}), which are stellar mass, star formation rate and kinematically selected based on the KMOS observations, from the KGES survey (Tiley et al. 2019, Gillman et al. in prep.). The OSIRIS spectropgraph is a lenslet integral field unit that uses the Keck Adaptive Optics System to observe from 1.0\,--\,2.5$\mu$m on the 10\,m Keck I Telescope. The AO correction is achieved using a combination of a Laser Guide Star (LGS) and Tip--Tilt Star (TTS) to correct for atmospheric turbulence down to 0.1 arcsec resolution in a rectangular field of view of order 4 arcsec $\times$ 6 arcsec \citep{Wizinowich2006}.

Observations were carried out on 2017 December 06 and 07. Each exposure was 900s, dithering by 3.2 arcsec in the Hn4, Hn3, and Hn1 filters to achieve good sky subtraction while keeping the galaxy within the OSIRIS field of view. Each OB consists of two AB pairs and for each target a total of four AB pairs were observed equating to 7.2ks in total. Each AB was also jittered by pre--defined offsets to reduce the effects of bad pixels and cosmic rays. 

We used the OSIRIS data reduction pipeline version 4.0.0 using rectification matrices taken on 2017 December 14 and 15, to reduce the OSIRIS observations. The pipeline removes crosstalk, detector glitches, and cosmic rays per frame, to later combine the data into a cube. Further sky subtraction and masking of sky lines was also undertaken in targets close to prominent sky lines, following procedures outlined in \cite{Davies2007}. Each reduced OB was then centred, trimmed, aligned and stacked with other OBs to form a co--added fully reduced data cube of an object. On average each final reduced data cube was a combination of four OBs.

In total 25 H\,$\alpha$ and 9 [O\,{\sc{iii}}] detections were made using the SINFONI, NIFS and OSIRIS spectrographs from $z$\,$\sim$\,0.8\,--\,3.33, full details of which is given in Appendix \ref{App:A}, Table \ref{Table:Sample}. A summary of the observations are given in Appendix \ref{App:B}, Table \ref{Tab:Obs} . 

\subsection{Point Spread Function Properties}
It is well known that the adaptive optics corrected point spread function diverges from a pure Gaussian profile  \citep[e.g.][]{Baena2011,Exposito2012,Schreiber2018}, with a non-zero fraction of power in the outer wings of the profile. In order to measure the intrinsic nebula emission sizes of the galaxies in our sample we must first construct the PSF for the integral field data using the standard star observations taken in conjunction with the science frames. We centre and median combine the standard star calibration images, deriving a median PSF for the $J$, $H$, and $K$ wavelength bands.

We quantify the the half-light radii of the these median PSFs using a three-component S\'ersic model, with S\'ersic indices fixed to be a Gaussian profile ($n$\,=\,0.5). The half-light radii, R$_{\rm h}$, of the PSF are derived using a curve-of-growth analysis on the three component S\'ersic model's two-dimensional light profile. We derive the median PSF  R$_{\rm h}$ for the $J$, $H$, and $K$ bands where  R$_{\rm h}$\,=\,0.18 arcsec \,$\pm$\,0.05 , 0.14 \,$\pm$\,0.03 and 0.09 \,$\pm$\,0/01 arcsec respectively. The integral field PSF half-light radii in kilo-parsecs are shown in Appendix \ref{App:B}, Table \ref{Tab:Obs}. We convolve half-light radii of the median PSF in each wavelength band with the intrinsic size of galaxies in our sample when extracting kinematic properties from the integral field data (e.g Section \ref{Sec:vdisp} and \ref{Sec:DynMod}). The median Strehl ratio achieved for our observations is 33 per cent  and the median encircled energy within 0.1 arcsec is 25 per cent (the approximate spatial resolution is 0.1 arcsec FWHM, 825\,pc at $z$\,$\sim$\,2.22, the median redshift of our sample).

\section{Analysis}\label{Sec:Analysis}
With the sample of 34 emission-line galaxies with adaptive optics assisted observations assembled, we first characterize the integrated properties of the galaxies. In the following section we investigate the stellar masses and star formation rates, sizes, dynamics, and their connection with the galaxy morphology, placing our findings in the context of the general galaxy population at these redshifts. We first discuss the stellar masses and star formation rates which we will also use in Section \ref{Sec:DynProp} when investigating how the dynamics evolve with redshift, stellar mass and star formation rate.

\subsection{Star Formation Rates and Stellar Masses}\label{Sec:MS}

Our targets are taken from some of the best--studied extragalactic fields with a wealth of ancillary photometric data available. This allows us to construct spectral energy distributions (SEDs) for each galaxy spanning from the rest-frame $UV$ to mid-infrared with photometry from the Ultra-Deep Survey \citep{Almaini2007}, COSMOS \citep{Muzzin2013} and SA22 \citep{Simpson2017}.

To measure the galaxy integrated properties we use the
{\sc{magphys}} code to fit the $UV$\,--\,8\,$\mu$m photometry \citep[e.g.][]{Cunha2008,daCunha2015}, from which we derive stellar masses and extinction factors (A$_{\rm v}$) for each galaxy. The full stellar mass range of our sample is $\log$(M$_{*}$[M$_{\odot}$])\,=\,9.0\,--\,10.9 with a median of $\log$(M$_{*}$[M$_{\odot}$])\,=\,10.1\,$\pm$\,0.2. We compare the stellar masses of our objects to those previously derived in \cite{Sobral2013}, finding a median ratio of 
M$^{\rm \textsc{magphys}}_{*}$\, / M$^{\rm \textsc{sobral}}_{*}$\,=\,1.07\,$\pm$\,0.23, 
indicating the {\sc{magphys}} stellar masses are slightly higher than those derived from simple interpretation of galaxy colours alone. However we employ a homogeneous stellar mass uncertainty of $\pm$0.2 dex throughout this work, which should conservatively account for the uncertainties in stellar mass values derived from SED fitting of high-redshift star--forming galaxies \citep{Mobasher2015}.

The star formation rates of $z$\,$<$\,3 galaxies in our sample were derived from the H\,$\alpha$ emission--line fluxes presented in \cite{Sobral2013}. We correct the H\,$\alpha$ flux assuming a stellar extinction of A$_{\rm H{\alpha}}$\,=\,0.37, 0.33, and 0.07 for $z$\,=\,0.84, 1.47, and 2.23, the median derived from {\sc{magphys}} SED fitting. Correcting to a Chabrier initial mass function and following \cite{Wuyts2013} to convert between stellar and gas extinction and the methods outlined  \cite{Calzetti2000}, we derive extinction corrected star formation rates for each galaxy. The uncertainties on the star formation rates are derived from bootstrapping the 1$\sigma$ uncertainties on the H\,$\alpha$ emission--line flux outlined in \cite{Sobral2013}. For the nine [O\,{\sc{iii}}] sources in our sample, we adopt the SFRs and uncertainties derived in \cite{Khostovan2015}.
  
The median SFR of our sample is $\langle$SFR$\rangle$=22\,$\pm$\,4 M$_{\odot}$yr$^{-1}$ with a range from SFR=2\,--\,120 M$_{\odot}$yr$^{-1}$. However, our observational flux limits mean that the median star formation evolves with redshift with $\langle$SFR$\rangle$\,=\,6\,$\pm$\,1, 13\,$\pm$\,5, 38\,$\pm$\,8 $\&$ 25\,$\pm$\,10 M$_{\odot}$yr$^{-1}$ for $z$\,=\,0.84, 1.47, 2.23, and 3.33. The median star formation rate of our H\,$\alpha$--detected galaxies is comparable, within uncertainties, to the knee of the HiZELS star formation rate function at each redshift (SFR$^{*}$) with SFR$^{*}$\,=\, 6, 10, and 25 M$_{\odot}$yr$^{-1}$ at $z$\,=\,0.84, 1.47, and  2.23, as presented in \cite{Sobral2014}.

The stellar masses and star formation rates for the sample are shown in Figure \ref{Fig:MS}. As a comparison we also show the HiZELS population star formation and stellar masses, derived in the same way, and tracks of constant specific star formation rate (sSFR) with sSFR = 0.1, 1, and 10\,Gyr$^{-1}$. A clear trend of increasing star formation rate at fixed stellar mass with redshift is visible. We note that the galaxies in our sample at $z$\,=\,1.47 typically have the highest stellar masses, and as shown by \cite{Cochrane2018}, the HiZELS population at $z$\,=\,1.47 is at higher L/L$^{*}$  than the $z$\,=\,0.84 or $z$\,=\,2.23 samples.
The star formation rate and stellar mass for each galaxy are shown in Appendix \ref{App:A}, Table \ref{Table:Sample}.
We also show the distribution of the rest-frame ($U-V$) colour as a function of the rest-frame ($V-J$) colour for our sample in Figure \ref{Fig:MS}. The HiZELS population is shown for comparison, indicating that our galaxies cover the full range of the HiZELS population colour distribution. Based on the above, we conclude that the galaxies in our sample at $z$\,=\,0.84, 2.23 and $\&$ 3.33 are representative of the SFR--stellar mass relation at each redshift, whilst galaxies at $z$\,=\,1.47 lie slightly above this relation.

\subsection{Galaxy Sizes}\label{Sec:Sizes}

Next we turn our attention to the sizes of the galaxies in our sample. 
All of the galaxies in the sample were selected from the extragalactic deep fields, either UDS, COSMOS or SA22. Consequently there is a wealth of ancillary broad--band data from which the morphological properties of the galaxy can be derived \citep{Stott2013,Paulino2017}. 
The observed near \,--\,infrared emission of a galaxy is dominated by the stellar continuum. At our redshifts, the observed near \,--\,infrared samples the rest frame 0.4\,--\,0.8\,$\mu$m 
emission and is always above the 4000$\AA$ break and so is less likely to be affected by sites of ongoing intense star formation. Therefore parametric fits to the near \,--\,infrared photometry are more robust than 
H\,$\alpha$ measurements for measuring the `size' of a galaxy. For just over half the sample (21 galaxies) we exploit \emph{HST} imaging, the majority of which is in the near-infrared (F140W,  F160W) or optical (F606W) bands at 0.12 arcsec resolution. The remainder is in the F814W band at 0.09 arcsec  resolution. All other galaxies, in SA22 and UDS, have ground based \emph{K}--band imaging with sampling of 0.13 arcsec per pixel  and a PSF  of 0.7 arcsec FWHM from the UKIRT Infrared Deep Sky Survey (UKIDDS; \citealt{Lawrence2007}).

To measure the observed stellar continuum size and galaxy morphology, we first perform parametric single S\'ersic fits to the broad--band photometric imaging of each galaxy. To account for the PSF of the image, we generate a PSF for each image from a stack of normalized unsaturated stars in the frame. 
We build two-dimensional S\'ersic models of the form
\begin{equation}\label{Eqn:Sersic}
{\rm I(R)=I_{\rm e}exp\left(- b_{n}\left[\left(\frac{R}{R_h}\right)^{(1/n)} - 1\right]\right)},
\end{equation} 
and use the MPFIT function \citep{Markwardt2009} to convolve the PSF and model in order to optimise the S\'ersic parameters including the axial ratio \citep{Sersic1963}.

\begin{figure}
	\centering
	\includegraphics[width=1\linewidth]{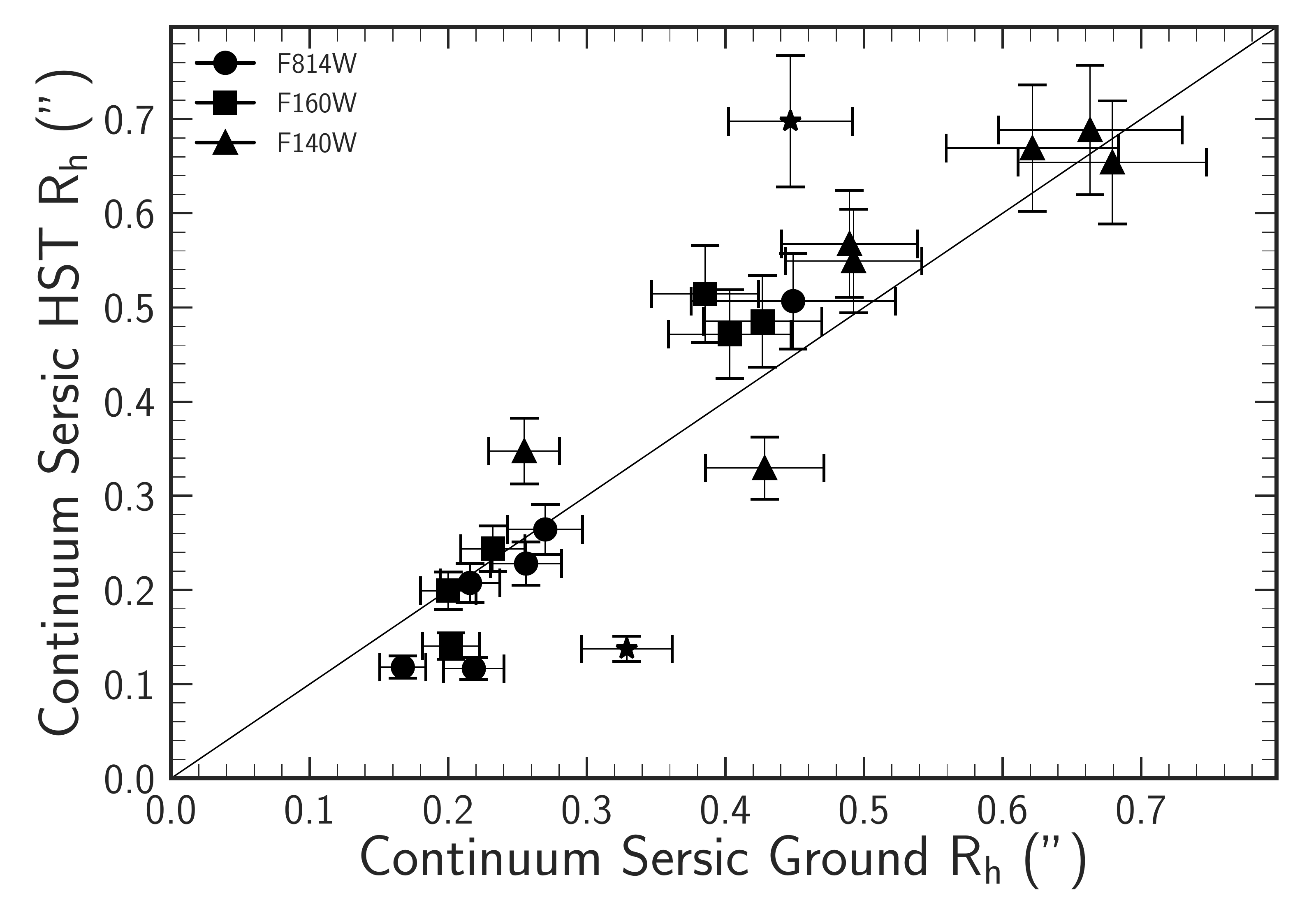}
	\caption{The half-light radius derived from S\'ersic function fits to both ground--based and \emph{HST} data in near--infrared bands, for 21 galaxies in our sample. The marker shape represents the \emph{HST} filter, star points indicate galaxies where ground and \emph{HST} photometry show different morphological features or defects. The majority of sizes show good agreement with $\langle$R\SPSB{G}{h}/R\SPSB{HST}{h}\quad$\rangle$=0.97 $\pm$ 0.05, independent of the band of the observation.}
	\label{Fig:sizes}
\end{figure}

Since the galaxies can be morphologically complex and to provide a non-parametric comparison to the S\'ersic half-light radii, we also derive half--light radii numerically within an  aperture two times the Petrosian radius (2R$_{\text{p}}$) of the galaxy. The Petrosian radius is derived by integrating the broadband image light directly and is defined by  R$_{\text{p}}$=1.5R$_{\eta=0.2}$ where R$_{\eta=0.2}$ is the radius (R) at which the surface brightness at R is one--fifth of the surface brightness within R \citep[e.g.][]{Conselice2002}. This provides a non--parametric measure of the size that is independent of the mean surface brightness. The half--light radius, R$_{\text{h}}$, is then defined as the radius at which the flux is one--half of that within 2R$_{\text{p}}$ deconvolved with the PSF.

For the 21 galaxies with $HST$ imaging, we measure R$_{\text{h}}$ in both ground-- and  \emph{HST}--based photometry, both parametrically (Figure \ref{Fig:sizes}) and non-parametrically. To test how well we recover the sizes in ground-based measurements alone, we compare the ground based continuum half--light radii to the  $HST$ continuum half--light radii, deriving a median ratio of $\langle$R\SPSB{G}{h}/R\SPSB{HST}{h}\quad$\rangle$\,=\,0.97\,$\pm$\,0.05. Applying the same parametric fitting procedure to the remaining galaxies we derive half-light radii for all 34 galaxies with $\langle$\,$R_{\text{h}}$\,$\rangle$\,=\,0.43\,$\pm$\,0.06 arcsec, which equates to 3.55\,$\pm$\,0.50\,kpc at $z$\,=\,2.22 (the median redshift of the sample). Numerically we derive a median of $\langle$\,$R_{\text{h}}$\,$\rangle$\,=\,0.55$\,\pm$\,0.04 arcsec (4.78\,$\pm$\,0.41\,kpc at the $z$\,=2\,.22), with $\langle$R\SPSB{S\'ersic}{h}\qquad $\rangle$/R\SPSB{Numerical}{h}\qquad \ \ \quad $\rangle$\,=\,0.82\,$\pm$\,0.04, indicating that the non-parametric fitting procedure broadly reproduces the parametric half-light radii. The median continuum half--light size derived for our sample from S\'ersic fitting is comparable to that obtained by \cite{Stott2013} for HiZELS galaxies out to $z$\,=\,2.23, with $\langle$\,$R_{\text{h}}$\,$\rangle$\,=\,3.6\,$\pm$\,0.3\,kpc.

We further test the reliability of the recovered sizes (and their uncertainties), by randomly generating 1000 S\'ersic models with 0.5\,$<$\,$n$\,$<$\,2 and 0.1 arcsec\,$<$\,R$_{\text{h}}$\,$<$\,1 arcsec. These models are convolved with the UDS image PSF and Gaussian random noise is added appropriate for the range in total signal to noise for our observations. Each model is then fitted to derive `observed' model parameters. We recover a median size of $\langle$R\SPSB{True}{h}\quad/R\SPSB{Obs}{h}\quad$\rangle$\,=\,0.99\,$\pm$\,0.05 and  S\'ersic index $\langle$n$_{\text{True}}$/n$_{\text{Obs}}$\,$\rangle$\,=\,1.05\,$\pm$\,0.07. This demonstrates our fitting procedures accurately derive the intrinsic sizes of the galaxies in our sample. From this point forward we take the parametric S\'ersic half-light radii as the intrinsic R$_{\text{h}}$ of each galaxy.

As a test of the expected correlation between continuum size and the extent of nebular emission \citep[e.g.][]{Bournard2008,Schreiber2011}, we calculate the H\,$\alpha$ ([O\,{\sc{iii}}] for galaxies at $z$\,$>$\,3) half--light radii of the galaxies in the sample. We follow the same procedures as for the continuum stellar emission, but using narrow--band images generated from the integral field data. We model the PSFs, using a stack of unsaturated stars that were observed with the spectrographs at the time of the observations using a multi-component S\'ersic ($n$\,=\,0.5) model. 

We derive both parametric and non-parametric half-light radii from S\'ersic fitting and numerical analysis within 2R$_{\text{p}}$.  For the full sample of 34 galaxies, the median parametric nebula half-light radii is $\langle$R\SPSB{Nebula}{h}\qquad\ $\rangle$\,=\,0.31\,$\pm$\,0.06 arcsec with $\langle$R\SPSB{S\'ersic}{h}\qquad  /R\SPSB{Numerical}{h}\quad\qquad \ $\rangle$\,=\,0.93\,$\pm$\,0.04. The nebula emission sizes on average are consistent with the continuum stellar size, with $\langle$R\SPSB{Continuum}{h}\qquad\qquad / R\SPSB{Nebula}{h}\qquad\ $\rangle\,=\,$1.15\,$\pm$\,0.19. We note that the low-surface brightness of the outer regions of the high-redshift galaxies may account for the apparent $\sim$10 per cent smaller nebula sizes in our sample.

\subsection{Galaxy Inclination and Position angles}

To derive the inclination of the galaxies in our sample we first measure the ratio of semiminor (b) and major (a) axis from the parametric S\'ersic model. We derive an uncertainty on the axial ratio of each galaxy by bootstrapping the fitting procedure over an array of initial conditions. For galaxies that are disc-like, the axial ratio is related to the inclination by
\begin{equation}\label{Eqn:Inc}
\text{cos}^\text{2}(\theta_{\text{inc}})=\frac{\left(\frac{\text{b}}{\text{a}}\right)^\text{2}-\text{q}_\text{0}^\text{2}}{\text{1}-\text{q}_\text{0}^\text{2}},
\end{equation} 
where $\theta_{\text{inc}}$\,=\,0 represents a face-on galaxy. The value of q$_0$, which accounts for the fact that galaxy discs are not infinitely thin, depends on the galaxy type, but is typically in the range of q$_0$\,=\,0.13\,--\,0.20 for rotationally supported galaxies at $z$\,$\sim$\,0 \citep[e.g.][]{Weijmans2014}. We adopt q$_0$\,=\,0.2 to be consistent with other high redshift integral field surveys (KROSS; \citealt{Harrison2017}; KMOS3D, \citealt{Wisnioski15}). The full range of axial ratios in the sample is $b/a$\,=\,0.2\,--\,0.9 with $\langle$b/a$\rangle$\,=\,0.69$\,\pm$\,0.04 corresponding to a median inclination for the sample of $\langle$\,$\theta_{\rm inc}$\,$\rangle$\,=\,48$^\circ$\,$\pm$\,3$^\circ$. 

\begin{figure*}
	\centering
	\includegraphics[width=\linewidth]{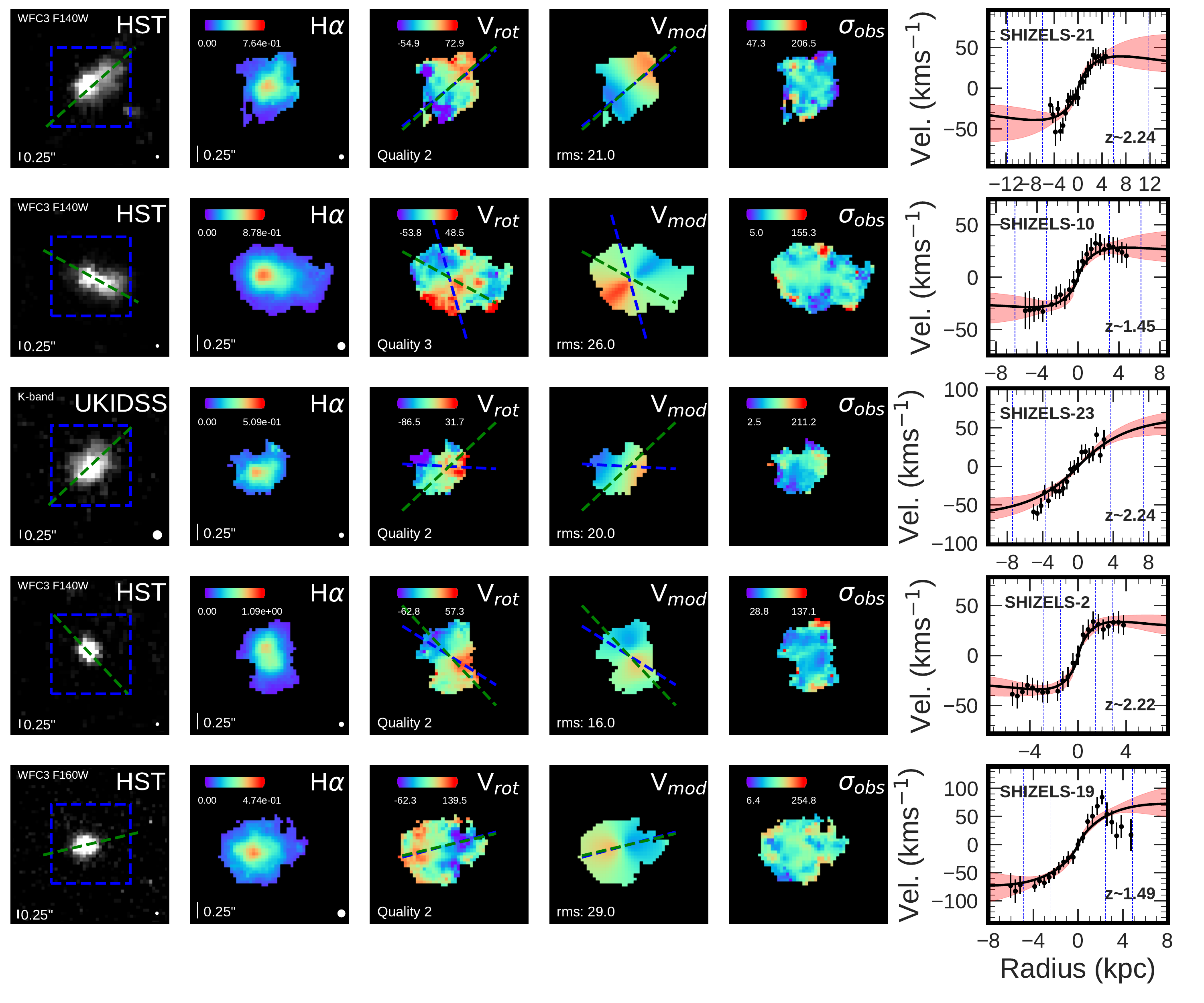}
	\caption{Example of spatially resolved galaxies in our sample. From left to right; broad--band photometry of the galaxy (left), with PA$_{\rm im}$ (green dashed line) and data cube field of view (blue dashed square). H\,$\alpha$ or [O\,{\sc{iii}}] flux map, velocity map, velocity model, and velocity dispersion map, derived from the emission--line fitting. PA$_{\rm vel}$  (blue dashed line) and PA$_{\rm im}$ (green dashed line) axes plotted on the velocity map and model. Rotation curve extracted about the kinematic position axis (right). The rotation curve shows lines of R$_\text{h}$ and 2R$_\text{h}$ derived from S\'ersic fitting, as well as the 1$\sigma$ error region (red) of rotation curve fit (black line).} 
	\label{Fig:Kin}
\end{figure*}

\subsection{Emission--Line Fitting}\label{Sec:DynProp}

Next we derive the kinematics, rotational velocity, and dispersion profiles of the galaxies by performing emission--line fits to the spectrum in each data cube. 

For the H\,$\alpha$ and [N\,{\sc{ii}}] doublet (25) sources, we fit a triple Gaussian profile to all three emission lines simultaneously, whilst for [O\,{\sc{iii}}] emitters a single Gaussian profile is used when we model the [O\,{\sc{iii}}] $\lambda$5007 emission--line.  We do not have significant detections of the $\lambda$4959 [O\,{\sc{iii}}] or $\lambda$4862 H$\beta$ emission--line. The fitting procedure uses a five or six parameter model with redshift, velocity dispersion, continuum and emission--line amplitude as free parameters. For the H\,$\alpha$ emitting galaxies we also fit the [N\,{\sc{ii}}]/H\,$\alpha$ ratio, constrained between 0 and 1.5. The FWHM of the emission lines are coupled, the wavelength offsets fixed, and the flux ratio of the [N\,{\sc{ii}}] doublet ${\rm (\frac{[N\,{\textsc{ii}}]\lambda 6583}{[N\,{\textsc{ii}}]\lambda 6548})}$ fixed at 2.8 \citep{Osterbrock2006}. We define the instrumental broadening of the emission lines from the intrinsic width of the OH sky lines in each galaxy's spectrum,  by fitting a single Gaussian profile to the sky line. The instrumental broadening of the OH sky lines in the $J$, $H$, and $K$ bands are $\sigma_{\rm int}$\,=\,71\,km\,s$^{-1}$\,$\pm$\,2\,km\,s$^{-1}$, 50\,km\,s$^{-1}$\,$\pm$\,5\,km\,s$^{-1}$, and 39\,km\,s$^{-1}$\,$\pm$\,1\,km\,s$^{-1}$, respectively. The initial parameters for spectral fitting are estimated from spectral fits to the galaxy integrated spectrum summed from a 1 arcsecond aperture centred on the continuum centre of the galaxy. 

We fit to the spectrum in 0.15 arcsec\,$\times$\,0.15 arcsec (3\,$\times$\,3 spaxels) spatial bins, due to the low signal to noise in individual spaxels, and impose a signal to noise threshold of S/N$\,\geq$\,5 to the fitting procedure. If this S/N is not achieved, we bin the spectrum over a larger area until either the S/N threshold is achieved or the binning limit of 0.35 arcsec $\,\times\,0.35 arcsec$ is reached ($\sim$1.5\,$\times$ the typical AO-corrected PSF width). In Figure \ref{Fig:Kin} we show example H\,$\alpha$ and [O\,{\sc{iii}}] intensity, velocity, and velocity dispersion for five galaxies in the sample.

\subsection{Rotational Velocities}\label{Sec:DynMod}

We use the H\,$\alpha$ and [O\,{\sc{iii}}] velocity maps to identify the kinematic major axis for each galaxy in our sample. We rotated the velocity maps around the continuum centre in 1$^\circ$ steps, extracting the velocity profile in 0.15 arcsecond wide ‘slits’ and calculating the maximum velocity gradient along the slit. We bootstrap this process, adding Gaussian noise to each spaxel's velocities of the order of the velocity error derived from emission--line fitting. The position angle with the greatest bootstrap median velocity gradient was identified as being the major kinematic axis (PA$_{\rm vel}$), as shown by the blue line in Figure \ref{Fig:Kin}.
\begin{figure*}
    \centering
    \includegraphics[width=1\linewidth]{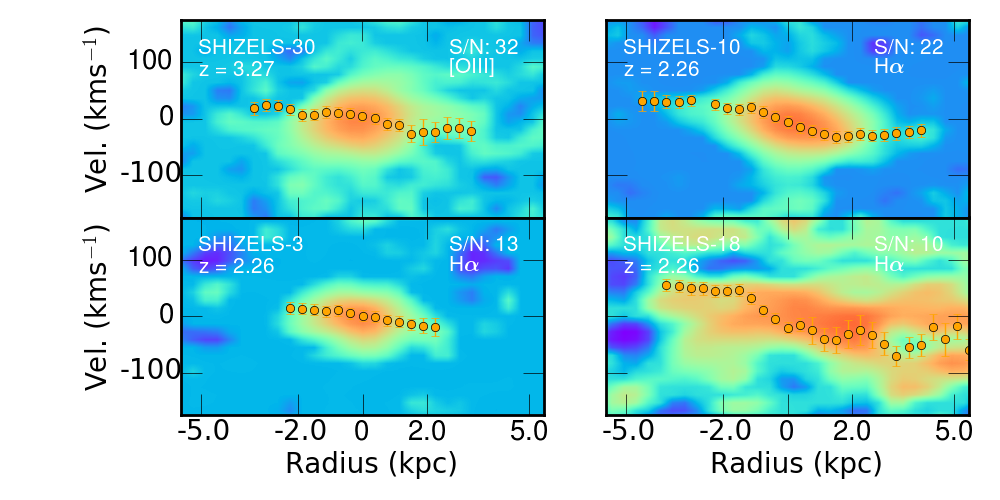}
    \caption{The position-velocity diagrams of four galaxies in the sample extracted from a slit about the kinematic major axis of each galaxy. The galaxies shown are selected from bins of emission--line S/N derived from the galaxies' integrated spectrum. We overlay each galaxy's ionised gas rotation curve as derived in Section \ref{Sec:DynProp} for comparison. The redshift, emission--line, and S/N of each position-velocity map is shown, with upper left to bottom right as high to low galaxy integrated S/N.}
\label{Fig:PV}
\end{figure*}

By extracting the velocity profile of the galaxies in our sample about the kinematic major axis, we are assuming the galaxy is an infinitely thin disc with minimal non-circular motions and is kinematically `wellbehaved'. We note however that this may not be true for all the galaxies in the sample, with some galaxies having significant non-circular motions, leading to an underestimate of the rotation velocity and an overestimate of the velocity dispersion in these galaxies.

The accuracy of the velocity profile extracted for each galaxy depends on the accuracy to which the kinematic major axis is identified. To quantify the impact on the rotation velocity profile of deriving an incorrect kinematic position angle, we extract the rotation profiles of our galaxies about their broad--band semimajor axes as well as their kinematic axis. On average we find minimal variation between V$_{\rm rot, BB}$(r) and  V$_{\rm rot, KE}$(r) with $\langle$\,V$_{\rm rot, BB}$(r)/V$_{\rm rot, KE}$(r)\,$\rangle$\,=\,0.94\,$\pm$\,0.15. 

In order to minimise the impact of noise on our measurements, we also fit each emission--line rotation velocity curve ($v$) with a combination of an exponential disc ($v_\text{D}$) and dark matter halo ($v_\text{H}$). We use these models to extrapolate the data in the outer regions of the galaxies' velocity field, as opposed to interpreting the implications of the individual model parameters. For the disc dynamics we assume that the baryonic surface mass density follows an exponential profile \citep{Freeman1970a} and the halo term can be modelled as a modified Navarro, Frenk \& White (NFW) profile \citep{Navarro1997}. The halo velocity model converges to the NFW profile at large distances and, for suitable values of $r_0$, it can mimic the NFW or an isothermal profile over the limited region of the galaxy that is mapped by the rotation curve. The dynamics of the galaxy are described by the following disc and halo velocity components
\begin{equation*}
v^2 = v_D^2 + v_H^2,
\end{equation*}
\begin{equation*}
v^2_D(x) =\frac{1}{2}\frac{GM_{d}}{R_{d}}(3.2 x)^2 (I_0 K_0 − I_1 K_1),
\end{equation*}
\begin{equation*}
v^2_H(r)= \frac{6.4 G \rho_0 r_0^3}{r}\left(\ln(1+\frac{r}{r_0})-\tan^{-1}(\frac{r}{r_0})+\frac{1}{2}
	\ln[1+(\frac{r}{r_0})^2]\right),
\end{equation*}
where $x=R/R_{\rm d}$ and $I_{\rm n}$ and $K_{\rm n}$ are the modified Bessel functions computed at 1.6$x$ with M$_{\rm d}$ and R$_{\rm d}$ as the disc mass and disc scale length respectively. In fitting this model to the rotation profiles, there are strong degeneracies between R$_{\rm d}$, $\rho_0$ and $r_0$. To derive a physically motivated fit, we modified the dynamical model to be a function of the dark matter fraction, disc scale radius and disc mass. 
Using the stellar mass, derived in Section \ref{Sec:Analysis}, as a starting parameter for the disc mass, enables the fitting routine to converge. The dynamical centre of the galaxy was allowed to vary in the fitting procedure by having velocity and radial offsets as free parameters constrained to $\pm$\,20\,km\,s$^{-1}$ and $\pm$\,0.1 arcsec. The dark matter fraction in galaxy with a given disc and dark matter mass is given by
\begin{equation*}
f_{DM}=\frac{M_{\text{DM}}}{M_{\rm d}+M_{\rm DM}},
\end{equation*}
\noindent
where the dark matter mass and disc mass are derived from;
\begin{equation*}
M_{DM}({<}R)=\int_{0}^{R}\rho(r)4 \pi r^2 dr = \int_{0}^{R}\frac{4 \pi \rho_0 r_0^3 r^2}{(R+r_0)(R^2+r_0^2)}dr,
\end{equation*}
\begin{equation*}
M_d({<}R)=\int_{0}^{R}e^{-\frac{r}{R_d}}2\pi r dr,
\end{equation*}

The dynamical model therefore contains  five free parameters, M$_{\text{d}}$, $R_{\text{d}}$, f$_{\text{DM}}$, V$_{\text{off}}$ and r$_{\text{off}}$ where V$_{\text{off}}$ and r$_{\text{off}}$ are velocity and radial offsets for the rotation curve to allow for continuum centre uncertainties. We use the {\sc{mcmc}} package designed for {\sc{python}} ({\sc{emcee}}; \citealt{Foreman-Mackey2013}), to perform Markov Chain Monte Carlo sampling with 500 walkers, initial burn-in of 250 and final steps to convergence of 500. We then use a $\chi^2$ minimisation method to quantify the uncertainty on the rotational velocity extracted from the model. The  1$\sigma$ error is defined as the region in parameter space where the $\delta \chi^2$ = $|\chi^2_{\text{best}}$ - $\chi^2_{\text{params}}|$\,$\leq$\,number of parameters. Prior to the {\sc{mcmc}} procedure we apply the radial and velocity offsets to the rotation to reduce the number of free parameters and centre the profiles. The parameter space for 1$\sigma$ uncertainty is thus $\delta \chi^2$\,$\leq$\,3. Taking the extremal velocities derived within the $\delta \chi^2$\,$\leq$\,3 parameter space provides the uncertainty on V$_{\rm rot}$. The rotation velocities and best fit dynamical models are shown in Figure \ref{Fig:Kin}. The full samples kinematics are shown in Appendix D. To show the full extent of the quality of data in our sample, we derive position-velocity diagrams for each galaxy. In Figure \ref{Fig:PV} we show one position-velocity diagram from each quartile of galaxy integrated signal to noise with the galaxies' ionised gas rotation curve overlaid.  

\begin{figure*}
	\centering
	\includegraphics[width=1\linewidth]{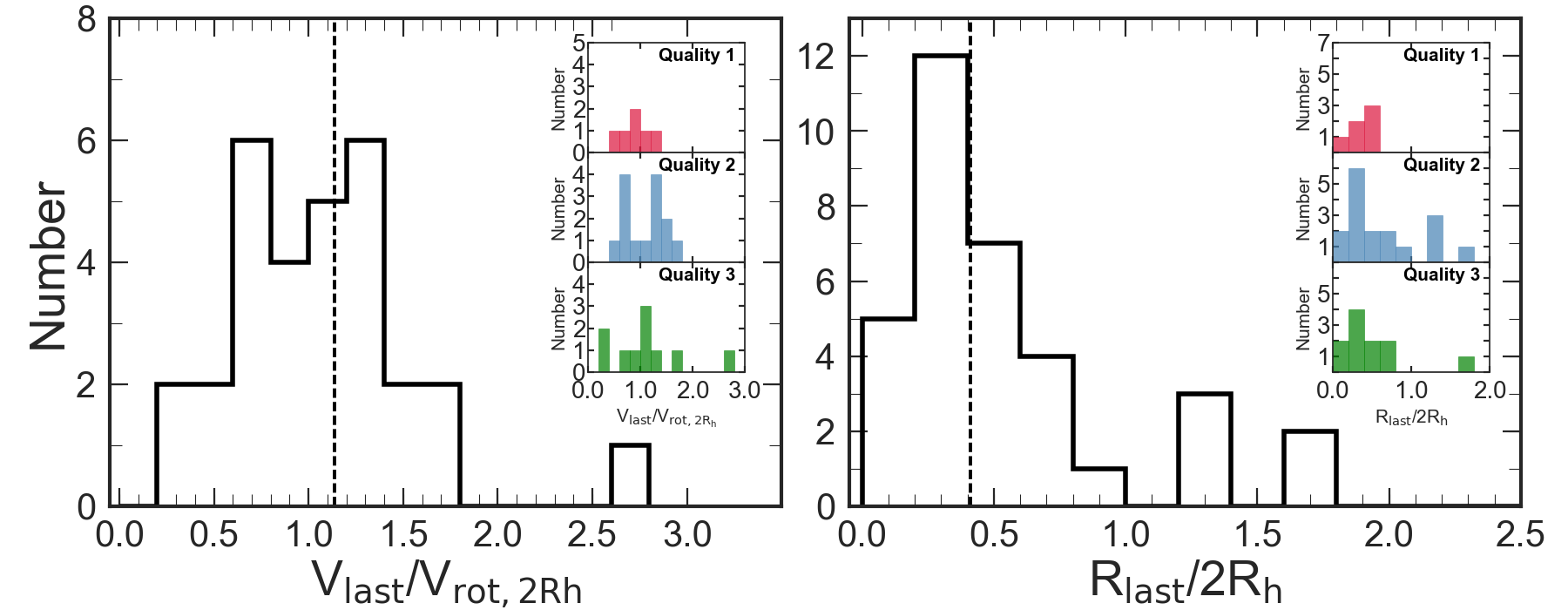}
	\caption{\emph{Left:} histogram of the ratio of the last rotational velocity data point to the velocity at 2R$_{\text{h}}$. \emph{Right:} histogram of the ratio of the radius of the last data point on the rotation curve to 2R$_{\text{h}}$. Inset histograms show the distribution for the kinematic sub-classes (Section \ref{Sec:Qual}). The dashed line indicates the median in both figures, where $\langle$R$_{\rm last}$/2R$_{\rm h}$\,$\rangle$\,=\,0.42\,$\pm$\,0.04 and $\langle$V$_{\rm last}$/V$_{\rm rot, R_h}$\,$\rangle$\,=\,1.14\,$\pm$\,0.11. On average extracting the rotational velocity at 2R$_{\rm h}$ requires extrapolation of the model beyond the last data point, leading to a decrease in velocity of $\sim$\,14 per cent.}
\label{Fig:last_hist}
\end{figure*}

Next we measure the rotation velocities of our sample at 2R$_{\rm h}$ (=\,3.4 R$_{\rm d}$ for an exponential disc) \citep[e.g.][]{Miller2011}. For each galaxy we convolve R$_{\rm h}$ with the PSF of the integral field unit observation and extract velocities from the rotation curve. At a given radii our measurement is a median of the absolute values from the low and high components of the rotation curve. Finally we correct for the inclination of the galaxy, as measured in Section \ref{Sec:Sizes}. On average the extraction of V$_{\rm rot,}$\SB{2R$_{\rm h}$} from each galaxy's rotation curve requires extrapolation from the last data point (R$_{\rm last}$) to 2R$_{\rm h}$ in our sample, where the median ratio is $\langle$\,$\rm {R_{last}}/\text{2R}_{\text{h}}$\,$\rangle$\,=\,0.42$\,\pm$\,0.04. However for the sample, the average V$_{\rm rot,}$\SB{2R$_{\rm h}$} is $\sim$14 per cent smaller than the velocity of the last data point (V$_{\rm last}$) with $\langle$V$_{\rm last}$/V$_{\rm rot,}$$_{\rm 2R_h}$\,$\rangle$\,=\,1.14\,$\pm$\,0.11 which is within 1$\sigma$. 
%This indicates the rotation curves turns over on average within 2R$_{\rm h}$. 
Figure \ref{Fig:last_hist} shows the distribution of radial and velocity ratios. 

To quantify the impact of beam--smearing on the rotational velocity measurements, we follow the methods of \cite{Johnson2018}, and derive a median ratio of $\langle$\,R$_{\rm d}$\,/\,R\SPSB{PSF}{h}\quad$\rangle$\,=\,2.17\,$\pm$\,0.18 which equates to an average rotational velocity correction of 1 per cent. We derive the correction for each galaxy in the sample and correct for beam--smearing effects. Appendix \ref{App:C}, Table \ref{Table:sizes}  displays the inclination, beam--smearing--corrected rotation velocity (V$_{\rm rot,}$\SB{2R$_{\rm h}$}) for each galaxy. The full distribution of R$_{\rm d}$\,/\,R\SPSB{PSF}{h}\quad is shown in Appendix E. 

The median inclination beam--smearing corrected rotation velocity in our is sample is $\langle$V$_{\rm rot,}$\SB{2R$_{\rm h}$}$\rangle$\,=\,64\,$\pm$\,14\,km\,s$^{-1}$, with the sample covering a range of velocities from V$_{\rm rot,}$\SB{2R$_{\rm h}$}\,=\,17\,--\,380\,km\,s$^{-1}$. 
The SINS/ZC-SINF AO survey (\citealt{Schreiber2018}) of 35 star--forming galaxies at $z$\,$\sim$\,2 identify a median rotation velocity of $\langle$V$_{\rm rot}$\,$\rangle$\,=\,181\,km\,s$^{-1}$, with a range of V$_{\rm rot}$\,=\,38\,--\,264\,km\,s$^{-1}$. This is approximately a factor of 3 larger than our sample, although we note their sample selects galaxies of higher stellar mass with $\log(M_{*}[M_{\odot}]$)\,=\,9.3\,--\,11.5 whereas our selection selects lower mass galaxies.

\subsection{Kinematic Alignment}

\begin{figure}
	\centering
	\includegraphics[width=1\linewidth]{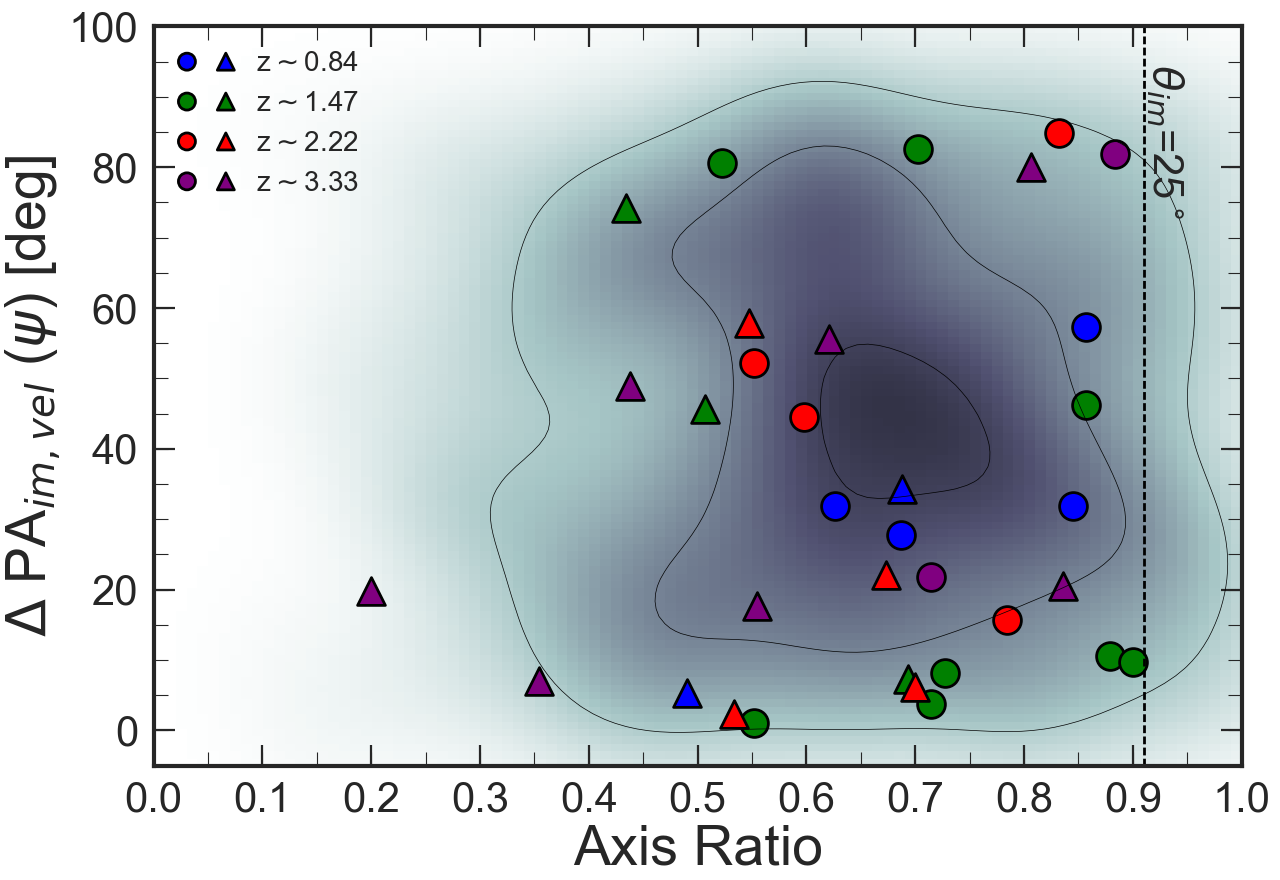}
	\caption{The absolute misalignment between the kinematic and morphological axes ($\Psi$) as a function of semiminor (b) to semimajor (a) axis ratio for the galaxies in our sample derived from S\'ersic fitting. Our sample is coloured by redshift as Figure \ref{Fig:MS}, and the KMOS Redshift One Spectroscopic Survey (KROSS) is shown for comparison as the grey shaded region. The \emph{circles} indicate galaxies with V$_{\rm rot, 2R_h}/\sigma_{\rm median}$\,$>$\,1 whilst \emph{triangles} highlight galaxies with V$_{\rm rot, 2R_h}/\sigma_{\rm median}$\,$<$\,1. The majority of galaxies in our sample are moderately inclined with $\langle$b/a$\rangle$=0.68\,$\pm$0.04 showing kinematic misalignment of $\Psi<$48$^{\circ}$.}
	\label{Fig:dpa_ba}
\end{figure}

The angle of the galaxy on the sky can be defined as the morphological position angle (PA$_{\rm im}$) or the kinematic position angle (PA$_{\rm vel}$). High-redshift integral field unit studies \citep[e.g.][]{Wisnioski15,Harrison2017} use the misalignment between the two position angles to provide a measure of the kinematic state of the galaxy. The (mis)alignment is defined such that: 
\begin{equation}\label{Eqn:psi}
\sin{\Psi}=|\sin{(PA_{\rm im}-PA_{\rm vel}})|,
\end{equation} 
where $\Psi$ takes values between 0$^\circ$ and 90$^\circ$. In Figure \ref{Fig:dpa_ba} we show $\Psi$ as a function of the image axial ratio for the sample compared to the KROSS survey of $\sim$700 star--forming galaxies at $z$\,$\sim$\,0.8. The sample covers a range of position angle misalignment, with $\langle$\,$\Psi$\,$\rangle$=31.8$^\circ$\,$\pm$\,5.7$^\circ$, 10.52$^\circ$\,$\pm$\,19.8$^\circ$, 33.2$^\circ$\,$\pm$\,15.2$^\circ$, and 21.8$^\circ$\,$\pm$\,17.5$^\circ$ at $z$\,=\,0.84, 1.47, 2.22, and 3.33 respectively. 
This is larger than that identified in KROSS at $z$\,$\sim$\,0.8 (13$^\circ$), but at all redshifts comparable to or within the criteria of $\Psi \leq$30$^\circ$ imposed by \cite{Wisnioski15}, to define a galaxy as kinematically `discy'. This indicates that the average galaxy in our sample is on the boundary of what is considered to be a kinematically `discy'. A summary of the morphological properties for our sample is shown in Appendix \ref{App:C}, Table \ref{Table:sizes}. Example broad--band images of our sample are shown in the left--panel of Figure \ref{Fig:Kin}, with the appropriate PA$_{\rm im}$ and integral field spectrograph field of view.  The kinematic PA for the sample is derived in Section \ref{Sec:DynProp}. We will use this criteria, together with other dynamical criteria later, to define the most disc-like systems.

\subsection{Two-dimensional Dynamical Modelling}

To provide a parametric derivation and test of the numerical kinematic properties derived for each galaxy, we model the broadband continuum image and two-dimensional velocity field with a disc and halo model. The model is parametrized in the same way as the one-dimensional kinematic model used to interpolate the data points in each galaxy's rotation curve (Section \ref{Sec:DynMod}) but takes advantage of the full two-dimensional extent of the galaxy's velocity field. To fit the dynamical models to the observed images and velocity fields, we again use an {\textsc{MCMC}} algorithm. We first use the imaging data to estimate the size, position angle, and inclination of the galaxy disc. Then, using the best-fitting parameter values from the imaging as a first set of prior inputs to the code, we simultaneously fit the imaging and velocity fields. We allow the dynamical centre of the disc and position angle (PA$_{\rm vel}$) to vary, but require that the imaging and dynamical centre lie within 1 kpc (approximately the radius of a bulge at $z\sim$1; \citealt{Bruce2014}). We note also that we allow the morphological and dynamical major axes to be independent. The routine converges when no further improvement in the reduced chi-squared of the fit can be achieved within 30 iterations. For a discussion of the model and fitting procedure, see \cite{Swinbank2017}.

For the sample of 34 galaxies the average of the ratio of kinematic positional angle derived from  the velocity map to numerical modelling is $\langle$PA$_{\rm vel(Slit)}$/PA$_{\rm vel(2D)}$\,$\rangle$=0.97\,$\pm$\,0.09, whilst the morphological position angles agree on average with $\langle$PA$_{\rm im(S\acute{e}rsic)}$/PA$_{\rm im(2D)}$\,$\rangle$=1.10\,$\pm$\,0.14. 
We compare the velocity field generated from the fitting procedure (see Figure \ref{Fig:Kin} for examples), to the observed field for each galaxy derived from emission--line fitting (Section \ref{Sec:DynProp}). We derive a velocity--error--weighted rms based on the residual for each galaxy and normalize this by the galaxy's rotational velocity (V$_{\rm rot,}$\SB{2R$_{\rm h}$}). On average the sample is well described by the disc and halo model, with the median rms of the residual images being $\langle$rms$\rangle$=22\,$\pm$\,1.42

\subsection{Velocity Dispersions}\label{Sec:vdisp}

To further classify the galaxy dynamics of our sources we also make measurements of the velocity dispersion of the star--forming gas ($\sigma_0$). High--redshift star--forming galaxies are typically highly turbulent clumpy systems, with non-uniform velocity dispersions  \citep[e.g.][]{Genzel2006, Kassin2007,Stark2008, ForsterSchreiber2009, Jones2010,Johnson2018}. The effects of beam--smearing on our sample are reduced compared to non-AO observations due to the high AO resolution although we still apply a correction. First we measure the velocity dispersion of each galaxy by taking the median of each velocity dispersion map, examples of which are shown in Figure \ref{Fig:Kin}, in an annulus between R$_{\rm h}$ and 2R$_{\rm h}$. This minimizes the effects of beam--smearing towards the centre of the galaxy as well as the impact of low surface brightness regions in the outskirts of the galaxy. We also measure the velocity dispersion from the inner regions of the dispersion map as well as the map as a whole, finding excellent agreement between all three quantities, to within on average 3 per cent.

To take into account the impact of beam--smearing on the velocity dispersion of the galaxies in our sample we follow the methods of \cite{Johnson2018}. We measure the ratio of galaxy stellar continuum disc size (R$_{\rm d}$) to the half--light radii of the PSF of the AO observations deriving a median ratio of $\langle$\,R$_{\rm d}$\,/\,R\SPSB{PSF}{h}\quad$\rangle$\,=\,2.17\,$\pm$\,0.18 which equates to an average velocity dispersion correction of $\sim$\,4 per cent. We derive the correction for each galaxy in the sample and correct for beam--smearing effects.

The average velocity dispersion for our sample is $\langle$\,$\sigma_{\rm median}$\,$\rangle$\,=\,85\,$\pm$\,6\,km\,s$^{-1}$, with the full range of $\sigma_{\rm median}$\,=\,40\,--\,314\,km\,s$^{-1}$. This is similar to KROSS at $z$\,$\sim$\,0.8 which has $\langle$\,$\sigma_{\rm median}$\,$\rangle$\,=\,83\,$\pm$\,2\,km\,s$^{-1}$ but much higher than the KMOS$^{\rm 3D}$ survey, which identified a decrease in the intrinsic velocity dispersion of star--forming galaxies by a factor of 2 from 50\,km\,s$^{-1}$ at $z$\,$\sim$\,2.3 to 25\,km\,s$^{-1}$ at $z$\,$\sim$\,0.9 \citep{Wisnioski15}. The evolution of velocity dispersion with cosmic time is minimal in our sample with $\langle$\,$\sigma_{\rm median}$\,$\rangle$\,=\,79$\,\pm$\,15\,km\,s$^{-1}$, 87\,$\pm$\,10\,km\,s$^{-1}$, 79\,$\pm$\,12\,km\,s$^{-1}$, and 83\,$\pm$\,27\,km\,s$^{-1}$ at $z$\,=\,0.84, 1.47, 2.23, and 3.33 respectively. The KMOS Deep Survey \citep{Turner2017} identified a stronger evolution in velocity dispersion with $\sigma_{\rm int}$\,=\,10\,-–\,20\,km\,s$^{-1}$ at $z$\,$\sim$\,0, 30\,-–\,60\,km\,s$^{-1}$ at $z$\,$\sim$\,1, and  40\,-–\,90\,km\,s$^{-1}$ at $z$\,$\sim$\,3 in star--forming galaxies. This indicates that the lower redshift galaxies in our sample are more turbulent than the galaxy samples discussed in \cite{Turner2017}. We note however, that the different selection functions of the observations will influence this result.

\begin{figure*}
	\centering
	\includegraphics[width=\linewidth]{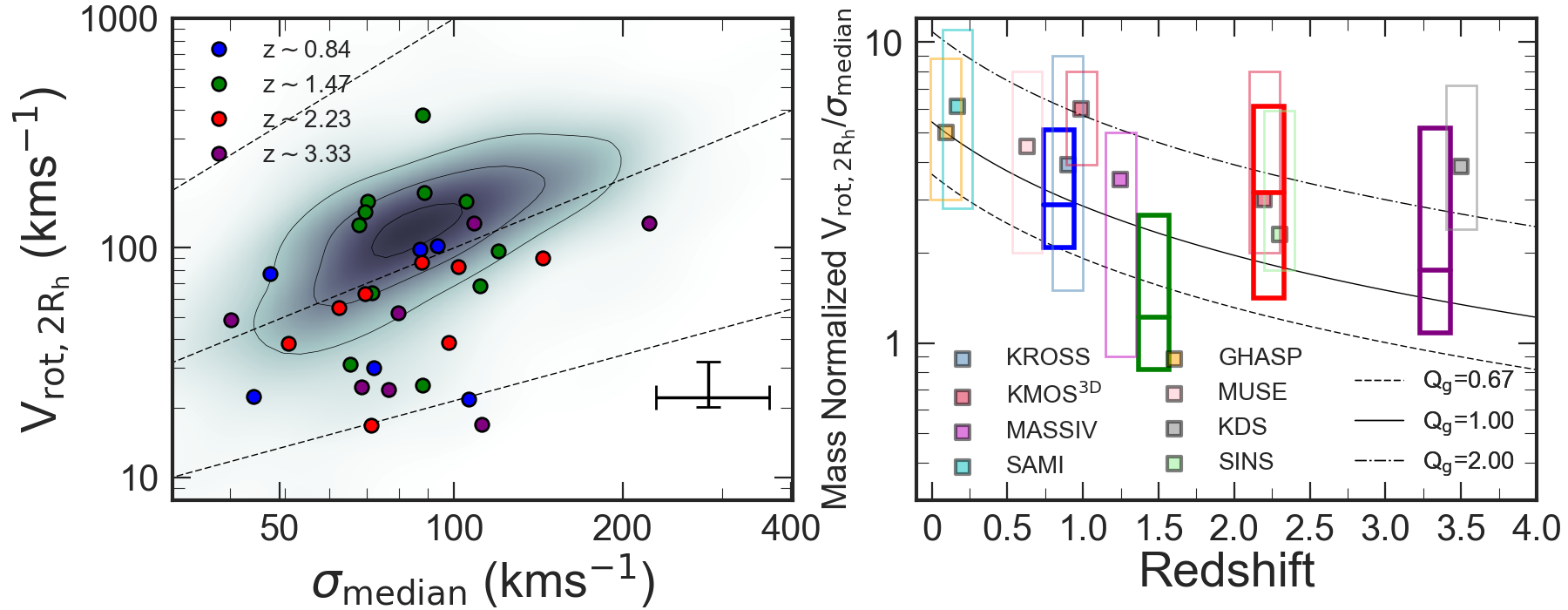}
	\caption{\emph{Left}: distribution of velocity V$_{\rm rot, 2R_ h}$ and $\sigma_{\rm median}$ in our sample, coloured by spectroscopic redshift as in Figure \ref{Fig:MS}. The KROSS $z$\,$\sim$\,0.8 survey is shown for comparison by the shaded region. Lines of 1.5V$_{\rm  rot}$ /$\sigma_{\rm median}$, V$_{\rm  rot}$ /$\sigma_{\rm median}$ and V$_{\rm  rot}$ /1.5$\sigma_{\rm median}$ shown for reference. \emph{Right}: mass normalised V$_{\rm  rot, 2R_ h}$ /$\sigma_{\rm median}$ as function of redshift, the 16th and 84th percentile shown by the extent of the box, median as a solid line at each redshift. We also show eight comparison surveys of star--forming galaxies from 0.09\,$<$\,$z$\,$<$\,3.5 selected from the literature with median values shown by the squares. We plot tracks of V$_{\rm  rot, 2R_ h}$ /$\sigma_{\rm median}$ as function of redshift, for different Toomre disc stability criterion (Q$_{\rm g}$; \citealt{Toomre1964}) following the procedures of \citet{Johnson2018}. 
	The majority of the sample has a mass normalised  V$_{\rm  rot, 2R_ h}$ /$\sigma_{\rm median}$\,$>$1, with an indication of a slight evolution in the dominate dynamical support process with cosmic time, with V$_{\rm  rot, 2R_ h}$ /$\sigma_{\rm median}$ increasing at lower redshift.}
	\label{Fig:Vsig}
\end{figure*}

To measure whether the galaxies in our sample are `dispersion dominated' or `rotation dominated' we take the ratio of rotation velocity (V$_{\rm rot, 2R_h}$) to intrinsic velocity dispersion ($\sigma_{\rm median}$), following \cite{Weiner2006} and \cite{Genzel2006}. Taking the full sample of 34 galaxies, we find a median ratio of rotational velocity to velocity dispersion, across all redshift slices of $\langle$V$_{\rm rot,2R_h}$/$\sigma_{\rm median}$\,$\rangle$\,=\,0.82\,$\pm$\,0.13 with $\sim$\,32 per cent having  V$_{\rm rot,2R_h}/\sigma_{\rm median}$\,$\rangle$\,1 (Figure \ref{Fig:Vsig}). This is significantly lower than other high redshift integral field unit studies such as KROSS, in which 81 per cent of its $\sim$\,600 star--forming galaxies having V$_{\rm rot, 2R_h}$/$\sigma_{\rm 0}$\,$\rangle$\,1 with a $\langle$V$_{\rm rot, 2R_h}/\sigma_{0}$\,$\rangle$\,=\,2.5\,$\pm$\,1.4. We note that the median redshift of the KROSS sample is $\langle$\,$z$\,$\rangle$\,=\,0.8, compared to $\langle$\,$z$\,$\rangle$\,=\,2.22 for our sample. \cite{Johnson2018} identified that galaxies of stellar mass $10^{10}$M$_{\odot}$ show a decrease in V$_{\rm rot, 2R_h}/\sigma_{0}$ from $z$\,$\sim$\,0 to $z$\,$\sim$\,2 by a factor $\sim$\,4.

The SINS/ZC-SINF AO survey of 35 star--forming galaxies at $z$\,$\sim$\,2 identify a median V$_{\rm  tot}/\sigma_{\rm 0}$\,=\,3.2 ranging from  V$_{\rm tot}/\sigma_{\rm 0}$\,=\,0.97\,--\,13 \citep{Schreiber2018}.  In our sample at $z$\,=\,0.84, 1.47, 2.23, and 3.33 the median ratio is $\langle$\,V$_{\rm rot, 2R_h}/\sigma_{\rm median}$\,$\rangle$\,=\,1.26\,$\pm$\,0.43, 1.75\,$\pm$\,0.90, 1.03\,$\pm$\,0.20, and 0.52\,$\pm$\,0.22 respectively. This indicates that on average the dynamics of the $z$\,$\sim$\,3.33 galaxies in our sample are more dispersion driven. \cite{Turner2017} identified a similar result with KMOS Deep Survey galaxies at $z$\,$\sim$\,3.5, finding a median value of V$_{\rm C}/\sigma_{\rm int}$\,=\,0.97\,$\pm$\,0.14.

In order to compare our sample directly to other star--forming galaxy surveys, we must remove the inherent scaling between stellar mass and V/$\sigma$, by mass normalising each comparison sample to a consistent stellar mass, for which we use M$_{*}$\,=\,10$^{10.5}$M$_{\odot}$, following the procedures of \cite{Johnson2018}. In Figure \ref{Fig:Vsig} we show the mass normalised V/$\sigma$ of our sample as a function of redshift as well as eight comparison samples taken from the literature. GHASP (\citealt{Epinat2010}; $z$\,=\,0.09), SAMI (\citealt{Bryant2015}; $z$\,=\,0.17), MASSIV (\citealt{Epinat2012}; $z$\,=\,1.25), KROSS (\citealt{Stott2016}; $z$\,=\,0.80), KMOS3D (\citealt{Wisnioski15};  $z$\,=\,1\, and \,2.20), SINS (\citealt{Cresci2009}; $z$\,=\,2.30), and KDS (\citealt{Turner2017}; $z$\,=\,3.50). We overplot tracks of V$_{\rm  rot, 2R_ h}$ /$\sigma_{\rm median}$ as function of redshift, for different Toomre disc stability criterion (Q$_{\rm g}$; \citealt{Toomre1964}) following the procedures of \cite{Johnson2018} and \cite{Turner2017}, normalised to the median V/$\sigma$ of the GHASP Survey at $z$\,=\,0.093. The galaxies in our sample align well with the mass--normalized comparison samples from the literature, with a trend of increasing V/$\sigma$ with increasing cosmic time, as star--forming galaxies become more rotationally dominated.

\subsection{Circular Velocities}\label{Sec:Vcirc}
It is well known that high--redshift galaxies are highly turbulent systems with heightened velocity dispersions in comparison to galaxies in the local Universe \citep[e.g.][]{ForsterSchreiber2006, ForsterSchreiber2009, Genzel2011, Swinbank2012a, Wisnioski15}. It is therefore necessary to account for the contribution of pressure support from turbulent motions to the circular velocity of high--redshift galaxies. As shown in \citet{Burkert2016}, if we assume the galaxies in our sample consist of an exponential disc with a radially constant velocity dispersion, the true circular velocity of a galaxy (V$_{\rm circ}$(r)) is given by

\begin{equation} \label{Eqn:Vcirc} 
\rm V_{\rm circ}^2(\rm r)=V_{\rm rot}^2(r)+2\sigma_{0}^2(\frac{\rm r}{ \rm R_d}),
\end{equation}
where R$_{\rm d}$ is the disc scale length and $\sigma_{0}$ is the intrinsic velocity dispersion of the galaxy. For a galaxy with V$_{\rm rot}$/$\sigma_{0}$\,$\geq$\,3 the contribution from turbulent motions is negligible and  V$_{\rm circ}$(r)\,$\approx$\,V$_{\rm rot}$(r). All the galaxies in our sample have V$_{\rm rot}$/$\sigma_{0}$\,$<$\,3. For each object we convert the inclination--corrected rotational velocity profile to a circular velocity profile. Following the same methods used to derive the rotational velocity of a galaxy (Section \ref{Sec:DynMod}), we fit one--dimensional dynamical models to the circular velocity profiles of each galaxy and extract the velocity at two times the stellar continuum half--light radii of the galaxy (V$_{\rm circ}$(r\,=\,2R$_{\rm h}$)). The ratio of V$_{\rm circ}$(r\,=\,2R$_{\rm h}$) to V$_{\rm rot}$(r\,=\,2R$_{\rm h}$) for each galaxy is shown in Appendix \ref{App:C}, Table \ref{Table:sizes}. The median circular velocity to rotational velocity ratio for galaxies in our sample is  $\langle$\,V$_{\rm circ}$(r\,=\,2R$_{\rm h}$)/V$_{\rm rot}$(r\,=\,2R$_{\rm h}$)\,$\rangle$\,=\,3.15\,$\pm$\,0.41 ranging from V$_{\rm circ}$(r\,=\,2R$_{\rm h}$)/V$_{\rm rot}$(r\,=\,2R$_{\rm h}$)\,=\,1.17\,--\,12.91.

\subsection{Sample Quality}\label{Sec:Qual}
Our sample of 34 star--forming galaxies covers a broad range in rotation velocity and velocity dispersion. Figure  \ref{Fig:Kin} and Figure \ref{Fig:Vsig} demonstrate there is dynamical variance at each redshift slice, with a number of galaxies demonstrating more dispersion--driven kinematics. To constrain the effects of these galaxies on our analysis, we define a subsample of galaxies with high signal to noise, rotation--dominated kinematics and `discy' morphologies. 

We note that if we were to split the sample by galaxy integrated signal to noise rather than morpho-kinematic properties, we would not select `discy' galaxies with rotation--dominated kinematics as the best--quality objects. Splitting the sample into three bins of signal to noise with S/N $\leq$ 14 (low), S/N $>$ 14 and S/N $\leq$ 23 (medium), and S/N $>$ 24 (high), we find 12, 11, and 11 galaxies in each bin, respectively, with the low and median S/N bins having a median redshift of $z$\,=\,1.47\,$\pm$\,0.17 and \,1.45\,$\pm$\,0.54 whilst the highest S/N bin has a median redshift of  $z$\,=\,2.24\,$\pm$\,0.38. All three signal to noise bins and have median rotation velocities, velocity dispersion and specific angular momentum values within 1$\sigma$ of each other, therefore not distinguishing between `discy' rotation--dominated galaxies and those with more dispersion driven dynamics.

The morpho-kinematic criteria that define our three subsamples are
\begin{itemize}
    \item Quality 1: V$_{\rm rot, 2R_h}/\sigma_{\rm median}$\,$>$\,1 and $\Delta$PA\SB{im,vel}$\Psi$\,$<$\,30\SP{$\circ$} \\
    \item Quality 2: V$_{\rm rot, 2R_h}/\sigma_{\rm median}$\,$>$\,1 or $\Delta$PA\SB{im,vel}$\Psi$\,$<$\,30\SP{$\circ$} \\
    \item Quality 3: V$_{\rm rot, 2R_h}/\sigma_{\rm median}$\,$<$\,1 and $\Delta$PA\SB{im,vel}$\Psi$\,$>$\,30\SP{$\circ$} \\
\end{itemize}

Of the 34 galaxies in the sample, 11 galaxies have V$_{\rm rot, 2R_h}/\sigma_{\rm med}$\,$>$\,1 and 17 have $\Delta$PA\SB{im,vel}$\Psi$\,$<$30\SP{$\circ$}. We classify 6 galaxies that pass both criteria as `Quality 1' whilst galaxies that pass either criteria are labelled `Quality 2' (17 galaxies). The remaining 11 galaxies that do not pass either criterion are labelled `Quality 3'.

The following analysis is carried out on the full sample of 34 galaxies as well as just the `Quality 1 ' and `Quality 2' galaxies. In general we draw the same conclusions from the full sample as well the sub--samples, indicating the more turbulent galaxies in our sample do not bias our interpretations of the data. In each of the following sections we remark on the properties of `Quality 1 ' and `Quality 2' galaxies.

\begin{figure}
	\centering
	\includegraphics[width=\linewidth]{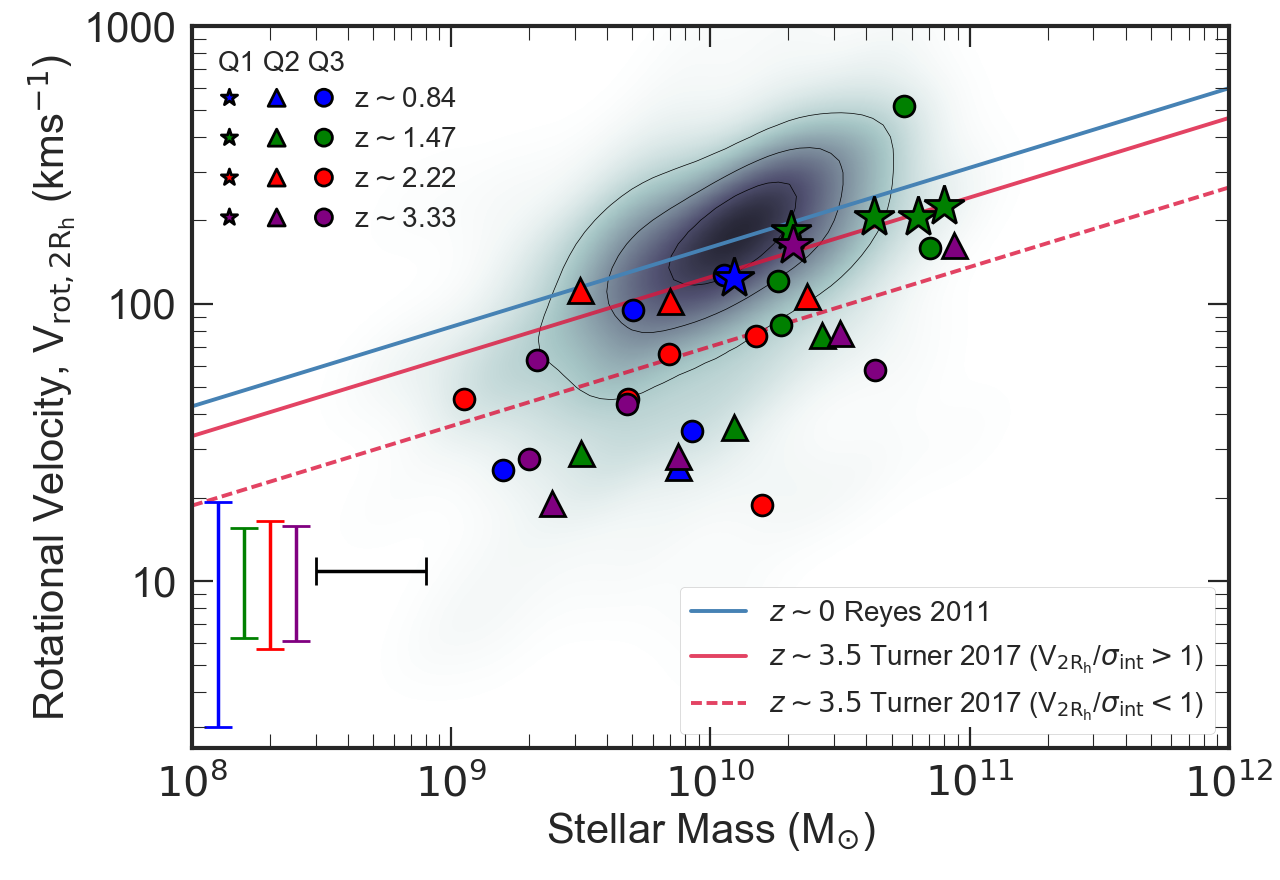}
	\caption{Rotation velocity extracted from the rotation curve at 2R$_{\rm h}$ as a function of stellar mass derived from SED fitting as described in Section \ref{Sec:Analysis}, formally known as the Stellar Mass Tully--Fisher relation. The sample is coloured by spectroscopic redshift, as in Figure \ref{Fig:MS}, whilst the blue shaded region represents the KROSS $z$\,$\sim$\,0.8 sample (\citet{Harrison2017}). The \emph{stars} represent `Quality 1' targets (V$_{\rm rot, 2R_h}/\sigma_{\rm med}$\,$>$\,1 and $\Delta$PA\SB{im,vel}$\Psi$\,$<$\,30\SP{$\circ$}), \emph{circles} `Quality 2' (V$_{\rm rot, 2R_h}/\sigma_{\rm med}$\,$>$\,1 or $\Delta$PA\SB{im,vel}$\Psi$\,$<$\,30\SP{$\circ$}) and \emph{triangles} `Quality 3' galaxies (V$_{\rm rot, 2R_h}/\sigma_{\rm med}$\,$<$\,1 and $\Delta$PA\SB{im,vel}$\Psi$\,$>$\,30\SP{$\circ$}). We also show $z$\,$\sim$\,0 tracks from \citet{Reyes2011}, z\,$\sim$\,3.5 tracks for rotation--dominated (V$_{\rm rot, 2R_h}$/$\sigma_{\rm int}$\,$>$\,1) and dispersion--dominated (V$_{\rm rot, 2R_h}/\sigma_{\rm int}$\,$<$\,1) galaxies in the KMOS Deep Survey (KDS) from \citet{Turner2017}. There is a clear distinction between the different sub--samples, with `Quality 1' galaxies having higher rotation velocity for a given stellar mass, aligning with the KROSS sample. `Quality 3' targets have lower rotation velocities, aligning more with  V$_{\rm rot, 2R_h}/\sigma_{\rm int}$\,$<$\,1 KMOS Deep Survey $z$\,$\sim$\,3.5 track, whilst `Quality 2' targets on average lie in between, with intermediate rotation velocities for a given stellar mass. The median uncertainty on rotational velocity  at each redshift is shown in the lower left corner as well as the uncertainty of the stellar mass. The $z$\,$\sim$1.47 `Quality 3' galaxy, with V$_{\rm rot, 2R_h}$\,$\sim$\,380\,km\,s$^{-1}$ has low inclination of $\sim$25$^{\circ}$, hence large line-of-sight velocity correction.}
	\label{Fig:TF}
\end{figure}

\subsection{Rotational velocity versus stellar mass}\label{Sec:TF}
The stellar mass `Tully-Fisher relationship', (TFR; Figure \ref{Fig:TF}), represents the correlation between the rotational velocity (V$_{\rm  rot, 2R_ h}$ ) and the stellar mass (M$_{*}$) of a galaxy (\citealt{TF1977}; \citealt{Bell2001}). The relationship demonstrates the link between total mass (or `dynamical mass')\footnote{For rotationally-dominated galaxies \cite{Tiley2019}} of a galaxy, which can be probed by how rapidly the stars and gas are rotating,  and the luminous (i.e. stellar) mass. 

In Figure \ref{Fig:TF} we plot V$_{\rm rot, 2R_h}$ as a function of stellar mass for our sample as well as a sample of $z$\,$<$\,0.1 star--forming galaxies from \citet{Reyes2011} using spatially resolved H\,$\alpha$ kinematics. The KROSS survey at $z$\,$\sim$\,0.8 is also indicated \citep{Harrison2017}. 
We over plot two tracks from the KMOS Deep Survey (KDS; \citealt{Turner2017}), with median redshift of $z$\,$\sim$\,3.5. The KDS sample is split into `rotation-dominated' systems (V$_{\rm rot, 2R_h}/\sigma_{\rm int}$\,$>$\,1) and `dispersion-dominated' systems (V$_{\rm rot, 2R_h}/\sigma_{\rm int}$\,$<$\,1), for which we show both tracks.

Figure \ref{Fig:TF} shows a distinction between `Quality 1' and `Quality 2\,/\,3' galaxies. `Quality 1' galaxies, which have the most disc-like properties have higher rotation velocity for a given stellar mass with a $\langle$V$_{\rm rot, 2R_h}$\,$\rangle$\,=\,151\,km\,s$^{-1}$\,$\pm$\,13\,km\,s$^{-1}$, and align with the rotational velocities of the KROSS sample. The median rotation velocity of `Quality 2 \& 3' galaxies is $\langle$V$_{\rm  2R_h}$\,$\rangle$\,=\,53\,km\,s$^{-1}$\,$\pm$\,10\,km\,s$^{-1}$, occupying similar parameter space to the V$_{\rm rot, 2R_h}/\sigma_{\rm int}$\,$<$\,1 KMOS Deep Survey $z$\,$\sim$\,3.5 track. This is a consequence of construction, as `Quality 1' galaxies have a median $\langle$V$_{\rm rot, 2R_h}/\sigma_{\rm med}$\,$\rangle$\,=\,1.74\,$\pm$\,0.30 whilst 'Quality 2 \& 3' sources have $\langle$V$_{\rm rot, 2R_h}/\sigma_{\rm med}$\,$\rangle$\,=\,0.62\,$\pm$\,0.11

The Tully-Fisher relation provides a method to constrain galaxy dynamical masses however due to degeneracies and ambiguity in the evolution of the intercept and slope of the relationship with cosmic time \citep[e.g.][]{Ubler2017,Tiley2018}, and the strong implications of sample selection this becomes increasingly challenging. There is discrepancy amongst other high--redshift star--forming galaxy studies \citep[e.g.][]{Conselice2005,Flores2006,DiTeodoro2016,Pelliccia2017} finding no evolution in the intercept or slope of Tully--Fisher relation. Even with the inclusion of non-circular motions through gas velocity dispersions via the kinematic estimator S$_{0.5}$ \citep[e.g.][]{Kassin2007,Gnerucci2011} no evolution across $\sim$\,8\,Gyr of cosmic time is found. Whilst other studies \citep[e.g.][]{Miller2012,Sobral2013b} identify evolution in the stellar mass zero point of $\Delta$M$^*$\,=\,0.02\,$\pm$\,0.02 dex out to $z$\,=\,1.7.

We have demonstrated that the galaxies in our sample exhibit properties that are typical for `main--sequence' star--forming galaxies from $z$\,=\,0.8\,--\,3.5 and show good agreement with other high-redshift integral field surveys when the sample selection is well matched \citep[e.g.][]{Ubler2017,Harrison2017,Turner2017}. For the remainder of this work we focus on a fundamental property of the galaxies in our sample; their angular momentum, which incorporates the observed velocity, galaxy size and stellar mass.

\begin{figure*}
	\centering
	\includegraphics[width=\linewidth]{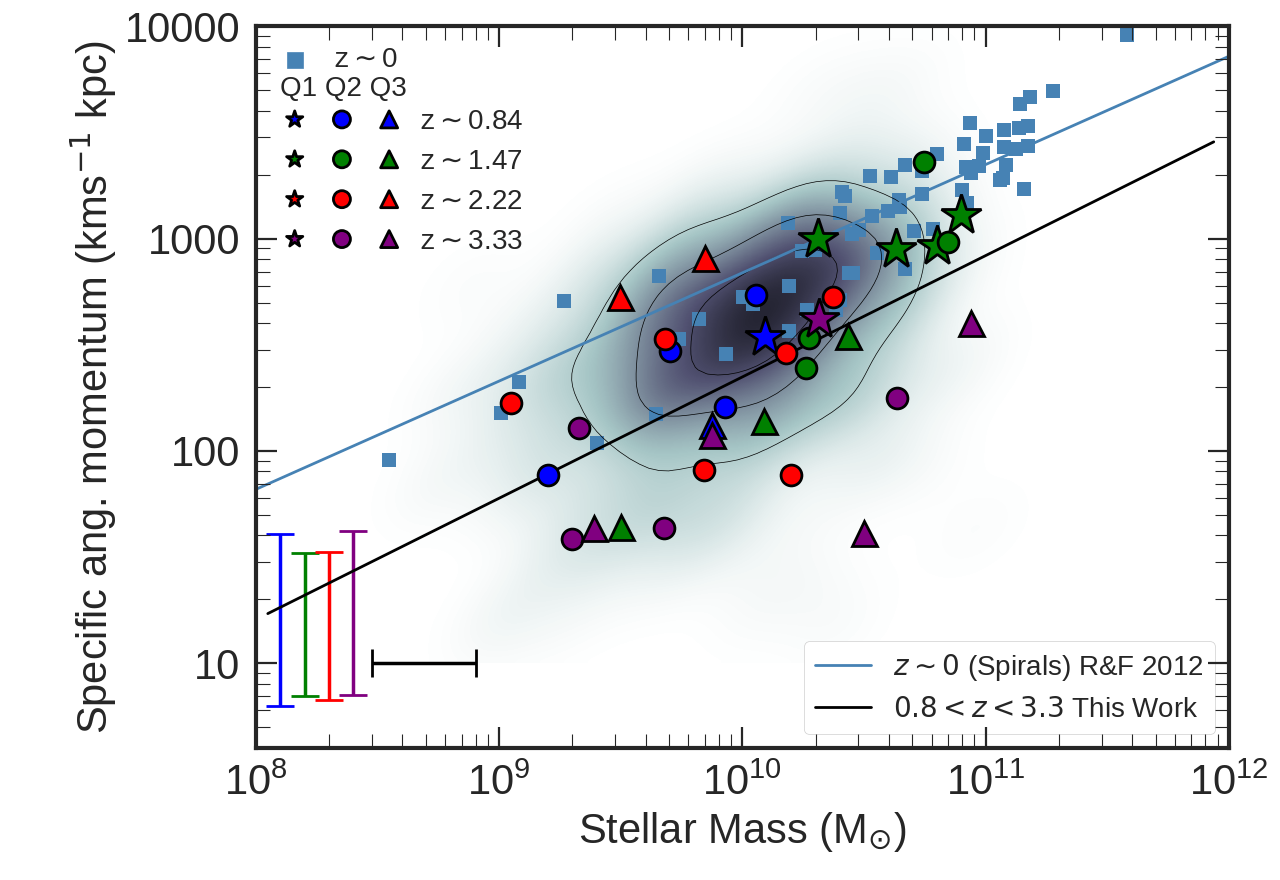}
	\caption{Specific stellar angular momentum as measured at 2R$_{\rm h}$ as a function of stellar mass. The sample coloured by spectroscopic redshift as as shown in Figure \ref{Fig:MS}, and the blue shaded regions represents the KROSS $z$\,$\sim$\,0.8 sample \citep{Harrison2017}. The \emph{stars} represent `Quality 1' targets (V$_{\rm rot, 2R_h}/\sigma_{\rm med}$\,$>$\,1 and $\Delta$PA\SB{im,vel}$\Psi$\,$<$\,30\SP{$\circ$}), \emph{circles} `Quality 2' (V$_{\rm rot, 2R_h}/\sigma_{\rm med}$\,$>$\,1 or $\Delta$PA\SB{im,vel}$\Psi$\,$<$\,30\SP{$\circ$}) and \emph{triangles} `Quality 3' galaxies (V$_{\rm rot, 2R_h}/\sigma_{\rm med}$\,$<$\,1 and $\Delta$PA\SB{im,vel}$\Psi$\,$>$\,30\SP{$\circ$} ). The $z$\,$\sim$\,0 \citet{Romanowsky2012} comparison sample is shown, with the fit to the data of the form $\log_{10}(j_*)$\,=\,$\alpha$+$\beta(\log_{10}(M_*/M_{\odot})-10.10)$, with $\alpha$\,=\,2.89 and $\beta$\,=\,0.51, whilst for KROSS ($z$\,$\sim$\,1) $\alpha$\,=\,2.58 and $\beta$\,=\,0.62.  Our sample appears in good agreement with other $z$\,$\sim$\,1 samples, having lower specific stellar angular momentum for a given stellar mass than galaxies at $z$\,$\sim$\,0, with a $\alpha$\,=\,2.41 and $\beta$\,=\,0.56. The median uncertainty on specific angular momentum  at each redshift is shown in the lower left corner as well as the uncertainty of the stellar mass.}
\label{Fig:JMS}
\end{figure*}

\section{Angular Momentum}\label{Sec:AnD}

With a circular velocity, stellar mass, and size derived for each galaxy, we can now turn our attention to analysing  the angular momentum properties of our sample. First we investigate the galaxy stellar specific angular momentum of the disc. We then take advantage of the high resolution of the data, and study the distribution of angular momentum within each galaxy.

\subsection{Total Angular Momentum}

We start by deriving the stellar specific angular momentum ($j_{*}$=J$_{*}$\,/\,M$_{*}$) for the 34 star--forming galaxies in our sample. This quantity, unlike other relations between stellar mass and circular velocity, comprises of three uncorrelated variables with a mass scale and a length scale times a rotation--velocity scale (\citealt{Fall1980}, \citealt{Fall1983}). The stellar specific angular momentum also removes the inherent scaling between the total angular momentum and mass. It is derived from

\begin{equation}\label{Eqn:AngMom} 
j_{*}=\frac{\rm J_*}{\rm M_{*}}=\frac{\int(\textbf{r} \times \bar{\textbf{v}}(r) ) \rho_{*}(r) \rm d^3 \textbf{r}}{\int \rho_{*}(r) \rm d^3 \textbf{r}},
\end{equation}
where \textbf{r} and $\bar{\textbf{v}}$ are the position and mean-velocity vectors (with respect to the centre of mass of the galaxy) and $\rho(r)$ is the three--dimensional density of the stars and gas \citep{Romanowsky2012}. 

In order to compare between observations and empirical models (or numerical models, as we will in Section \ref{Sec:EAGLE}), this expression can be simplified to be a function of the intrinsic circular rotation velocity of the star--forming gas and the stellar continuum half--light radius.  These intrinsic properties of the galaxy are correlated to the observable rotation velocity and disc scale length by the inclination of the galaxy and the PSF of the observations. As derived by \cite{Romanowsky2012}, this expression can be expanded to incorporate non-exponential discs. The specific angular momentum can be written as function of inclination and S\'ersic index\footnote{See \cite{Romanowsky2012} and \cite{Glazebrook2014} for the full derivation and discussion of this approach.} 
\begin{equation}\label{Eqn:AngMom3}
j_{*} = k_{\text{n}}C_{\text{i}}v_{\text{s}}R_{\text{h}},
\end{equation}
Where $v_{\text{s}}$ is the rotation velocity at 2\,$\times$ the half-light radii (R$_{\text{h}}$), C$_{i}$ is the correction factor  for inclination, assumed to be $\sin^{-1}$($\theta_{inc}$) (see Appendix A of \citealt{Romanowsky2012}) and k$_{n}$ is a numerical coefficient that depends on the S\'ersic index, $n$, of the galaxy and is approximated as:
\begin{equation}\label{Eqn:AngMom4}
k_{n} = 1.15+0.029n+0.062n^2, 
\end{equation}
We derive the specific stellar angular momentum of all 34 galaxies in our sample, adopting the appropriate S\'ersic index for each galaxy as measured in Section \ref{Sec:Sizes}, and for comparison we compare this to the specific angular momentum of the galaxies from the KROSS survey at $z\sim$\,0.8 (derived in the same way), as a function of stellar mass in Figure \ref{Fig:JMS}. We also show the specific angular momentum of $z$\,$\sim$\,0 disc galaxies from \cite{Romanowsky2012}. 
The full range of specific stellar angular momentum in the sample is $j_{*}$\,=\,40\,--\,2200\,km\,s$^{-1}$kpc with a median value of $\langle$\,$j_{*}$\,$\rangle$\,=\,294\,$\pm$\,70km\,s$^{-1}$kpc. 

\begin{figure}
	\centering
	\includegraphics[width=\linewidth]{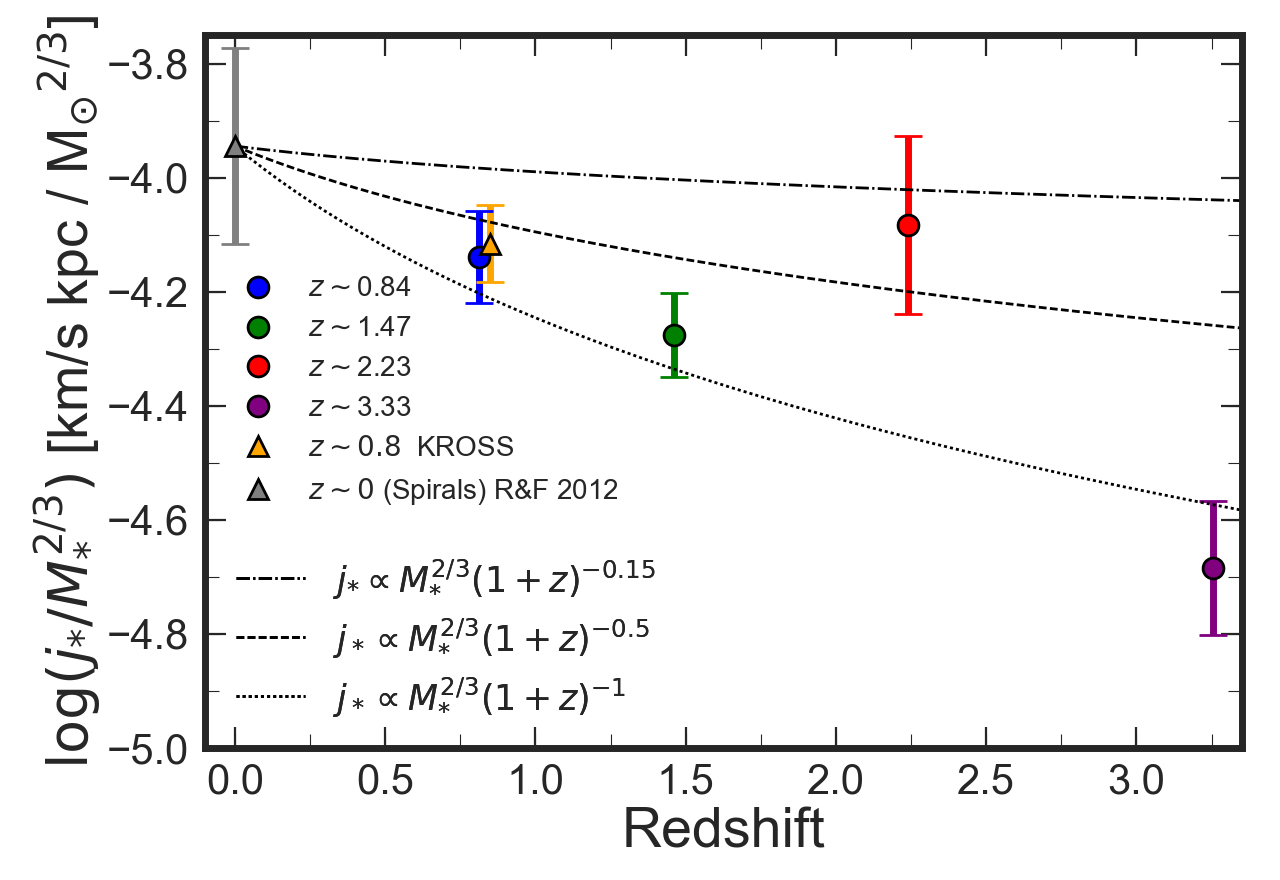}
	\caption{The redshift evolution of j$_*$/M$_*^{2/3}$ from $z$\,$\sim$\,0 to $z$\,$\sim$\,3.3. We show our sample coloured by redshift, as Figure \ref{Fig:MS}, as well the $z$\,=\,0 discs from \citet{Romanowsky2012} and the $z$\,=\,0.8 KROSS sample from \citet{Harrison2017}. We overlay tracks of j$_*$/M$_*^{2/3}$\,$\propto$\, (1+$z$)$^{-n}$, with $n$\,=\,0.15\,--\,1 as derived in \citet{Obreschkow2015}. Our galaxies show good agreement with other high redshift samples, and overall demonstrate a trend of decreasing j$_*$/M$_*^{2/3}$ with increasing redshift.}
	\label{Fig:jm_EVO}
\end{figure}

The $j_{*}$\,--\,M$_{*}/M_{\odot}$ relation can also be quantified by the relation $\log_{10}(j_*$)=$\alpha$+$\beta$(log$_{10}$(M$_*$/M$_{\odot}$)\,--\,10.10). For the $z$\,$\sim$\,0 sample, as derived in \cite{Romanowsky2012}, $\alpha$\,=\,2.89 and $\beta$\,=\,0.51. We fit the same model to our sample and derive $\alpha$\,=\,2.41\,$\pm$\,0.05 and $\beta$\,=\,0.56\,$\pm$\,0.03. This demonstrates that our sample has low specific angular momentum for a given stellar mass but with approximately the same dependence on stellar mass. This evolution in intercept was also identified in KROSS at $z$\,$\sim$\,0.8 with  $\alpha$\,=\,2.55 and $\beta$\,=\,0.62 \citep{Harrison2017}. 

We note however that other integral field studies of high--redshift star--forming galaxies such as \citet{Contini2016} and \citet{Marasco2019} find no evolution in the intercept of the specifc stellar angular momentum and stellar mass relation for high redshift galaxies. Both these studies model the integral field data in three dimensions using a model data cube. 
In addition \cite{Marasco2019} derive the specific stellar angular momentum of their sample directly from surface-brightness profiles of the galaxy as opposed to the approximations of angular momentum given in Equation \ref{Eqn:AngMom3}. 

One prediction of $\Lambda$CDM, is that the relation between the mass and angular momentum of dark matter haloes evolves with time \citep{Mo1998}. In a simple, spherically symmetric halo in a matter-dominated Universe, the specific angular momentum, $j_{\rm h}$\,=\,J$_{\rm h}$/M$_{\rm h}$ should scale as $j_{\rm h}$\,=\,M$_{\rm h}^{2/3}$(1+$z$)$^{-1/2}$  and if the ratio of stellar-to-halo mass is independent of redshift, then the specific angular momentum of baryons should scale as $j_{\rm *}$\,$\,\propto\,$M$_{\rm *}^{2/3}(1+z)^{-1/2}$ \citep{Behroozi2010,Munshi2013}. At $z$\,$\sim$\,3 this simple model predicts that the specific angular momentum of discs should be a factor of $\sim$2 lower than at $z$\,=\,0. However, this `closed box' model does not account for gas inflows or outflows, which can significantly affect the angular momentum of galaxy discs, with the redistribution of low--angular--momentum material from the central regions to the halo and the accretion of higher angular momentum material at the edges of the disc. This model further assumes the halo lies in a matter--dominated Universe, which only occurs at $z$\,$\gtrapprox$\,1. At lower redshifts the correlation is expected to be much weaker with 
$j_{\rm *}$\,$\,\propto\,$M$_{\rm *}^{2/3}(1+z)^{-0.15}$ \citep{Catelan1996, Obreschkow2015}. To search for this evolution in our sample, we derive j$_*$/M$_*^{2/3}$ at each redshift slice (Figure \ref{Fig:jm_EVO}) and compare to the KROSS $z$\,$\sim$\,0.8 sample as well the \cite{Romanowsky2012} disc sample at $z$\,$\sim$\,0. We find that galaxies in our sample between $z$\,=\,0.8\,--\,3.33 follow the scaling of j$_*$/M$_*^{2/3}$\,$\propto$\, (1+$z$)$^{-n}$ well, with lower specific angular momentum for a given stellar mass at higher redshift. Future work on larger non-AO samples of high-redshift star--forming galaxies, such as the KMOS Galaxy Evolution Survey (KGES), will explore this correlation further (e.g. Gillman et al. in prep)

To understand the angular momentum evolution of the galaxies in our sample, we can go beyond a measurement of size and asymptotic rotation speed and take advantage of the resolved dynamics. Next we investigate how the radial distribution of angular momentum changes as a function of stellar mass and redshift to constrain how the internal distribution of angular momentum might affect the morphology of galaxies.

\subsection{Radial Distribution of Angular Momentum}

To quantify the angular momentum properties of the galaxies in our sample and to provide empirical constraints on the evolution of main--sequence galaxies, from turbulent clumpy systems at high redshift with high velocity dispersion, to the well--ordered `Hubble'--type galaxies seen in the local Universe, we can measure their internal dynamics. This is made possible with our adaptive optics sample of galaxies, with $\sim$kpc resolution integral field observations. In this section we discuss the method and show results for the construction of one dimensional radial angular momentum profiles of each galaxy. 

We analyse the total stellar angular momentum distribution in the `Quality 1\,$\&$\,2' galaxies, galaxies with V$_{\rm rot, 2R_h}/\sigma_{\rm median}$\,$>$\,1 or $\Delta$PA\SB{im,vel}$\Psi$\,$<$\,30\SP{$\circ$} in our sample, as opposed to the specific stellar angular momentum in order to account for the evolution of the stellar mass distribution in galaxies with cosmic time. We focus on `Quality 1\,$\&$\,2' galaxies as these are the galaxies that most resemble star--forming kinematically stable `rotationally supported' galaxies in our sample.

We infer how the angular momentum distribution changes by extracting the radius that encompasses 50 per cent of the total (R$_{\rm J50}$). We explore how this radius evolves as a function of redshift and to aid the interpretation compare it to fixed--mass and evolving--mass evolution tracks of R$_{\rm J50}$ derived from a suitably selected sample of galaxies drawn from the {\sc{eagle}} hydro-dynamical cosmological simulation from 0.1\,$\leq$\,$\,z\,$\,$\leq$\,3.5 \citep{Joop2015,Crain2015}.

\begin{figure}
	\centering
	\includegraphics[width=\linewidth]{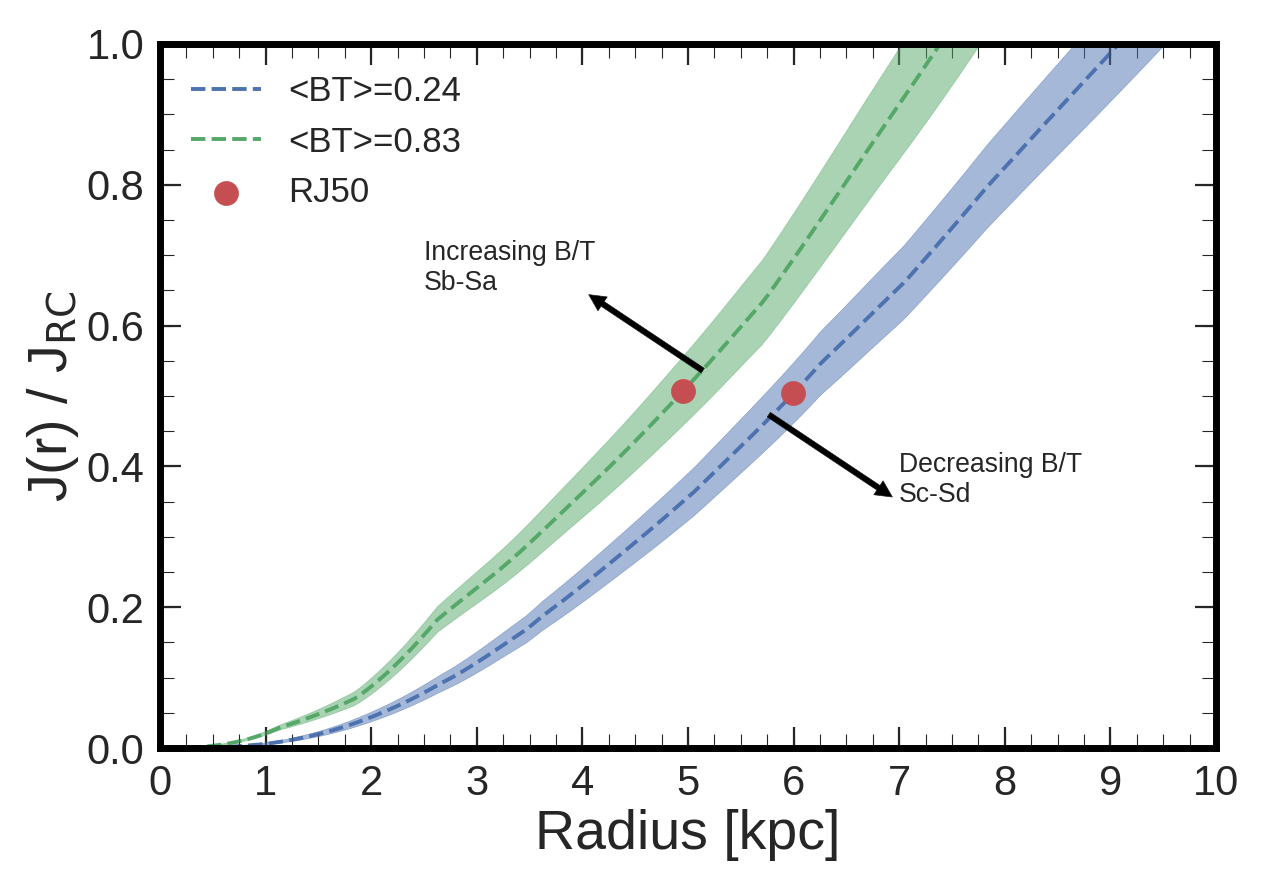}
	\caption{The total stellar angular momentum as a function of radius, normalized by the rotation curve estimate of the total angular momentum (Equation \ref{Eqn:AngMom3}) for EAGLE galaxies with stellar mass $\rm \log({10.5/M_{\odot}})$ at $z$\,=\,0.1. We define two subsamples of galaxies using the B/T values defined in \citet{Trayford2018}. We require B/T\,$>$\,0.6 for a galaxy to be defined as bulge dominated, identifying a median B/T value for these galaxies of $<$\,B/T\,$\rangle$\,=\,0.83 that resemble Sb-Sa early-type galaxies. We also define a sample of disc-dominated galaxies, with the criteria B/T\,$<$\,0.4. These galaxies  align more with Sc-Sd late-type galaxies and have a median B/T value of $\langle$\,B/T\,$\rangle$\,=\,0.24. On average EAGLE galaxies of the same stellar mass, but with a more bulge--dominated morphology have a smaller radii containing 50 per cent of the angular momentum (R$_{\rm J50}$).}
\label{Fig:Jprofile}
\end{figure}

\begin{figure}
	\centering
	\includegraphics[width=\linewidth,trim={1.5cm 0.5cm 25cm 0},clip]{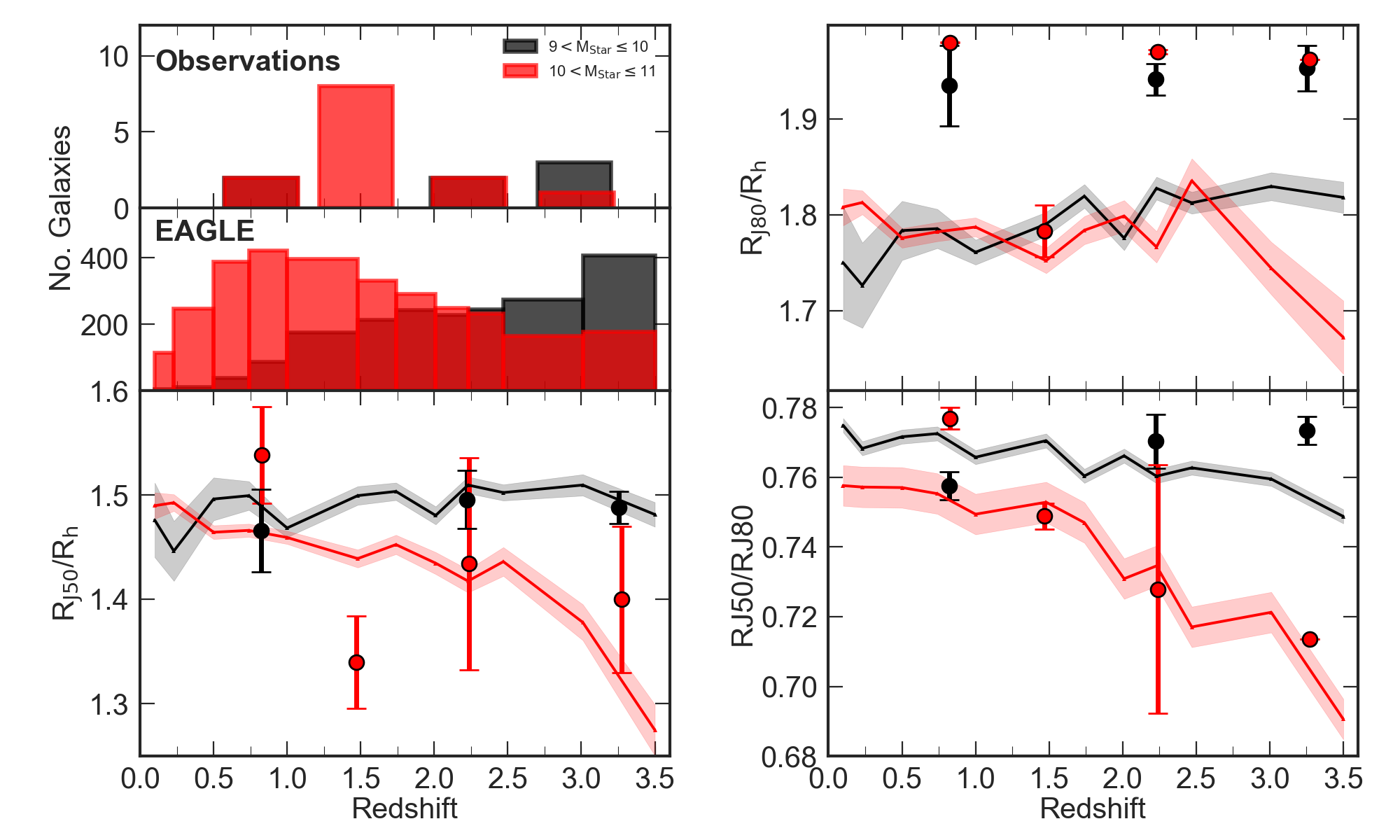}
	\caption{\emph{Top}: the distributions of galaxies in each redshift slice for our sample and {\sc{eagle}}. \emph{Bottom:}  the radius (R$_{\rm J50}$) within which 50 per cent  of the galaxy's angular momentum is contained, normalized by the half--light radius of the galaxy, as a function of redshift. Coloured points indicate the galaxies in our sample split into two stellar mass bins. The tracks show the median and 1$\sigma$ evolution of {\sc{eagle}} galaxies in the same redshift and stellar mass bins. R$_{\rm J50}$/R$_{\rm h}$ in lower stellar mass galaxies shows no evolution with cosmic time whilst for higher mass galaxies a tentative evolution in the observational sample is seen. In {\sc{eagle}} a similar trend is visible with higher stellar mass galaxies showing an increase in R$_{\rm J50}$/R$_{\rm h}$ increasing by $\sim$16 per cent from z\,$\sim$\,3.5 to z\,$\sim$\,0.1.} 
	\label{Fig:mfix}
\end{figure}

\subsubsection{Angular Momentum Profile}

We derive a stellar mass profile for each galaxy from the broad--band photometry, as shown in Appendix \ref{App:B}, Table \ref{Tab:Obs}. We first construct a one--dimensional surface brightness profile for each galaxy by placing elliptical apertures on the broad--band photometry of the galaxy. 
We measure the surface brightness within each aperture (deconvolving the profile with the broad--band PSF). We assume mass follows light, with the total stellar mass derived from the SED fitting, as for most objects with $HST$ coverage we only have single--band photometry and so are unable to measure (or include) mass--to--light gradients.

We use the circular velocity profiles as derived in Section \ref{Sec:Vcirc} in order to account for the pressure support from the turbulent gas in the galaxies in our sample as well as to align more accurately with the dynamical rotation curves of the {\sc{eagle}} galaxies (Section \ref{Sec:EAGLE}). We combine these with the stellar mass profiles. For each galaxy we measure the integrated stellar angular momentum as a function of radius J(r), which is then normalized against the total angular momentum estimate (Equation \ref{Eqn:AngMom3}). 

We then extract the radii at the which profile reaches 50 per cent of its total. Since galaxy sizes also evolve with redshift \citep[e.g.][]{Roy2018}, we normalize by the galaxy's half--light radius, in order to remove this intrinsic scaling. An example of the angular momentum profiles for a sample of {\sc{eagle}} galaxies at $z$\,$\sim$\,0.1 is shown in Figure \ref{Fig:Jprofile}.

To remove the implicit scaling between stellar mass and angular momentum distribution, we split the galaxies in our observed sample at each redshift slice in our sample into two stellar mass bins, 9$<$\,$\log$(M$_{*}$[M$_{\odot}$])$\leq$10 and 10$<$\,$\log$(M$_{*}$[M$_{\odot}$])$\leq$11. In Figure \ref{Fig:mfix} we show how R$_{\rm J50}$ for both low-- and high--stellar--mass galaxies evolves with cosmic time. In the lowest stellar mass bin, the distribution of angular momentum remains constant  whilst for the higher stellar mass galaxies (10$<$\,$\log$(M$_{*}$[M$_{\odot}$])$\leq$11) there is a weak trend with redshift, with $\langle$R$_{\rm J50}$$_{\rm z \sim 3.5}$/R$_{\rm J50}$$_{\rm z \sim 0.84}$\,$\rangle$=0.91\,$\pm$\,0.01. If the radius which encloses 50 per cent of the angular momentum in the galaxy has increased with cosmic time, relative to the size of the galaxy, this would suggest there is more angular momentum at larger radii in low--redshift galaxies i.e the angular momentum in the galaxies has grown outwards with cosmic time.

In order to understand further the tentative trend that R$_{\rm J50}$/R$_{\rm h}$ increases in galaxies with stellar mass 10$<$\,$\log$(M$_{*}$[M$_{\odot}$])$\leq$11, as suggested by our observational sample, we make a direct comparison to the {\sc{eagle}} hydrodynamical simulation which provides a significant comparison sample across a broad range of redshift.

\subsubsection{EAGLE Comparison}\label{Sec:EAGLE}

To understand the context of the evolution of  angular momentum in our sample, we make a direct comparison to the Evolution and Assembly of GaLaxies and their Environments ({\sc{eagle}}) hydrodynamical simulation \citep{Joop2015,Crain2015}. 

The {\sc{eagle}} simulation follows the evolution of dark matter, stars, gas and black holes in a 10$^6$ Mpc$^3$ cosmological volume from $z$\,$\sim$\,10 to $z$\,$\sim$\,0, recreating the local Universe galaxy stellar mass function and colour--magnitude relations to high precision. It therefore provides a useful test bed to understand the observational biases and further interpret the angular momentum distributions in our galaxies. 

Prior to making a comparison between the angular momentum properties of {\sc{eagle}} galaxies and our observational sample, we first test the accuracy of using the {\sc{eagle}} rotation curves as an estimate of the total angular momentum of the galaxy. The angular momentum of {\sc{eagle}} galaxies can be derived directly from the sum of angular momentum of each star particle (J$_{\text{ps}}$) assigned to the galaxy, where

\begin{equation}\label{Eqn:Jparticle}
J_{\text{ps}}=\sum_{i} m_i r_i \times v_i,
\end{equation}

The rotation curves in {\sc{eagle}} galaxies, as derived in \cite{Schaller2015}, are generated by assuming circular motion for all the bound material in a galaxy's halo. The simulated galaxies match the observations exceptionally well, in
terms of both the shape and the normalization of the curves \citep[for a full comparison to observations, see][]{Schaller2015,Schaye2015}.

In order to test whether our estimates of the total angular momentum from the rotation curves (J$_{\text{RC}}$) using Equation \ref{Eqn:AngMom3} are in good agreement with the particle angular momentum, we derive J$_{\text{RC}}$  for each {\sc{eagle}} galaxy using Equation \ref{Eqn:AngMom3} and \ref{Eqn:AngMom4} (with $n$\,=\,1).

In galaxies with high stellar particle angular momentum, J$_{\text{RC}}$ on average accurately estimates the total angular momentum of the galaxy with $\langle$\,J$_{\rm ps}$/J$_{\rm rc}$$\rangle$\,=\,0.69\,$\pm$\,0.05. We select galaxies in {\sc{eagle}} where J$_{\text{ps}}$\,$<$J$_{\text{RC}}$\,$<$\,2J$_{\text{ps}}$ and adopt  J$_{\text{ps}}$ as the estimate of the total angular momentum of {\sc{eagle}} galaxies.

\begin{figure}
	\centering
	\includegraphics[width=\linewidth,trim={1.5cm 0.5cm 25cm 0},clip]{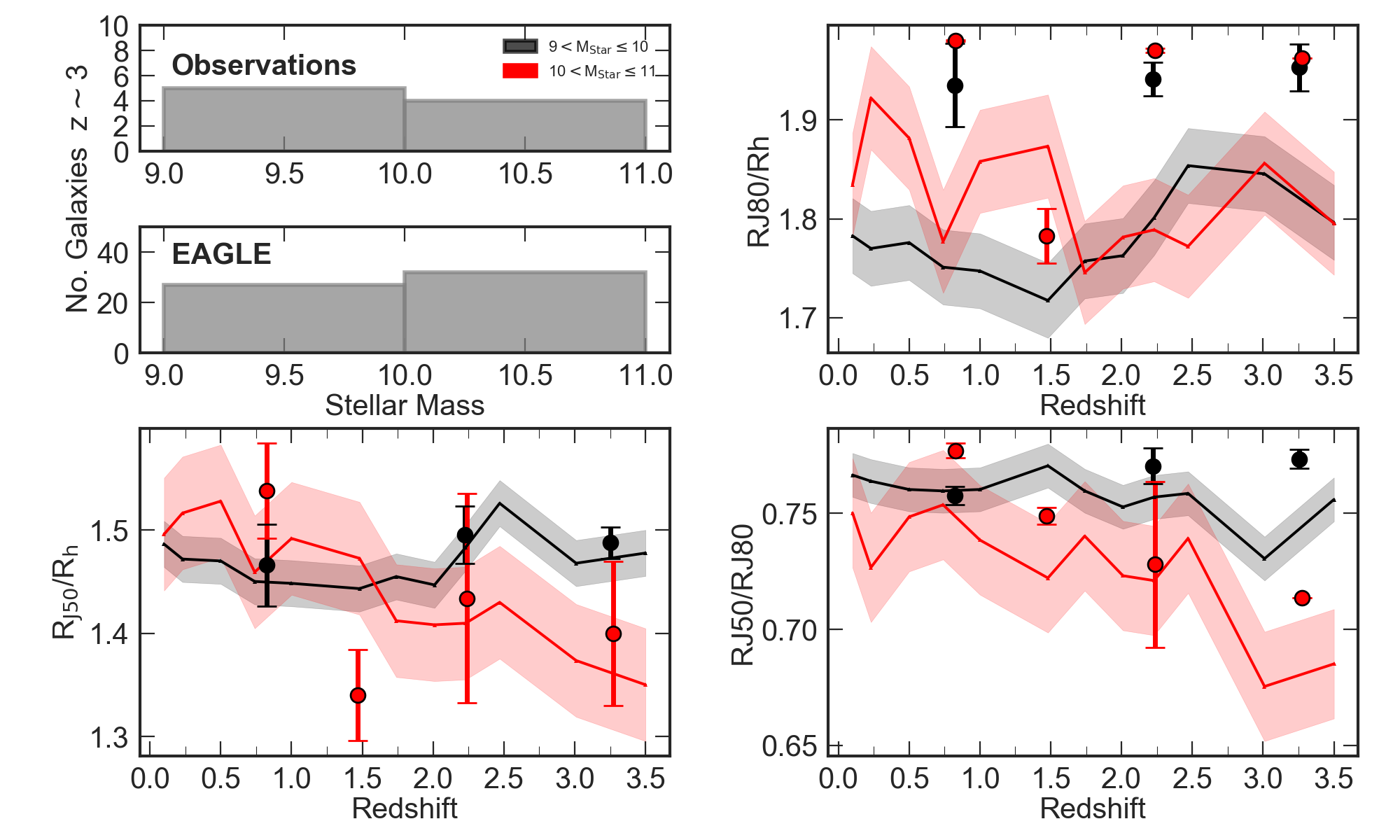}
	\caption{\emph{Top}: The stellar mass distribution of our sample and  {\sc{eagle}} galaxies at $z$\,=\,3. \emph{Bottom:} The radius (R$_{\rm J50}$) within which 50 per cent of the galaxy's angular momentum is contained, normalised by the half--light radius of the galaxy  as a function of redshift. Coloured points indicate the galaxies in our sample split into two stellar mass bins. The tracks show the median and 1$\sigma$ evolution of {\sc{eagle}} galaxies selected by stellar mass at $z\,=\,3$. For the {\sc{eagle}} galaxies, we apply the stellar mass and star formation criteria at z\,$\sim$\,3 and trace the galaxies back to z$\sim$\,0.1 using the {\sc{eagle}} merger trees, thus incorporating the mass evolution of galaxies. The galaxies in our sample have the mass criteria applied at their redshift and therefore shouldn't be compared directly to the tracks. We see similar evolution as the fixed--mass tracks (Figure \ref{Fig:mfix}) with R$_{\rm J50}$/R$_{\rm h}$. increasing by $\sim$11 per cent from z\,$\sim$\,3.5 to z\,$\sim$\,0.1. and minimal evolution in the lower stellar mass galaxies.}
	\label{Fig:mevo}
\end{figure}

\subsubsection{Fixed Mass Evolution}

To compare directly the angular momentum properties of {\sc{eagle}} galaxies to those of our sample, we first match the selection function of the observations at each redshift snapshot in {\sc{eagle}}. We select galaxies in {\sc{eagle}} with stellar masses between $\log$(M$_{*}$[M$_{\odot}$])=\,9\,--\,11 and star formation rates SFR[M$_{\odot}$yr$^{-1}$]\,=\,2\,--\,120, which covers the range of our sample. 

Following the same procedures as for the observations, we derive one-dimensional angular momentum profiles for each galaxy and measure R$_{\rm J50}$ (Figure \ref{Fig:Jprofile}). We do this for all {\sc{eagle}} galaxies from 0.1\,$\leq$\,$\,z\,$\,$\leq$\,3.5. We split the sample into the two stellar mass bins, applying the mass and star formation selection of the observations  at each redshift snapshot. In Figure \ref{Fig:mfix} we plot median tracks of R$_{\rm J50}$ (normalized by the half stellar mass radius) as a function of redshift.

\begin{figure*}
	\centering
	\includegraphics[width=\linewidth]{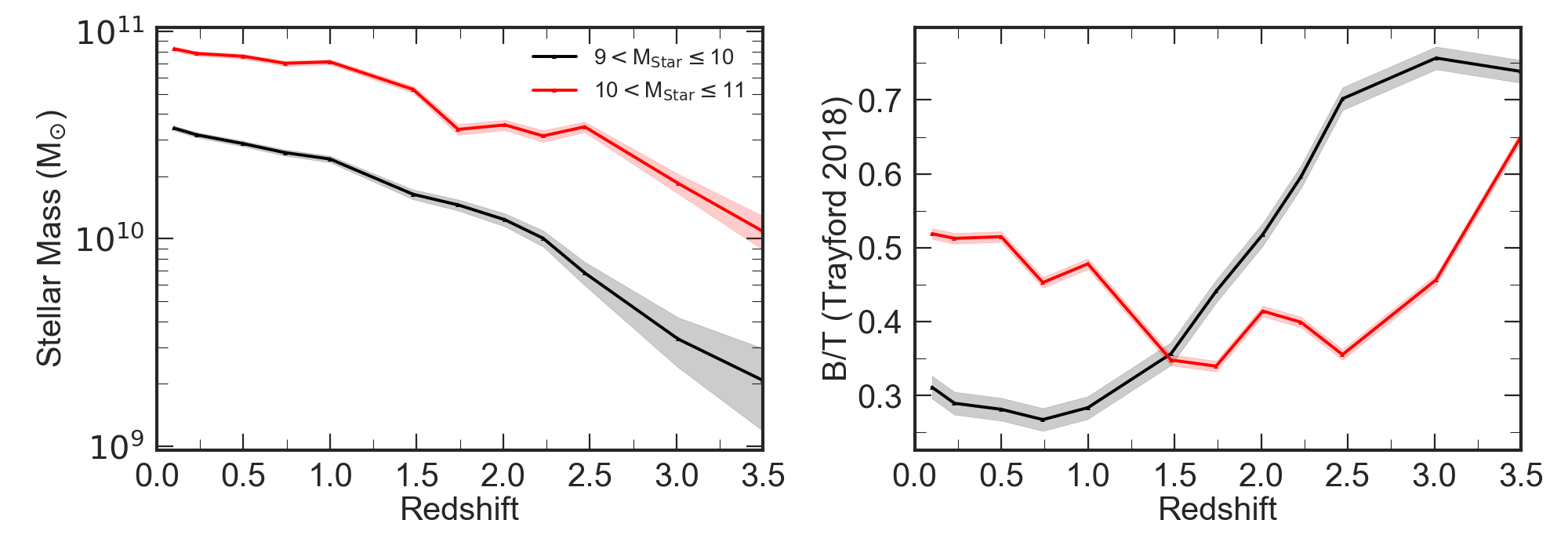}
	\caption{The stellar mass (\emph{left}) and  bulge-total fraction (\emph{right}) as a function of redshift for {\sc{eagle}} galaxies selected at $z$\,=\,3 with $\log$(M$_{*}$[M$_{\odot}$])=\,9\,--\,11 and star formation rates SFR[M$_{\odot}$yr$^{-1}$]\,=\,2\,--\,120. We split the galaxies into stellar mass bins of 9$<$\,$\log$(M$_{*}$[M$_{\odot}$])$\leq$10 and 10$<$\,$\log$(M$_{*}$[M$_{\odot}$])$\leq$11. Both stellar bins show comparable evolution in stellar mass with redshift. The B/T values for galaxies with stellar mass between  10$<$\,$\log$(M$_{*}$[M$_{\odot}$])$\leq$11 indicate the formation pseudo-bulges at $z$\,$<$\,1.5 compared to 9$<$\,$\log$(M$_{*}$[M$_{\odot}$])$\leq$10 stellar mass galaxies which continue to maintain there bulge-total fractions beyond $z$\,$<$\,1.0 }
\label{Fig:Mstar_BT_z}
\end{figure*}

The evolution of {\sc{eagle}} galaxies' angular momentum distribution agrees well with the evolution in our sample. {\sc{eagle}} predicts little evolution in the lowest stellar mass bin, with R$_{\rm J50}$ remaining approximately constant from $z$\,=\,3.5 to $z$\,=\,0.1. The higher stellar mass galaxies show an evolution from R$_{\rm J50}$$_{z\sim3.5}$\,=\,1.27\,$\pm$\,0.02 to R$_{\rm J50}$$_{z\sim0.1}$\,=\,1.48\,$\pm$\,0.01, an increase of $\sim$16 per cent. The distribution of angular momentum in high--stellar--mass galaxies is growing outwards with increasing cosmic time. A galaxy of stellar mass 10$^{10.5}$M$_{\odot}$ at $z$\,=\,3.5 will have a more concentrated angular momentum distribution, normalized to its half--light radius, than a 10$^{10.5}$M$_{\odot}$ galaxy at $z$\,=\,0.1. 
This evolution in the angular momentum distribution could be driven by a number of physical processes. The accretion of high--angular--momentum material to the outer regions of the galactic disc would act to increase the total angular momentum and thus R$_{\rm J50}$ of the galaxy.

Over the cosmic time between $z$\,=\,3 and $z$\,=\,0.1 ($\sim$\,10\,Gyr) galaxies grow in stellar mass \citep[e.g.][]{Baldry2012,Behroozi2013B,Furlong2015M,Roy2018}. Based on the {\sc{eagle}} simulation \citep{Crain2015,Schaye2015}, a galaxy in our $z$\,=\,3 sample would grow by a factor $\sim$\,10 in stellar mass (factor of $\sim$\,3 for a $z$\,=\,2 galaxy and a factor of $\sim$\,1.5 for a $z$\,=\,1 galaxy).  The gain in stellar mass dominates the stellar mass that is in place at higher redshift. Thus we expect that the changes in galaxy angular momentum and its distribution arise primarily from the accretion of new star--forming gas. As the angular momentum of the infalling gas grows with time, the recently formed stellar population will have a higher angular momentum compared to the total stellar population. \citep[e.g.][]{Catelan1996,Obreschkow2015}. 

The removal of low--angular--momentum material via nucleated outflows driven by stellar winds would redistribute the angular momentum in the galaxy. If the evolution of the angular momentum is being driven by nucleated outflows from across the galactic disc, we expect a similar increase in R$_{\rm J50}$ with low--angular--momentum material being removed. In situ bulge formation at the centre of galaxies, increasing the fraction of low--angular--momentum material, would alter the angular momentum profile of the galaxy. We note that we are studying the angular momentum evolution of star--forming gas associated with young massive stars. The older stars may  have their orbit perturbed over time to form the galaxy's bulge. This complicates the interpretation of R$_{\rm J50}$/R$_{\rm h}$, but leads to a model in which the stellar bulge--to--total (B/T) ratio of the galaxy may be an effective measure of its past to current star formation rate. Recently, \cite{Liang2018} identified that the impact of bulge formation on a galaxy's angular momentum distribution depends on the significance of the bulge, with very high B/T galaxies maintaining their original angular momentum distribution. 

It is important to remember, however, that the galaxy sample we identify at higher redshift does not evolve into the galaxy sample at $z$\,=\,0.  Many of the $z$\,=\,3 galaxies with stellar masses $\sim$\,10$^{10.5}$M$_{\odot}$ are likely to be $\sim$\,10$^{11}$M$_{\odot}$ at $z$\,$\sim$\,0 and will evolve into passive  elliptical galaxies, perhaps at the centres of galaxy groups. These galaxies may become passive due to the the impact of black holes \citep[e.g.][]{Bower2006,Bower2017,Davies2018}. Other galaxies may merge with larger central group galaxies and disappear from observational samples entirely. A galaxy of stellar mass $\sim$\,10$^{9.5}$M$_{\odot}$ at $z$\,$\sim$\,3 is likely to be $\sim$\,10$^{10}$M$_{\odot}$ at $z$\,$\sim$\,0 and thus more likely to evolve into late-type `disc' galaxy at low-redshift. Instead of the observations tracing individual galaxies, we are viewing a sequence of snapshots of the star--forming population at each epoch, and exploring how the angular momentum evolves in this sense.

The selection function used in observations and {\sc{eagle}} comparison from  $z$\,=\,3.5\,--\,0.1 for the radii derived in Figure \ref{Fig:mfix} are not selecting the same descendent populations. To understand whether the evolution of R$_{\rm J50}$ is driven by the accretion of new material or bulge formation, we need to study the galaxies as they evolve. {\sc{eagle}} allows us to follow the evolution of individual galaxies through cosmic time, which is what we now finally focus on. 

\subsubsection{Evolving Mass Evolution}

One of the main advantages of a hydrodynamical simulation is having the ability to trace the evolution of individual galaxies across cosmic time. The mass evolution of a given galaxy can be traced as it evolves via secular processes and interactions with other galaxies. This is achieved using the merger trees output by the simulation \citep{McAlpine2016,Qu2017}. We can use this information to derive the evolution of R$_{\rm J50}$ from $z$\,=\,3.5 to $z$\,=\,0.1 in individual {\sc{eagle}} galaxies selected at high redshift. 

We derive the radius containing fifty percent of the galaxies angular momentum (R$_{\rm J50}$) for galaxies with $\log$(M$_{*}$[M$_{\odot}$])\,=\,9\,--\,11 and SFR$\geq$\,2\,M$_{\odot}$yr$^{-1}$ at $z$\,=\,3. In Figure \ref{Fig:mevo} we show the evolution of R$_{\rm J50}$ for these galaxies split into the two stellar mass bins at $z$\,$\sim$\,3 as well as our observational sample for reference. We note the data points should not be directly compared to the {\sc{eagle}} tracks due to differences in selection. 
The higher stellar mass {\sc{eagle}} galaxies in Figure \ref{Fig:mevo} show evolution in R$_{\rm J50}$ with R$_{\rm J50}$$_{z\sim3.5}$\,=\,1.23\,$\pm$\,0.05 to R$_{\rm J50}$$_{z\sim0.1}$\,=\,1.37\,$\pm$\,0.03, an increase of $\sim$11 per cent. 
The evolution of angular momentum, quantified by R$_{\rm J50}$, in {\sc{eagle}} galaxies with $\log$(M$_{*}$[M$_{\odot}$])\,=\,9\,--\,11 and SFR[M$_{\odot}$yr$^{-1}$]\,=\,2\,--\,120 at $z$\,=\,3 increases with cosmic time. The angular momentum in these galaxies is becoming less centrally concentrated as the galaxy evolves, as indicated in Figure \ref{Fig:mfix}. 

To understand the physical processes driving the increase of the R$_{\rm J50}$ relative to the half--light radius of higher stellar mass galaxies, we analyse the stellar mass growth and evolution of the stellar bulge--total (B/T) fraction in these galaxies (Figure \ref{Fig:Mstar_BT_z}). The stellar mass of the galaxy is extracted at each redshift snapshot in the  {\sc{eagle}} simulation. The bulge--to--total ratios are taken from \cite{Trayford2018}, where the disc fraction of the galaxy is defined as the prograde excess (the mass in co-rotation above what would be expected for a purely pressure-supported system) and the B/T is the complement of this. 

In {\sc{eagle}} star--forming galaxies with stellar mass between 10$<$\,$\log$(M$_{*}$[M$_{\odot}$])$\leq$11 at $z$\,=\,3.5 have significant bulge fractions: B/T\,=\,0.65\,$\pm$\,0.08. As the galaxies evolve with cosmic time their stellar mass grows through accretion of new material from the surrounding circumgalactic medium, increasing by a factor $\sim$5 by $z$\,=\,1.5. Their bulge fractions reduce to B/T\,=\,0.35\,$\pm$\,0.04 at $z$\,=\,1.5 and the radius containing 50 per cent of their stellar angular momentum (R$_{\rm J50}$) has increased by a $\sim$7 per cent relative to their half stellar mass radius in this period, indicating the presence of a more significant disc component in these galaxies from the recently accreted higher angular momentum material. Below $z$\,=\,1.5 the high stellar mass galaxies continue to accrete more material and the angular momentum continues to grow outwards with cosmic time, with R$_{\rm J50}$/R$_{\rm h}$ increasing by just $\sim$4 per cent from $z$\,=\,1.5  to $z$\,=\,0. The bulge fraction below $z$\,=\,1.5 however, begins to increase as these galaxies are massive enough to form pseudo-bulges, and resemble more Sa-Sb early-type morphologies. 

For lower stellar mass star--forming galaxies in  {\sc{eagle}} with 9$<$\,$\log$(M$_{*}$[M$_{\odot}$])$\leq$10 the distribution of stellar angular momentum remains roughly constant relative to the half stellar mass radius of the galaxies from $z$\,=\,3.5  to $z$\,=\,0. In this period, however, the galaxies' stellar mass has increased by a factor of 
$\sim$10 and the bulge fraction of the galaxies has significantly reduced from B/T\,=\,0.74\,$\pm$\,0.04 at $z$\,=\,3.5 to B/T\,=\,0.28\,$\pm$\,0.03 at $z$\,=\,1. From $z$\,=\,1 to $z$\,=\,0 the bulge-fraction of the galaxies remains relatively constant. This indicates that high redshift these lower stellar mass galaxies are compact and spheroidal and as they evolve they accrete new material from the  circumgalactic medium, which builds the disc component of the galaxies, driving them towards Sd-Sc late-type morphologies. Below $z$\,=\,1 the galaxies `settle' becoming more stable and maintain an approximately constant bulge fraction.

\section{Conclusions}\label{Sec:Conc}

We have presented H\,$\alpha$ and [O\,{\sc{iii}}] adaptive optics integral field observations of 34 star--forming galaxies from 0.8$\leq$\emph{z}$\leq$3.3 observed using the NIFS, SINFONI, and OSIRIS spectrographs. The sample has a median redshift of $\langle$\,$z$\,$\rangle$\,=\,2.22, and covers a range of stellar masses from $\log(M_{*}[M_{\odot}]$)\,=\,9.0\,--\,10.9, with `main-sequence' representative star formation rates of SFR$_{\rm H\alpha}$\,=\,2\,--\,120 \,M$_{\odot}$yr$^{-1}$. Our findings are summarized as follows,

\begin{description}
\item[$\bullet$] For 21 galaxies in our sample we measure continuum half--light sizes using $\emph{HST}$ photometry and ground--based broadband imaging from the parametric fitting of a single S\'ersic model. We find $\langle$R\SPSB{G}{h}/R\SPSB{HST}{h}\quad$\rangle$\,=\,0.97\,$\pm$\,0.05 (Figure \ref{Fig:sizes}). Applying the same fitting procedure to remainder of the sample we derive $\langle$R$_{\rm h}$\,$\rangle$\,=\,0.40\,$\pm$0.06 arcsec, $\sim$4kpc at the median redshift of the sample. We conclude the continuum sizes of the galaxies in our sample are comparable to other high-redshift star--forming galaxies such as those presented in \cite{Stott2013} and \cite{VanderWel2014}.

\item[$\bullet$] We identify that 11 ($\sim$32 per cent) of the galaxies in our sample have dynamics indicating they are supported by rotational gas kinematics, with rotational velocities that are the order of the intrinsic velocity dispersion. We measure a median $\langle$V$_{\rm rot, 2R_h}$/$\sigma_{\rm median}$\,$\rangle$\,=\,0.82\,$\pm$\,0.13 for the sample (Figure \ref{Fig:Vsig}). We compare the mass normalized V/$\sigma$ for our sample to that of other star--forming galaxy surveys, across a range of redshift, identifying that our sample follows a similar trend of increasing in V/$\sigma$ with cosmic time, as galaxies become more rotationally dominated.

\item[$\bullet$] We place our sample in the context of other integral field studies by exploring the relation between rotational velocity and stellar mass (Figure \ref{Fig:TF}). We identify no significant evolution in the relation since $z$\,$\sim$\,0. Our galaxies are consistent with the dispersion dominated KMOS Deep Survey at $z$\,$\sim$\,3.5 \citep{Turner2017} and other high-redshift surveys such as KROSS \citep{Tiley2019} and KMOS$^{\rm 3D}$ \citep{Ubler2018}. 

\item[$\bullet$] We combine the inclination-corrected rotational velocities, half--light sizes, and stellar masses, to investigate how the relationship between the specific stellar angular momentum and stellar mass in our sample evolves with cosmic time (Figure \ref{Fig:JMS}). We quantify the $j_{*}-M_{*}$ correlation with $\log(j_∗)$\,$\,=\,$\,$\alpha+\beta(\log(M_∗)$\,--\,10.10), finding $\alpha$\,=\,2.41\,$\pm$\,0.05 and $\beta$\,=\,0.56\,$\pm$\,0.03. The normalization of the $j_{*}$\,--\,$M_{*}$ relation for our sample is  smaller than other (non-AO) samples at $z$\,$\sim$\,1 and $z$\,$\sim$\,0 spiral galaxies.
 We derive the evolution of $j_{\rm *}$/M$_{\rm *}^{2/3}\propto(1+z)^{-n}$ for our sample (Figure \ref{Fig:jm_EVO}) identifying that the galaxies in our sample agree well with the prediction of $\Lambda$CDM with $n$\,=\,0.5\,--1.

\item[$\bullet$] Taking advantage of the $\sim$kpc resolution of our observations we investigate the radial distribution of angular momentum in each galaxy, deriving one-dimensional stellar angular momentum profiles. We quantify these profiles by the 50 per cent radii (R$_{\rm J50}$) and explore their median evolution with cosmic time for galaxies with stellar mass in the range 9$<$\,$\log$(M$_{*}$[M$_{\odot}$])$\leq$10 and 10$<$\,$\log$(M$_{*}$[M$_{\odot}$])$\leq$11. We identify in the higher stellar mass bin a tentative trend of increasing R$_{\rm J50}$/R$_{\rm h}$ with cosmic time (Figure \ref{Fig:mfix}).

\item[$\bullet$] We note, however, that the analysis we have undertaken on a sample of high redshift star--forming galaxies is limited by assumptions we have made. Throughout the analysis we assumed our galaxies resemble kinematically well--behaved `discs' and that the sample is representative of the high--redshift population. However, it is well known that peculiar galaxies become the dominant morphological population at higher redshift with galaxies having much higher velocity dispersions comparable to their rotational component. We therefore rely on hydrodynamical simulations to verify the conclusions we have drawn from the data. 

\item[$\bullet$] To confirm the trend of R$_{\rm J50}$/R$_{\rm h}$ in higher stellar mass galaxies increasing with cosmic time, we make a direct comparison to the ({\sc{eagle}})  hydrodynamical  simulation. We first test the validity of using the {\sc{eagle}} rotation curves as derived in \cite{Schaller2015} to estimate the stellar angular momentum of {\sc{eagle}} galaxies. We find good agreement between J$_{\rm RC}$ as derived from Equation \ref{Eqn:AngMom4} and the stellar particle angular momentum (J$_{\rm ps}$), suggesting that {\sc{eagle}} rotation curves can be used to accurately estimate the angular momentum of {\sc{eagle}} galaxies.

\item[$\bullet$] To compare to the observational sample we select galaxies in {\sc{eagle}} by mass and star formation rate that match the selection function of the observations. One-dimensional stellar angular momentum profiles are derived for each {\sc{eagle}} galaxy from which we measured the 50 per cent angular momentum radii (R$_{\rm J50}$). Splitting the {\sc{eagle}} sample into two stellar mass bins of 9$<$\,$\log$(M$_{*}$[M$_{\odot}$])$\leq$10 and 10$<$\,$\log$(M$_{*}$[M$_{\odot}$])$\leq$11, we identify a 16 per cent increase in R$_{\rm J50}$/R$_{\rm h}$ from $z$\,=\,3.5 to $z$\,=\,0.1 in higher stellar mass galaxies and minimal evolution in the lower stellar mass bin, as identified in the observations (Figure \ref{Fig:mfix}).

\item[$\bullet$] We note however that the selection function used in observations and {\sc{eagle}} comparison from $z$\,=\,3.5 to $z$\,=\,0.1 for the radii derived in Figure \ref{Fig:mfix}  are  not  selecting  the  same  descendent  populations. To understand how a galaxy's angular momentum distribution evolves with cosmic time we need to study galaxies as they evolve. Using the merger trees in {\sc{eagle}} we select galaxies at $z$\,=\,3 that match the selection function of our observations, and trace these galaxies through the simulation to $z$\,=\,0.1, measuring the radius containing 50 per cent of the stellar angular momentum (R$_{\rm J50}$) at each redshift snapshot (Figure \ref{Fig:mevo}). Splitting the sample into the two stellar mass bins, we identify an 11 per cent increase in R$_{\rm J50}$/R$_{\rm h}$ from $z$\,=\,3.5 to $z$\,=\,0.1 in higher stellar mass galaxies.

\item[$\bullet$] To understand the physical processes driving the increase in R$_{\rm J50}$/R$_{\rm h}$ in higher stellar mass galaxies, we explore the evolution of the stellar mass  and bulge-fraction as a function of cosmic time (Figure \ref{Fig:Mstar_BT_z}). Both high-- and low--stellar--mass galaxies show an increase in stellar mass by a factor of $\sim$10 from $z$\,=\,3.5 to $z$\,=\,0.1. The bulge fraction of galaxies with stellar mass 9$<$\,$\log$(M$_{*}$[M$_{\odot}$])$\leq$10, decreases from  B/T\,=\,0.74\,$\pm$\,0.04 at $z$\,=\,3.5 to  B/T\,=\,0.28\,$\pm$\,0.03 at $z$\,=\,1, remaining roughly constant to $z$\,=\,0.1. Higher stellar mass galaxies, those with stellar masses in the 10$<$\,$\log$(M$_{*}$[M$_{\odot}$])$\leq$11 at $z$\,=\,3, show a decrease in bulge fraction from  B/T\,=\,0.65\,$\pm$\,0.08 at $z$\,=\,3.5 to  B/T\,=\,0.35\,$\pm$\,0.04 at $z$\,=\,1.5, but with an increase below $z$\,=\,1.5 to  B/T\,=\,0.53\,$\pm$\,0.03 at $z$\,=\,0.1. The accretion of new material from the circumgalactic medium reduces the bulge fraction of both low-- and high--stellar--mass galaxies as they evolve with cosmic time. Below $z$\,=\,1 the low--mass galaxies become stable, with approximately constant bulge fractions and Sc-Sd late morphologies, whilst the higher stellar mass galaxies continue to increase their bulge fraction through the formation of pseudo--bulges, leading to more early-type morphologies.  
\newline

Overall our results show that high--stellar--mass main--sequence star--forming galaxies have a stronger evolution in angular momentum compared to low--stellar--mass galaxies. This process is likely to be driven by an internal redistribution of angular momentum from the accretion of new higher angular momentum material as well as other less dominant secular processes leading to the formation of pseudo-bulges. It is this process of redistributing the angular momentum, that coincides with changes in the galaxies' morphology, driving the galaxies towards the stable low-redshift discs that occupy the Hubble sequence. 

\end{description}
 
\section*{Acknowledgements}
We thank the anonymous referee for their comments and suggestions, which improved the content and clarity of the paper.
This work was supported by the Science and Technology Facilities 
Council (ST/L00075X/1). SG acknowledge the support of the Science and Technology
Facilities Council through grant ST/N50404X/1 for support. IRS acknowledge support from STFC (ST/P000541/1) and the ERC Advanced Grant DUSTYGAL (321334). E.I.\ acknowledges partial support from FONDECYT through grant N$^\circ$\,1171710.
JEG thanks the Royal Society for support via a University Research Fellowship. PNB is grateful for support from STFC via grants ST/M001229/1 and ST/R000972/1. ALT acknowledges support from STFC (ST/P000541/1) and ERC Advanced Grant DUSTYGAL (321334). J. M. acknowledges the support given by CONICYT Chile (CONICYT-PCHA/Doctorado-Nacional/2014-21140483). 

%The Acknowledgements section is not numbered. Here you can thank helpful
%colleagues, acknowledge funding agencies, telescopes and facilities used etc.

%%%%%%%%%%%%%%%%%%%%%%%%%%%%%%%%%%%%%%%%%%%%%%%%%%



\bibliographystyle{mnras}
\bibliography{Master} % if your bibtex file is called example.bib
% \section*{Supporting Information}

% The following appendices are available at \href{https://academic.oup.com/mnras}{MNRAS} online. \\

% \noindent
% \textbf{Appendix A: Integrated Galaxy Properties \\
% Appendix B: Integral Field Observations\\
% Appendix C: Morpho-Kinematic Properties.\\
% Appendix D: Kinematic Maps\\
% Appendix E: Beam-Smearing Correction}\\

% \noindent
% Please note: Oxford University Press is not responsible for the
% content or functionality of any supporting materials supplied by
% the authors. Any queries (other than missing material) should be
% directed to the corresponding author for the article.

%\begin{thebibliography}{99}
%\end{thebibliography}

%%%%%%%%%%%%%%%%%%%%%%%%%%%%%%%%%%%%%%%%%%%%%%%%%

%%%%%%%%%%%%%%%%% APPENDICES %%%%%%%%%%%%%%%%%%%%%
\onecolumn
\appendix
\centering
\begin{table*}
	\centering
  	\section{Integrated Galaxy Properties} \label{App:A}
 	\caption{(1) Target name,
 	(2) Previously published name, 1\,=\,\citet{Molina2017}, 2\,=\,\citet{Swinbank2012a}, 
 	(3-4) Right Ascension and Declination in J2000 coordinates, 
 	(5) Spectroscopic redshift derived from the near infra-red integral field spectrum. Galaxies at $z$\,$\leq$\,2.5 are detected in H$\alpha$, whilst those at $z$\,$\geq$\,3  have their kinematics traced by the [OIII] emission line \citet{Sobral2013,Sobral2015,Khostovan2015},
 	(6-8) Stellar properties derived using {\sc{MagPhys}} \citet{Cunha2008} using a \citet{Chabrier2003} IMF, the \citet{Calzetti2000} reddening law and either constant or exponentially declining SFRs. Uncertainties on stellar properties derived from SEDs are dominated by systematic model assumptions.}	
 	\label{Table:Sample}
	\begin{tabular}{llllccll}
		\hline
		Target & Published &R.A & Decl. & $z_{\text{spec}}$ & M$_{\text{H}}$ & $\log($M$_{\text{*}}$)  & SFR\SB{line} \\
		& Name &(J2000) & (J2000) &  & (AB mag) & (M$_{\odot}$) &  (M$_{\odot}$yr$^{-1}$) \\
        \hline
SHIZELS--5 & \textsuperscript{1}SA22--54 & 22:22:23.04 & +00:47:33.0 & 0.810 & $-$22.74 & 10.1 & 6 $\pm$ 1 \\
SHIZELS--6 & \textsuperscript{1}SA22--17 & 22:19:36.14 & +00:34:07.9 & 0.812 & $-$21.60 & 9.9 & 5 $\pm$ 2 \\
SHIZELS--13 & \textsuperscript{1}SA22--28 & 22:15:36.31 & +00:41:08.8 & 0.813 & $-$22.28 & 9.9 & 7 $\pm$ 1 \\
SHIZELS--15 & \textsuperscript{1}SA22-26 & 22:18:23.00 & +01:00:22.1 & 0.815 & $-$22.11 & 9.7 & 6 $\pm$ 2 \\
SHIZELS--4 & \textsuperscript{2}SHIZELS--4 & 10:01:55.29 & +02:14:03.3 & 0.830 & $-$20.88 & 9.2 & 2 $\pm$ 1 \\
SHIZELS--1 & \textsuperscript{2}SHIZELS--1 & 02:18:26.31 & $-$04:47:01.6 & 0.843 & $-$22.27 & 10.1 & 6 $\pm$ 1 \\
SHIZELS--16 & -- & 02:17:42.35 & $-$05:15:05.1 & 1.339 & $-$ & 10.4 & 17 $\pm$ 2 \\
SHIZELS--17 & \textsuperscript{1}COS--16 & 10:00:49.01 & +02:44:41.1 & 1.360 & $-$22.19 & 9.5 & 9 $\pm$ 3 \\
SHIZELS--10 & \textsuperscript{2}SHIZELS--10 & 02:17:39.02 & $-$04:44:41.4 & 1.447 & $-$22.62 & 10.1 & 9 $\pm$ 2 \\
SHIZELS--7 & \textsuperscript{2}SHIZELS--7 & 02:17:00.34 & $-$05:01:50.6 & 1.455 & $-$23.32 & 10.6 & 12 $\pm$ 1 \\
SHIZELS--8 & \textsuperscript{2}SHIZELS--8 & 02:18:20.96 & $-$05:19:07.5 & 1.460 & $-$23.66 & 10.3 & 16 $\pm$ 2 \\
SHIZELS--9 & \textsuperscript{2}SHIZELS--9\ & 02:17:12.99 & $-$04:54:40.7 & 1.462 & $-$24.01 & 10.8 & 26 $\pm$ 2 \\
SHIZELS--12 & \textsuperscript{2}SHIZELS--12\ & 02:19:01.45 & $-$04:58:15.0 & 1.467 & $-$23.90 & 10.7 & 21 $\pm$ 2 \\
SHIZELS--18 & -- & 02:17:34.20 & $-$05:10:16.7 & 1.470 & $-$22.34 & 10.3 & 49 $\pm$ 2 \\
SHIZELS--19 & \textsuperscript{1}COS--30 & 09:59:11.57 & +02:23:24.3 & 1.486 & $-$24.01 & 10.3 & 13 $\pm$ 2 \\
SHIZELS--11 & \textsuperscript{2}SHIZELS--11 & 02:18:21.23 & $-$05:02:48.9 & 1.492 & $-$25.69 & 10.9 & 23 $\pm$ 2 \\
SHIZELS--20 & -- & 09:59:37.96 & +02:18:02.1 & 1.620 & $-$22.35 & 10.8 & 33 $\pm$ 2 \\
SHIZELS--2 & -- & 02:19:25.50 & $-$04:54:39.6 & 2.223 & $-$22.14 & 9.8 & 18 $\pm$ 6 \\
SHIZELS--3 & --& 10:00:27.69 & +02:14:30.6 & 2.225 & $-$21.25 & 9.0 & 21 $\pm$ 3 \\
SHIZELS--21 & \textsuperscript{1}UDS--10 & 02:16:45.82 & $-$05:02:45.0 & 2.237 & $-$23.38 & 9.7 & 37 $\pm$ 4 \\
SHIZELS--22 & \textsuperscript{1}SA22--01 & 22:19:16.06 & +00:40:36.1 & 2.238 & $-$23.57 & 10.2 & 34 $\pm$ 2 \\
SHIZELS--23 & \textsuperscript{1}UDS--21 & 02:16:49.05 & $-$05:03:20.8 & 2.239 & $-$22.29 & 10.2 & 26 $\pm$ 5 \\
SHIZELS--24 & \textsuperscript{1}UDS--17 & 02:16:55.32 & $-$05:23:35.5 & 2.241 & $-$24.46 & 9.8 & 60 $\pm$ 3 \\
SHIZELS--14 & \textsuperscript{2}SHIZELS--14 & 10:00:51.58 & +02:33:34.1 & 2.242 & $-$25.35 & 9.5 & 81 $\pm$ 3 \\
SHIZELS--25 & \textsuperscript{1}SA22--02 & 22:18:58.93 & +00:05:58.3 & 2.253 & $-$23.48 & 10.4 & 40 $\pm$ 2 \\
SHIZELS--26 & -- & 02:17:03.88 & $-$05:16:19.5 & 3.227 & $-$24.74 & 10.9 & 28 $\pm$ 17 \\
SHIZELS--27 & -- & 09:57:59.05 & +02:38:19.7 & 3.238 & $-$22.35 & 9.3 & 17 $\pm$ 10 \\
SHIZELS--28 & -- & 02:18:21.37 & $-$05:19:16.7 & 3.252 & $-$23.42 & 9.9 & 26 $\pm$ 15 \\
SHIZELS--29 & -- & 09:59:28.00 & +02:44:34.0 & 3.253 & $-$22.35 & 9.7 & 92 $\pm$ 55 \\
SHIZELS--30 & -- & 09:59:20.40 & +02:25:21.1 & 3.256 & $-$19.82 & 9.4 & 39 $\pm$ 23 \\
SHIZELS--30 & -- & 09:59:36.39 & +02:17:44.0 & 3.263 & $-$20.60 & 9.3 & 14 $\pm$ 8 \\
SHIZELS--32 & -- & 02:17:45.85 & $-$05:25:45.4 & 3.273 & $-$22.00 & 10.5 & 113 $\pm$ 2 \\
SHIZELS--33 & -- & 9:57:51.526 & +02:36:37.9 & 3.278 & $-$24.42 & 10.5 & 121 $\pm$ 2 \\
SHIZELS--34 & -- & 02:17:11.66 & $-$04:54:44.7 & 3.300 & $-$23.03 & 10.3 & 53 $\pm$ 32 \\
\hline
	\end{tabular}
\end{table*}

\newpage

\begin{table*}
	\centering
	\section{Integral Field Observations.}\label{App:B}
 	\caption{(1) Target Name as per Table \ref{Table:Sample},
			(2) Spectroscopic H$\alpha$ or [O{\sc{iii}}] redshift derived from spectrum,
			(3-5) Extra-galactic Field, Wavelength band and Integral Field Spectrograph used for spectroscopic observation, 
			* = Laser Guide Star (LGS), otherwise Natural Guide Star (NGS),
            (6) Total on source integration time of integral field observations,
            (7) Integral field PSF size as measured from standard star observations in kpc,
            (8) Ancillary photometric data available for each target.}
	\begin{tabular}{llllllllll}
		\hline
		Target & $z_{\text{spec}}$ & Extra-galactic Field & Band & IFU & t$_{\text{exp}}$ & PSF R$_{\rm h}$ & Broadband \\
		&  & & & & (ks) & (kpc) &  \\
		\hline
SHIZELS--5 & 0.810 & SA22 & $J$ & SINFONI & 4.8 & 1.40 & UKIDSS $K$ \\
SHIZELS--6 & 0.812 & SA22 & $J$ & SINFONI & 4.8 & 1.40 & UKIDSS $K$ \\
SHIZELS--13 & 0.813 & SA22 & $J$ & SINFONI & 4.8 & 1.40 & UKIDSS $K$ \\
SHIZELS--15 & 0.815 & SA22 & $J$ & SINFONI & 4.8 & 1.40 & UKIDSS $K$ \\
SHIZELS--4 & 0.830 & COSMOS & $J$ & SINFONI & 7.2 & 1.41 & \emph{HST} F160W, F814W \\
SHIZELS--1 & 0.843 & UDS & $J$ & SINFONI & 7.2 & 1.42 & UKIDSS $K$ \\
SHIZELS--16 & 1.339 & UDS & $H$ & OSIRIS* & 7.2 & 1.20 & \emph{HST} F125W, F160W, F814W \\
SHIZELS--17 & 1.360 & COSMOS & $H$ & SINFONI & 7.2 & 1.20 & \emph{HST} F814W \\
SHIZELS--10 & 1.447 & UDS & $H$ & SINFONI & 9.6 & 1.20 & \emph{HST} F140W, F606W \\
SHIZELS--7 & 1.455 & UDS & $H$ & SINFONI & 9.6 & 1.20 & \emph{HST} F140W, F606W \\
SHIZELS--8 & 1.460 & UDS & $H$ & SINFONI & 7.2 & 1.20 & \emph{HST} F140W, F606W \\
SHIZELS--9 & 1.462 & UDS & $H$ & SINFONI & 9.6 & 1.20 & \emph{HST} F140W, F606W \\
SHIZELS--12 & 1.467 & UDS & $H$ & SINFONI & 9.6 & 1.20 & UKIDSS $K$ \\
SHIZELS--18 & 1.470 & UDS & $H$ & OSIRIS* & 7.2 & 1.20 & \emph{HST} F125W, F160W, F814W \\
SHIZELS--19 & 1.486 & COSMOS & $H$ & SINFONI & 7.2 & 1.20 & \emph{HST} F160W, F814W \\
SHIZELS--11 & 1.492 & UDS & $H$ & SINFONI & 7.2 & 1.20 & \emph{HST} F140W, F606W \\
SHIZELS--20 & 1.620 & COSMOS & $H$ & OSIRIS* & 7.2 & 1.21 & \emph{HST} F814W \\
SHIZELS--2 & 2.223 & UDS & $K$ & SINFONI & 14.4 & 0.74 & \emph{HST} F140W, F606W \\
SHIZELS--3 & 2.225 & COSMOS & $K$ & SINFONI & 4.8 & 0.74 & \emph{HST} F140W, F606W \\
SHIZELS--21 & 2.237 & UDS & $K$ & 2 NIFS \& SINFONI & 40.8 & 0.73 & \emph{HST} F140W, F606W \\
SHIZELS--22 & 2.238 & SA22 & $K$ & SINFONI & 9.6 & 0.73 & UKIDSS $K$ \\
SHIZELS--23 & 2.239 & UDS & $K$ & NIFS \& SINFONI & 27.6 & 0.73 & UKIDSS $K$ \\
SHIZELS--24 & 2.241 & UDS & $K$ & NIFS \& SINFONI & 27.6 & 0.73 & UKIDSS $K$ \\
SHIZELS--14 & 2.242 & COSMOS & $K$ & SINFONI & 12.0 & 0.73 & \emph{HST} F140W, F606W, F814W \\
SHIZELS--25 & 2.253 & SA22 & $K$ & SINFONI & 9.6 & 0.73 & UKIDSS $K$ \\
SHIZELS--26 & 3.227 & UDS & $K$ & SINFONI & 7.2 & 0.67 & \emph{HST} F125W, F160W, F814W \\
SHIZELS--27 & 3.238 & COSMOS & $K$ & SINFONI & 19.8 & 0.67 & \emph{HST} F814W \\
SHIZELS--28 & 3.252 & UDS & $K$ & SINFONI & 10.8 & 0.67 & UKIDSS $K$ \\
SHIZELS--29 & 3.253 & COSMOS & $K$ & SINFONI & 9.6 & 0.67 & \emph{HST} F814W \\
SHIZELS--30 & 3.256 & COSMOS & $K$ & SINFONI & 2.4 & 0.67 & \emph{HST} F814W \\
SHIZELS--30 & 3.263 & COSMOS & $K$ & SINFONI & 13.2 & 0.67 & \emph{HST} F160W, F814W \\
SHIZELS--32 & 3.273 & UDS & $K$ & SINFONI & 2.4 & 0.67 & UKIDSS $K$ \\
SHIZELS--33 & 3.278 & COSMOS & $K$ & SINFONI & 2.4 & 0.67 & \emph{HST} F814W \\
SHIZELS--34 & 3.300 & UDS & $K$ & SINFONI & 2.4 & 0.67 & UKIDSS $K$ \\
		\hline
	\end{tabular}
	\label{Tab:Obs}
	\end{table*}
\newpage

\begin{table*}
\centering
\section{Morpho-Kinematic Properties.} \label{App:C}
 \caption{(1) Target Name as per Table \ref{Table:Sample},
          (2-10) Morphological and Kinematic properties derived for our sample,
          (11) Qualify flag based on kinematic criteria (Section \ref{Sec:Qual}).}
\label{Table:sizes}
\begin{tabular}{lccclllllcc}
\hline	 
Target & R\SPSB{Sersic}{h} & S\'ersic Index & Axis Ratio & $\theta_{\rm inc}$ & PA$_{\text{vel}}$ & Vrot$_{\rm 2Rh}$ & \underline{Vcirc$_{\rm 2Rh}$} & $\sigma_{\text{median}}$ & \underline{Vrot$_{\rm 2Rh}$} & Quality   \\
   & (kpc) & (n) & & (deg) & (deg) & (km\,s$^{-1}$) & Vrot$_{\rm 2Rh}$  & (km\,s$^{-1}$) & $\sigma_{\text{median}}$ & Flag\\
\hline
SHIZELS--5 & 4.4 $\pm$ 1.5 & 0.9 $\pm$ 0.3 & 0.6 $\pm$ 0.1 & 52 $\pm$ 6 & 109 $\pm$ 44 & 101 $\pm$ 56 & 2.31 & 93 $\pm$ 9 & 1.1 $\pm$ 0.6 & 2 \\
SHIZELS--6 & 4.3 $\pm$ 2.1 & 0.6 $\pm$ 0.2 & 0.9 $\pm$ 0.1 & 31 $\pm$ 12 & 91 $\pm$ 16 & 22 $\pm$ 22 & 4.85 & 44 $\pm$ 4 & 0.5 $\pm$ 0.5 & 3 \\
SHIZELS--13 & 5.1 $\pm$ 1.2 & 0.9 $\pm$ 0.2 & 0.7 $\pm$ 0.1 & 47 $\pm$ 4 & 147 $\pm$ 30 & 29 $\pm$ 26 & 5.8 & 71 $\pm$ 7 & 0.4 $\pm$ 0.4 & 2 \\
SHIZELS--15 & 3.3 $\pm$ 2.6 & 0.6 $\pm$ 0.4 & 0.8 $\pm$ 0.2 & 33 $\pm$ 16 & 145 $\pm$ 7 & 77 $\pm$ 20 & 1.64 & 47 $\pm$ 4 & 1.6 $\pm$ 0.5 & 2 \\
SHIZELS--4 & 4.0 $\pm$ 2.9 & 0.9 $\pm$ 0.2 & 0.5 $\pm$ 0.1 & 62 $\pm$ 3 & 38 $\pm$ 46 & 21 $\pm$ 31 & 11.68 & 106 $\pm$ 10 & 0.2 $\pm$ 0.3 & 2 \\
SHIZELS--1 & 2.8 $\pm$ 0.2 & 1.2 $\pm$ 0.1 & 0.7 $\pm$ 0.1 & 47 $\pm$ 1 & 21 $\pm$ 65 & 98 $\pm$ 37 & 2.28 & 86 $\pm$ 8 & 1.1 $\pm$ 0.5 & 1 \\
SHIZELS--16 & 4.1 $\pm$ 0.8 & 1.6 $\pm$ 0.2 & 0.5 $\pm$ 0.1 & 60 $\pm$ 3 & 97 $\pm$ 34 & 63 $\pm$ 54 & 2.53 & 71 $\pm$ 7 & 0.9 $\pm$ 0.8 & 3 \\
SHIZELS--17 & 1.7 $\pm$ 1.0 & 2.0 $\pm$ 0.4 & 0.4 $\pm$ 0.1 & 67 $\pm$ 4 & 103 $\pm$ 46 & 25 $\pm$ 23 & 7.94 & 87 $\pm$ 8 & 0.3 $\pm$ 0.3 & 3 \\
SHIZELS--10 & 2.8 $\pm$ 0.6 & 2.0 $\pm$ 0.1 & 0.5 $\pm$ 0.2 & 61 $\pm$ 1 & 105 $\pm$ 23 & 30 $\pm$ 12 & 5.2 & 65 $\pm$ 6 & 0.5 $\pm$ 0.2 & 3 \\
SHIZELS--7 & 4.9 $\pm$ 0.5 & 1.4 $\pm$ 0.1 & 0.7 $\pm$ 0.1 & 44 $\pm$ 1 & 154 $\pm$ 59 & 159 $\pm$ 69 & 1.4 & 70 $\pm$ 7 & 2.3 $\pm$ 1.0 & 1 \\
SHIZELS--8 & 5.7 $\pm$ 0.4 & 0.6 $\pm$ 0.1 & 0.9 $\pm$ 0.1 & 28 $\pm$ 1 & 125 $\pm$ 20 & 143 $\pm$ 33 & 1.22 & 69 $\pm$ 6 & 2.1 $\pm$ 0.5 & 1 \\
SHIZELS--9 & 5.9 $\pm$ 0.6 & 0.8 $\pm$ 0.1 & 0.7 $\pm$ 0.1 & 46 $\pm$ 2 & 71 $\pm$ 3 & 125 $\pm$ 45 & 1.57 & 67 $\pm$ 6 & 1.8 $\pm$ 0.7 & 2 \\
SHIZELS--12 & 4.9 $\pm$ 0.2 & 0.6 $\pm$ 0.2 & 0.9 $\pm$ 0.1 & 31 $\pm$ 1 & 50 $\pm$ 31 & 379 $\pm$ 154 & 1.17 & 87 $\pm$ 8 & 4.4 $\pm$ 1.9 & 2 \\
SHIZELS--18 & 4.4 $\pm$ 0.5 & 0.6 $\pm$ 0.2 & 0.7 $\pm$ 0.1 & 47 $\pm$ 2 & 122 $\pm$ 12 & 68 $\pm$ 25 & 4.05 & 111 $\pm$ 11 & 0.6 $\pm$ 0.2 & 2 \\
SHIZELS--19 & 2.1 $\pm$ 0.5 & 0.9 $\pm$ 0.1 & 0.7 $\pm$ 0.2 & 45 $\pm$ 1 & 16 $\pm$ 6 & 96 $\pm$ 18 & 3.24 & 119 $\pm$ 11 & 0.8 $\pm$ 0.2 & 2 \\
SHIZELS--11 & 5.5 $\pm$ 0.6 & 2.0 $\pm$ 0.1 & 0.9 $\pm$ 0.1 & 26 $\pm$ 3 & 58 $\pm$ 19 & 174 $\pm$ 134 & 1.44 & 88 $\pm$ 8 & 1.9 $\pm$ 1.6 & 1 \\
SHIZELS--20 & 4.7 $\pm$ 3.1 & 1.0 $\pm$ 0.1 & 0.6 $\pm$ 0.1 & 58 $\pm$ 3 & 127 $\pm$ 18 & 159 $\pm$ 60 & 1.86 & 104 $\pm$ 10 & 1.5 $\pm$ 0.6 & 1 \\
SHIZELS--2 & 1.2 $\pm$ 0.2 & 0.9 $\pm$ 0.1 & 0.8 $\pm$ 0.3 & 39 $\pm$ 1 & 148 $\pm$ 10 & 54 $\pm$ 7 & 2.99 & 62 $\pm$ 6 & 0.9 $\pm$ 0.2 & 2 \\
SHIZELS--3 & 2.7 $\pm$ 0.7 & 2.0 $\pm$ 0.1 & 0.7 $\pm$ 0.3 & 49 $\pm$ 1 & 17 $\pm$ 71 & 38 $\pm$ 28 & 3.31 & 50 $\pm$ 5 & 0.7 $\pm$ 0.6 & 2 \\
SHIZELS--21 & 5.8 $\pm$ 1.1 & 2.0 $\pm$ 0.3 & 0.5 $\pm$ 0.1 & 59 $\pm$ 4 & 39 $\pm$ 3 & 38 $\pm$ 25 & 5.78 & 97 $\pm$ 9 & 0.4 $\pm$ 0.3 & 2 \\
SHIZELS--22 & 3.5 $\pm$ 3.4 & 0.8 $\pm$ 0.6 & 0.8 $\pm$ 0.2 & 34 $\pm$ 20 & 135 $\pm$ 48 & 16 $\pm$ 20 & 3.66 & 71 $\pm$ 7 & 0.2 $\pm$ 0.3 & 2 \\
SHIZELS--23 & 3.6 $\pm$ 1.0 & 1.2 $\pm$ 0.2 & 0.6 $\pm$ 0.1 & 58 $\pm$ 2 & 24 $\pm$ 80 & 63 $\pm$ 13 & 2.87 & 69 $\pm$ 6 & 0.9 $\pm$ 0.2 & 2 \\
SHIZELS--24 & 6.2 $\pm$ 2.0 & 2.0 $\pm$ 0.2 & 0.9 $\pm$ 0.1 & 28 $\pm$ 2 & 26 $\pm$ 64 & 82 $\pm$ 41 & 2.31 & 101 $\pm$ 10 & 0.8 $\pm$ 0.4 & 3 \\
SHIZELS--14 & 4.5 $\pm$ 0.7 & 1.6 $\pm$ 0.1 & 0.5 $\pm$ 0.1 & 58 $\pm$ 3 & 74 $\pm$ 21 & 90 $\pm$ 40 & 3.06 & 143 $\pm$ 14 & 0.6 $\pm$ 0.3 & 3 \\
SHIZELS--25 & 4.8 $\pm$ 3.0 & 0.6 $\pm$ 0.2 & 0.6 $\pm$ 0.2 & 54 $\pm$ 10 & 50 $\pm$ 35 & 86 $\pm$ 33 & 3.01 & 87 $\pm$ 8 & 1.0 $\pm$ 0.4 & 3 \\
SHIZELS--26 & 2.2 $\pm$ 2.2 & 2.0 $\pm$ 0.2 & 0.6 $\pm$ 0.1 & 55 $\pm$ 2 & 164 $\pm$ 58 & 127 $\pm$ 24 & 4.43 & 221 $\pm$ 22 & 0.6 $\pm$ 0.1 & 3 \\
SHIZELS--27 & 2.1 $\pm$ 1.4 & 0.6 $\pm$ 0.2 & 0.2 $\pm$ 0.1 & 90 $\pm$ 35 & 151 $\pm$ 43 & 52 $\pm$ 8 & 4.01 & 79 $\pm$ 7 & 0.7 $\pm$ 0.1 & 2 \\
SHIZELS--28 & 3.0 $\pm$ 6.2 & 2.0 $\pm$ 0.4 & 0.8 $\pm$ 0.2 & 33 $\pm$ 13 & 142 $\pm$ 14 & 24 $\pm$ 36 & 4.52 & 68 $\pm$ 6 & 0.4 $\pm$ 0.5 & 3 \\
SHIZELS--29 & 0.9 $\pm$ 0.5 & 1.5 $\pm$ 0.2 & 0.4 $\pm$ 0.3 & 72 $\pm$ 8 & 162 $\pm$ 29 & 37 $\pm$ 11 & 5.54 & 83 $\pm$ 8 & 0.4 $\pm$ 0.1 & 2 \\
SHIZELS--30 & 1.8 $\pm$ 2.4 & 0.6 $\pm$ 0.1 & 0.6 $\pm$ 0.1 & 53 $\pm$ 2 & 151 $\pm$ 21 & 17 $\pm$ 14 & 12.91 & 111 $\pm$ 11 & 0.2 $\pm$ 0.1 & 3 \\
SHIZELS--30 & 1.1 $\pm$ 0.5 & 2.0 $\pm$ 0.4 & 0.6 $\pm$ 0.1 & 58 $\pm$ 2 & 27 $\pm$ 51 & 24 $\pm$ 10 & 7.91 & 76 $\pm$ 7 & 0.3 $\pm$ 0.1 & 2 \\
SHIZELS--32 & 2.8 $\pm$ 0.1 & 1.9 $\pm$ 0.2 & 0.9 $\pm$ 0.1 & 28 $\pm$ 3 & 36 $\pm$ 31 & 48 $\pm$ 42 & 2.83 & 40 $\pm$ 7 & 1.2 $\pm$ 1.2 & 2 \\
SHIZELS--33 & 0.4 $\pm$ 2.1 & 2.0 $\pm$ 0.4 & 0.6 $\pm$ 0.3 & 55 $\pm$ 10 & 81 $\pm$ 12 & 64 $\pm$ 37 & 11.62 & 314 $\pm$ 31 & 0.2 $\pm$ 0.1 & 3 \\
SHIZELS--34 & 2.5 $\pm$ 0.9 & 1.7 $\pm$ 0.4 & 0.7 $\pm$ 0.1 & 45 $\pm$ 4 & 36 $\pm$ 35 & 128 $\pm$ 54 & 2.33 & 108 $\pm$ 10 & 1.2 $\pm$ 0.5 & 1 \\
\hline
\end{tabular}
\end{table*}

\FloatBarrier

\begin{figure*}
\section{Kinematic Maps} \label{App:kin}
\includegraphics[width=1.1\linewidth]{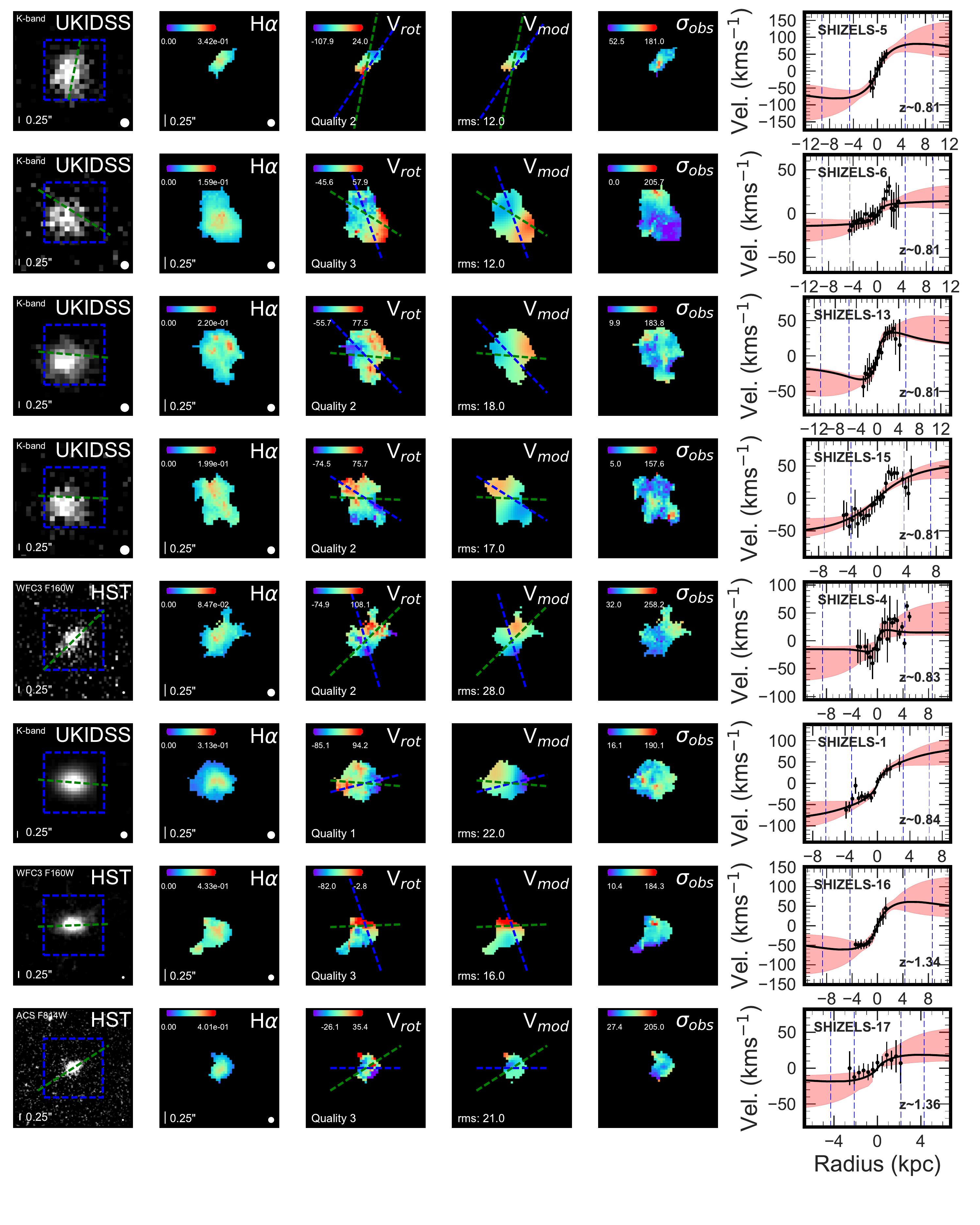}
\end{figure*}
\FloatBarrier

\begin{figure*}
\includegraphics[width=1.1\linewidth]{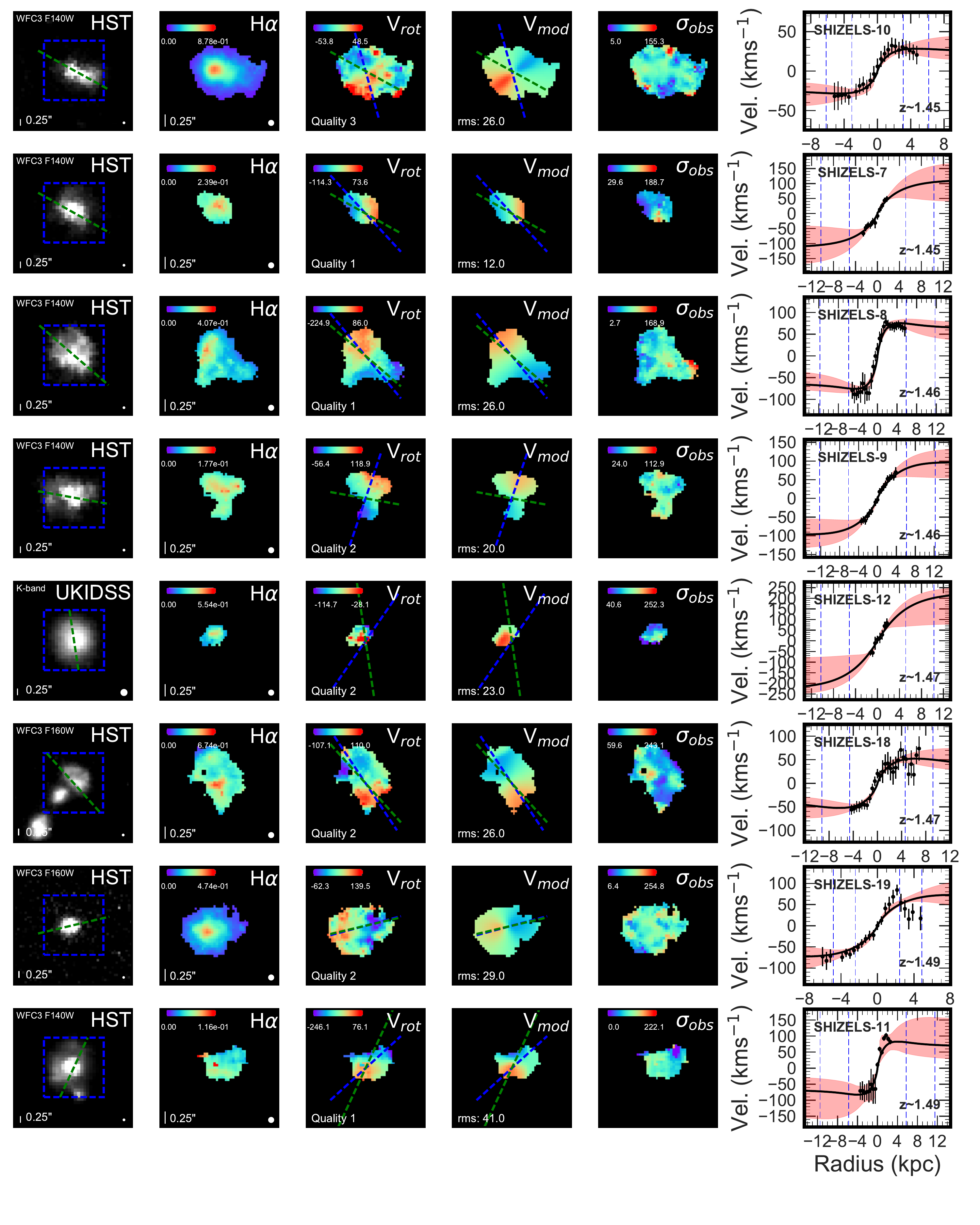}
\end{figure*}
\FloatBarrier

\begin{figure*}
\includegraphics[width=1.1\linewidth]{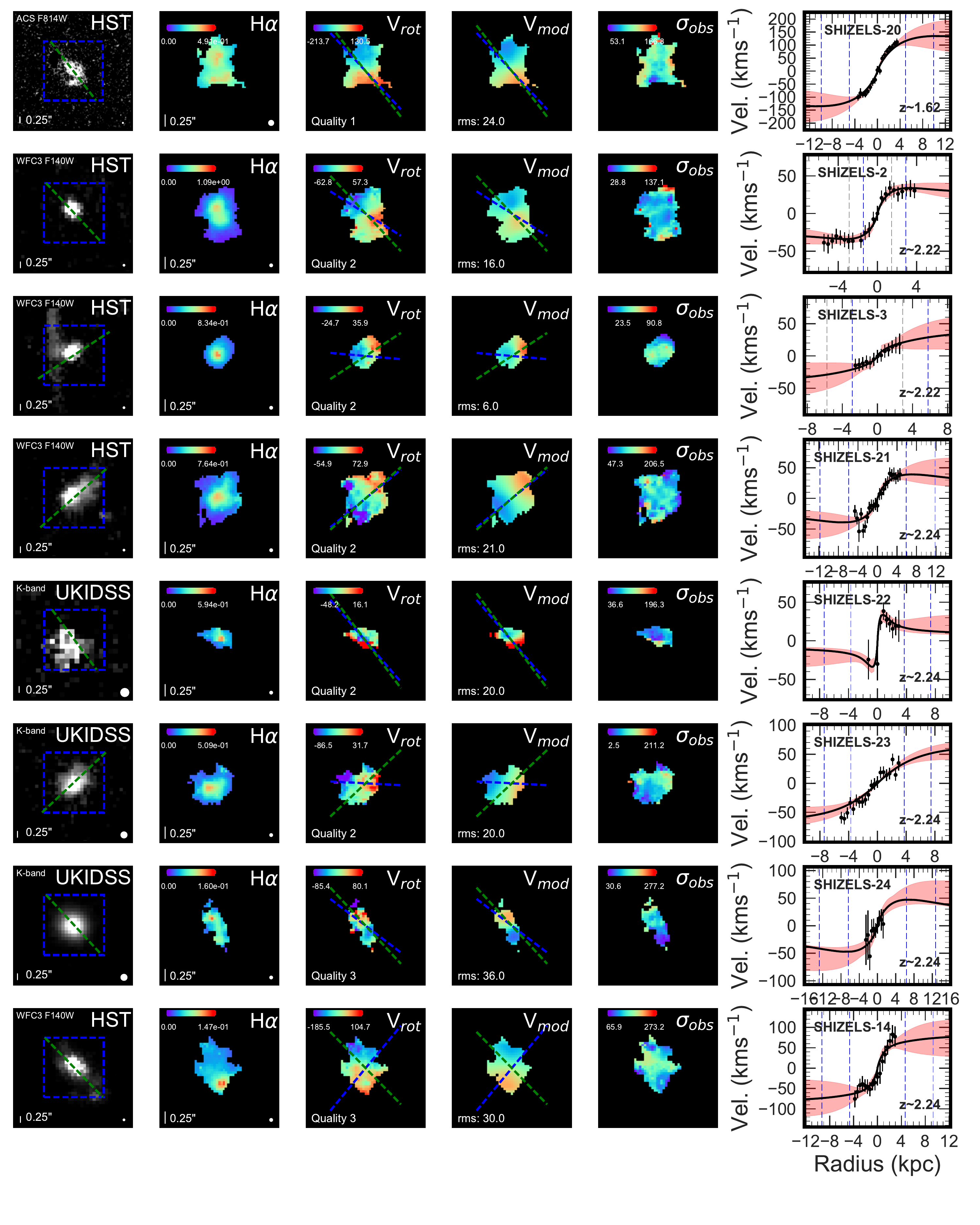}
\end{figure*}
\FloatBarrier

\begin{figure*}
\includegraphics[width=1.1\linewidth]{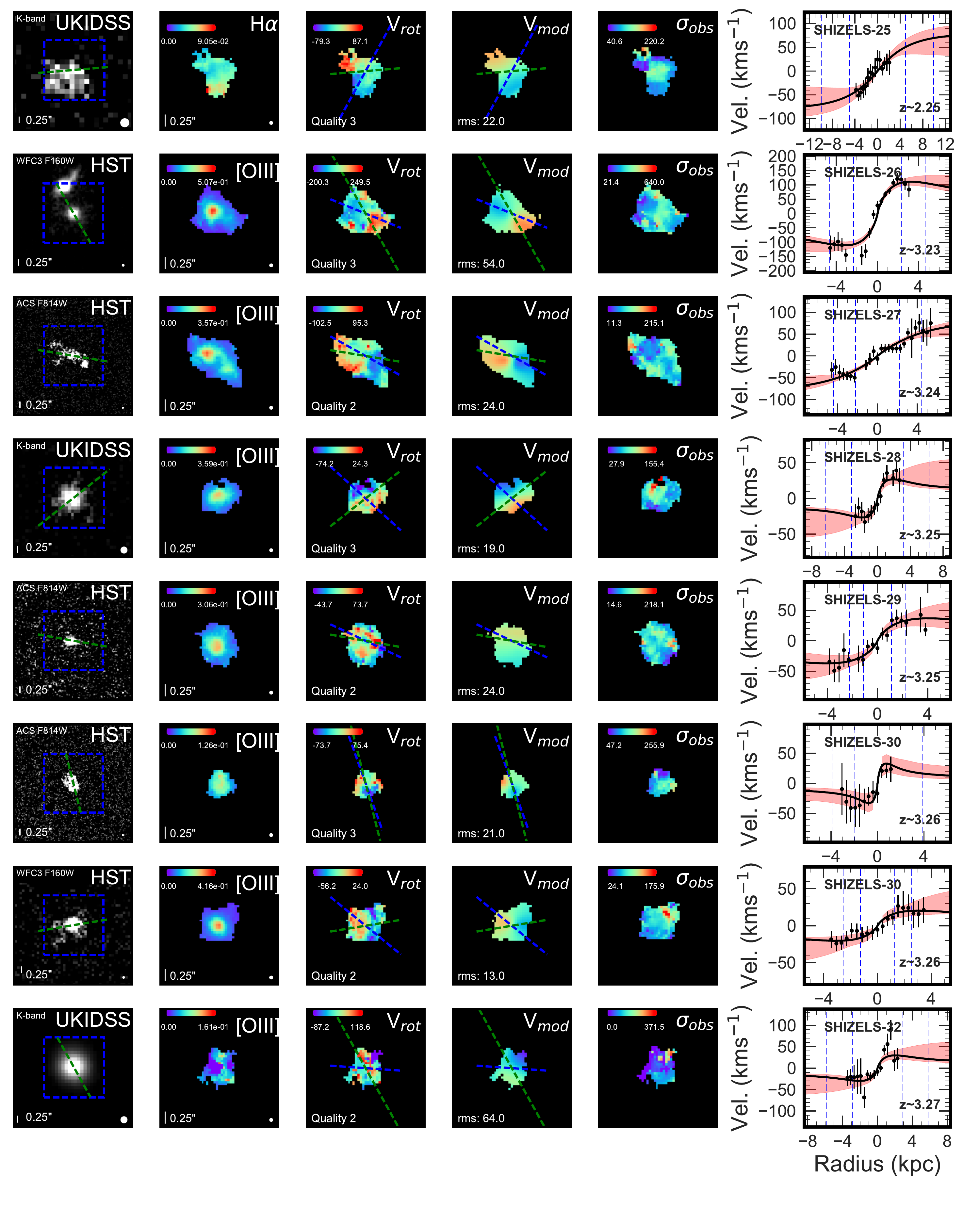}
\end{figure*}
\FloatBarrier

\includegraphics[width=1.1\linewidth,trim={0 27cm 0 0},clip,keepaspectratio]{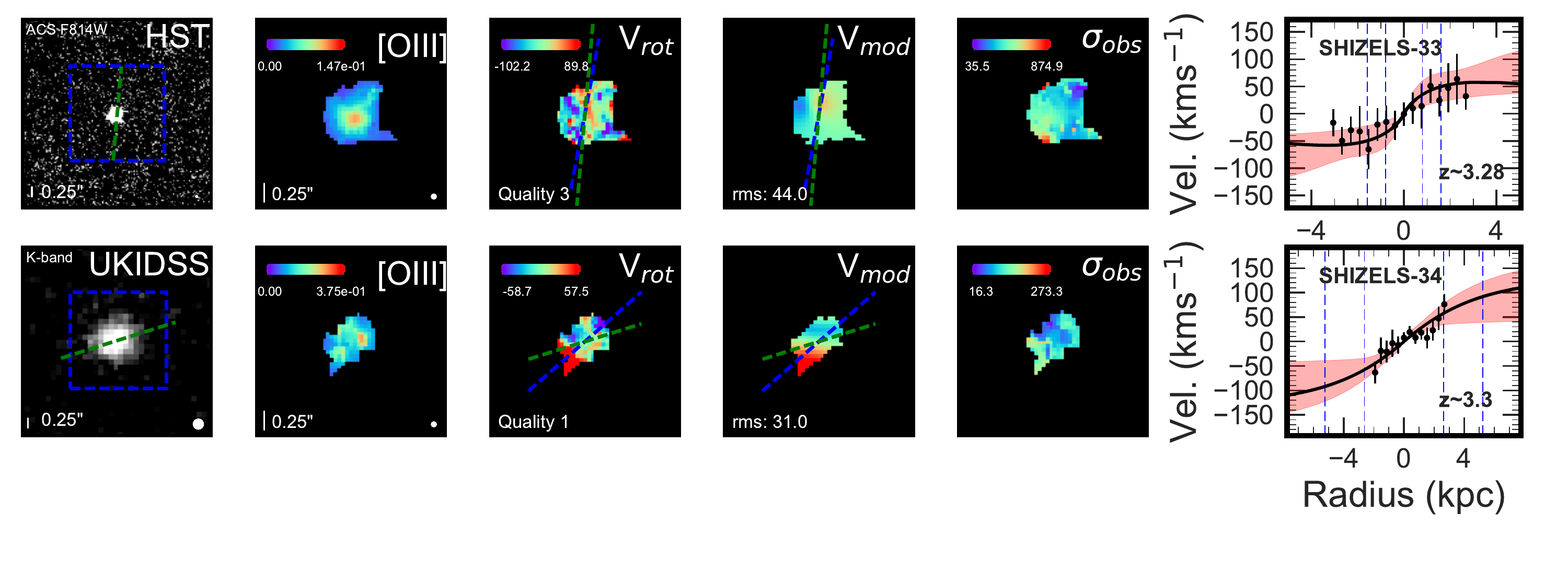}
\renewcommand\thefigure{D1}
\captionof{figure}{The spatially resolved galaxies in our sample order by redshift. From left to right; Broadband photometry of the galaxy (left), with PA$_{\rm im}$ (green dashed line) and data cube field of view (blue dashed square). H$\alpha$ or [O{\sc{iii}}] flux map, velocity map, velocity model and velocity dispersion map, derived from the emission line fitting. PA$_{\rm vel}$  (blue dashed line) and PA$_{\rm im}$ (green dashed line) axes plotted on the velocity map and model. Rotation curve extracted about kinematic position axis (right). Rotation curve shows lines of R$_\text{h}$ and 2R$_\text{h}$ derived from S\'ersic fitting, as well 1$\sigma$ error region (red) of rotation curve fit (black line).}
\label{Fig:K1}
\FloatBarrier

\centering
\section{Beam-Smearing Correction} \label{App:BS}
\includegraphics[width=1\linewidth]{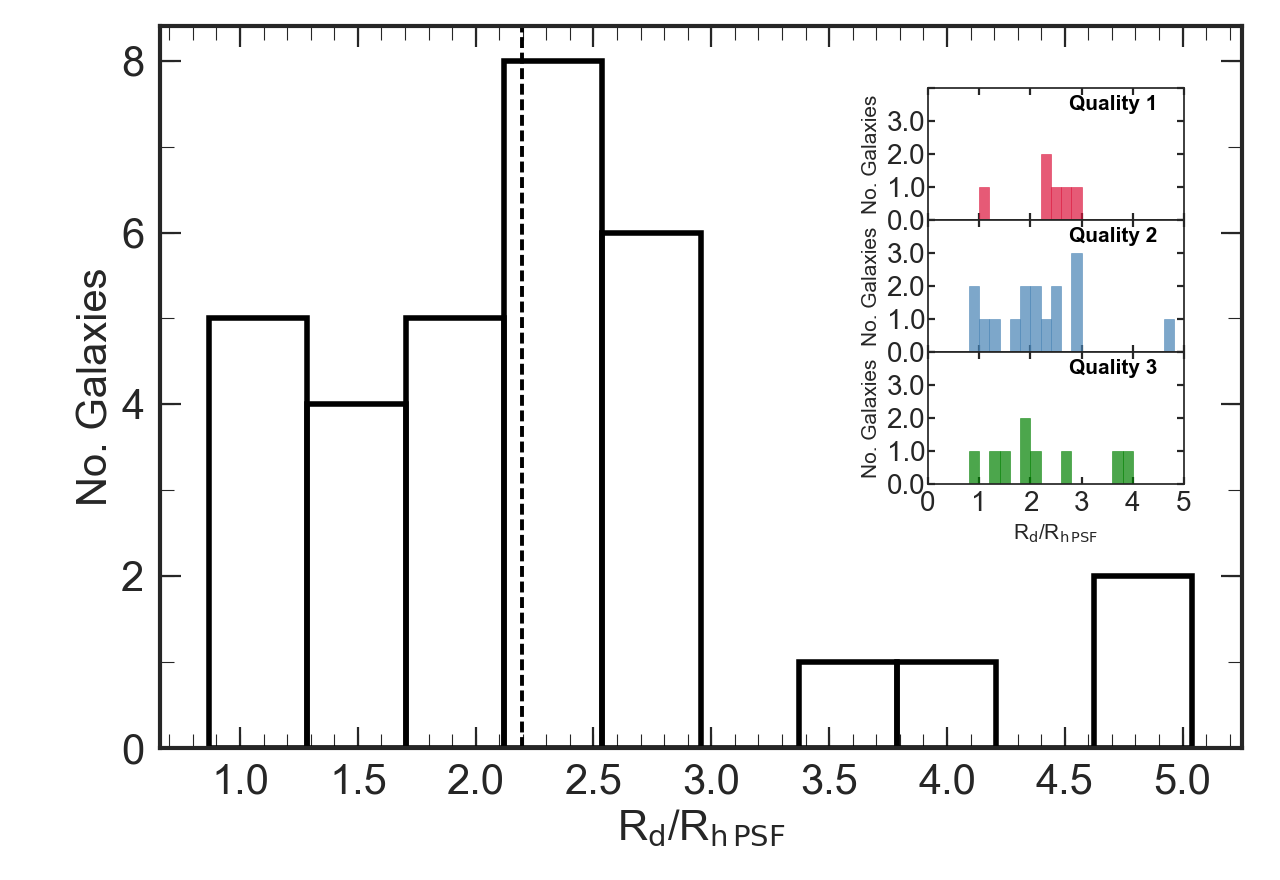}
\renewcommand\thefigure{E1}
\captionof{figure}{The ratio of $\rm R_d/R_{h\,PSF}$ for each galaxy in the sample, as well as for the individual kinematic classes. The median ratio of the sample, black dashed line, is $\langle$\,R$_{\rm d}$/R$_{\rm h\,PSF}$\,$\rangle$\,=\,2.17\,$\pm$\,0.18. For the sample the median ratio of rotation velocity is $\frac{v_{out}}{v_0}$\,=\,0.99, ranging from $\frac{v_{out}}{v_0}$\,=\,0.89\,--\,1.00 whilst the median ratio of velocity dispersion is $\frac{v_{out}}{v_0}$\,=\,1.04, ranging from $\frac{v_{out}}{v_0}$\,=\,1.00\,--\,1.11.}
\label{Fig:BS}
% \end{figure*}

\FloatBarrier

% %If you want to present additional material which would interrupt the flow of the main paper,
% %it can be placed in an Appendix which appears after the list of references.

%%%%%%%%%%%%%%%%%%%%%%%%%%%%%%%%%%%%%%%%%%%%%%%%%%

% Don't change these lines
%\bsp	% typesetting comment
\label{lastpage}
\end{document}